\documentclass[aps,11pt,prd,groupedaddress,nofootinbib,notitlepage,eqsecnum,preprintnumbers]{revtex4-2}
\usepackage[utf8]{inputenc}
\usepackage{hyperref}
\usepackage{xcolor}
\usepackage{graphicx}
\usepackage{amsmath,amssymb}
\usepackage{bm}
\usepackage{comment}
\usepackage[shortlabels]{enumitem}
\usepackage{overpic}
\usepackage{ulem}

\begin{document}
\title{An exact model for enhancing/suppressing primordial fluctuations}

\author{\textsc{Guillem Dom\`enech$^{a}$}}
    \email{{guillem.domenech}@{itp.uni-hannover.de}}
\author{\textsc{Gerson Vargas$^{b}$}}
    \email{{gerson.vargas}@{unmsm.edu.pe}}
\author{\textsc{Teófilo Vargas$^{b}$}}
    \email{{tvargasa}@{unmsm.edu.pe}}

\affiliation{$^a$Institute for Theoretical Physics, Leibniz University Hannover, Appelstraße 2, 30167 Hannover, Germany.}
\affiliation{$^b$ Grupo de Física Teórica and Grupo de Astronomía SPACE, Universidad Nacional Mayor de San Marcos, Avenida Venezuela s/n Cercado de Lima, 15081, Lima, Perú.}

\begin{abstract}
Enhancements of primordial curvature fluctuations in single field inflation often involve departures from attractor trajectories in the phase space. We study enhancement/suppression of primordial fluctuations in one of the simplest models with exact background solutions for arbitrary initial conditions: a single field inflationary model with a piecewise exponential potential. We then present close to exact analytical solutions for primordial fluctuations in a general transition between two slow-roll attractors, valid whether the first slow parameter increases or decreases. The main features in the primordial spectrum are determined by the ratio of exponents of the potential. We also discuss the imprint of such features in the induced GW spectrum. Lastly, we apply the $\delta N$ formalism to discuss non-Gaussianities and the tail of the probability distribution. We find that while non-Gaussianities are at most ${\cal O}(1)$ in the case of enhancement, they can be very large in the case of suppression. Our work can be easily generalized to multiple piecewise exponential potentials.
\end{abstract}

\maketitle

\section{Introduction \label{sec:intro}}

There is strong evidence that primordial density fluctuations were generated during a period of cosmic inflation in the very early universe \cite{Planck:2018vyg,Akrami:2018odb}. The amplitude of the primordial spectrum of curvature fluctuations, as measured by Cosmic Microwave Background (CMB) observations, is around $10^{-9}$ on the largest observable scales, and it is almost scale invariant \cite{Planck:2018vyg,Akrami:2018odb}. The spectrum of fluctuations on small scales though, which were generated towards the end of inflation, is for the moment mostly unconstrained. Two promising probes of small scale primordial fluctuations are Primordial Black Holes (PBHs) \cite{Hawking:1971ei,Carr:1974nx,Carr:1975qj,Khlopov:1985jw} and induced Gravitational Waves (GWs) \cite{Tomita,Matarrese:1992rp,Matarrese:1993zf,Ananda:2006af,Baumann:2007zm,Saito:2008jc,Saito:2009jt} (see also Refs.~\cite{Assadullahi:2009jc,Bugaev:2009zh,Bugaev:2009kq,Bugaev:2010bb,Inomata:2018epa,Sato-Polito:2019hws,Kalaja:2019uju,Gow:2020bzo,Kimura:2021sqz,Wang:2022nml,Dandoy:2023jot} for current and future constraints on the small scales primordial spectrum). However, for them to be detectable, the primordial spectrum of fluctuations must be enhanced by several orders of magnitude with respect to CMB scales.

There are numerous models of inflation in the literature capable of enhancing the primordial spectrum of fluctuations. Some examples are phases of ultra slow-roll, bumps in the potential, sudden turns in the inflationary trajectory, resonances during inflation, etcetera \cite{Kawasaki:1997ju,Frampton:2010sw,Kawasaki:2012wr,Namjoo:2012aa,Inomata:2017okj,Pi:2017gih,Cai:2018dkf,Cai:2018tuh,Cai:2019jah,Chen:2019zza,Ashoorioon:2019xqc,Chen:2020uhe,Garcia-Bellido:1996mdl,Yokoyama:1998pt,Kohri:2012yw,Clesse:2015wea,Cheng:2016qzb,Espinosa:2017sgp,Inomata:2017okj,Kannike:2017bxn,Garcia-Bellido:2017mdw,Cheng:2018yyr,Passaglia:2018ixg,Ando:2018nge,Espinosa:2018eve,Inomata:2018cht,Braglia:2020eai,Atal:2018neu,Liu:2020oqe,Ng:2021hll,Palma:2020ejf,Fumagalli:2020adf,Inomata:2021tpx,Gundhi:2020kzm,Zhou:2020kkf,Ragavendra:2020sop,Cai:2020qpu,Inomata:2022yte,Cai:2023uhc,Briaud:2023eae,Heydari:2023xts,Karam:2023haj,Karam:2023haj}. See Ref.~\cite{Ozsoy:2023ryl} for a recent review on inflation and PBHs. By now, there is also a good analytical understanding of the mechanism behind the enhancement of the primordial spectrum in single field inflationary models \cite{Leach:2001zf,Byrnes:2018txb,Carrilho:2019oqg,Ozsoy:2019lyy,Cole:2022xqc,Tasinato:2023ukp} and it often involves departures from attractor trajectories. More recently, the $\delta N$ formalism \cite{Starobinsky:1985ibc,Salopek:1990jq,Sasaki:1995aw,Wands:2000dp,Lyth:2004gb} (see Ref.~\cite{Abolhasani:2019cqw} for a recent book    ) is employed to explore the tail of the Probability Distribution Function (PDF) of primordial fluctuations \cite{Biagetti:2018pjj,Atal:2019cdz,Ezquiaga:2019ftu,Atal:2019erb,Pi:2021dft,Cai:2022erk,Abe:2022xur,Pi:2022ysn,Animali:2022otk}, which is crucial for PBH formation \cite{Kitajima:2021fpq}. For example, in ultra-slow-roll the PDF of primordial fluctuations may present an exponential tail, instead of the usual Gaussian distribution.

Sometimes, it is also useful to consider simple yet exact models. Beyond being toy models, they may help to further deepen our understanding and may be especially helpful to explore implications beyond the linear regime. Two examples of such models are Starobinsky’s piecewise linear potential \cite{Starobinsky:1992ts,Ivanov:1994pa} (also see Refs.~\cite{Suyama:2021adn,Pi:2022zxs} for a recent thorough analysis) and constant roll inflation \cite{Martin:2012pe,Motohashi:2014ppa,Motohashi:2019rhu} (see Ref.~\cite{Chataignier:2023ago} for reconstruction methods for the inflaton’s potential). In this work, we consider a model which is exact also for non-attractor trajectories: inflation with a piecewise exponential potential. Also known as power-law inflation in the attractor \cite{Lucchin:1984yf}, the exponential potential has known exact background solutions for arbitrary initial conditions \cite{Lucchin:1984yf,Russo:2004ym,Andrianov:2011fg} and exact solutions for linear perturbations along the attractor trajectory \cite{Lyth:1991bc} (see e.g. Refs.~\cite{Dudas:2010gi,Vanzan_2023} for applications in CMB). Although in the simplest case the linear spectrum of fluctuations may share some similarities with Starobinsky’s piecewise linear potential, as for small field excursions the exponential is well approximated by a linear potential, the fact that we have general background solutions enables us to follow any non-attractor trajectory and make use of the $\delta N$ formalism.

In addition to the interesting inflationary dynamics, PBHs and induced GWs are recently attracting considerable attention for their rich phenomenology. On one hand, PBHs may explain the dark matter \cite{Bellomo:2017zsr,Carr:2017jsz,Inomata:2017okj,Bartolo:2018rku,Bartolo:2018evs,Carr:2020xqk}, some of the LIGO, VIRGO, KAGRA binary black hole merger events \cite{Bird:2016dcv,Sasaki:2016jop,Wong:2020yig,Franciolini:2021tla} and the seeds of supermassive black holes \cite{Kawasaki:2012kn,Carr:2018rid}.\footnote{In addition to the collapse of primordial fluctuations, other PBH formation mechanisms include first-order phase transitions \cite{Crawford:1982yz,Kodama:1982sf}, the collapse of Q-balls \cite{Cotner:2016cvr,Cotner:2019ykd,Flores:2021jas} and long-range forces stronger than gravity \cite{Amendola:2017xhl,Flores:2020drq,Domenech:2023afs}.}  On the other hand, there is the possibility that the reported evidence of a GW background by PTAs \cite{NG15-SGWB,NG15-pulsars,EPTA2-SGWB,EPTA2-pulsars,EPTA2-SMBHB-NP,PPTA3-SGWB,PPTA3-pulsars,PPTA3-SMBHB,CPTA-SGWB,InternationalPulsarTimingArray:2023mzf} are induced GWs from primordial fluctuations \cite{Franciolini:2023pbf,Franciolini:2023wjm,Inomata:2023zup,Cai:2023dls,Wang:2023ost,Liu:2023ymk,Unal:2023srk,Figueroa:2023zhu,Yi:2023mbm,Zhu:2023faa,Firouzjahi:2023lzg,Li:2023qua,You:2023rmn,Balaji:2023ehk,HosseiniMansoori:2023mqh,Zhao:2023joc,Liu:2023pau,Yi:2023tdk,Bhaumik:2023wmw,Choudhury:2023hfm,Yi:2023npi,Harigaya:2023pmw,Basilakos:2023xof} (or the merger of supermassive PBHs \cite{Huang:2023chx,Gouttenoire:2023nzr,Depta:2023qst}). 

This paper is organized as follows. In \S~\ref{sec:PLreview} we review the general background dynamics in the exponential potential and apply it to general slow-roll to slow-roll transitions. In \S~\ref{sec:perturbations} we study linear perturbations in the piecewise exponential potential and derive close to exact solutions in the non-attractor regime. We provide analytical solutions to the primordial spectrum of fluctuations for both enhancement and suppression cases. We also study the imprint of the primordial spectrum in the induced GW spectrum. In \S~\ref{sec:deltaN} we have a close look at the $\delta N$ formalism and we use it to derive the PDF of non-linear curvature fluctuations. We conclude our work in \S~\ref{sec:conclusions}.

\section{General background solutions \label{sec:PLreview}}

We start by reviewing the general exact solutions of a scalar field $\phi$ in an exponential potential in a Friedmann–Lemaître–Robertson–Walker (FLRW) metric as in Refs.~\cite{Lucchin:1984yf,Russo:2004ym,Andrianov:2011fg}. The action in this model reads
\begin{align}\label{eq:action}
S=\int d^4 x \sqrt{-g}\left\{\frac{M^2_{\rm pl}}{2}R-\frac{1}{2}g^{\mu\nu}\partial_\mu\phi\partial_\nu\phi - V(\phi)\right\}\,,
\end{align}
where $M^2_{\rm pl}=1/(8\pi G)$ is the reduced Planck mass, $R$ is the Ricci scalar and $V(\phi)$ is the potential of the scalar field. Since we are interested in enhancing/suppressing primordial fluctuations in a fully analytical model, we consider a piecewise exponential potential given by
\begin{align}\label{eq:potential}
V(\phi)=V_\star\left\{
\begin{aligned}
&e^{-\lambda_1\phi/M_{\rm pl}}& \phi\leq \phi_\star\\
&e^{-\lambda_2(\phi-\phi_c)/M_{\rm pl}} & \phi>\phi_\star
\end{aligned}
\right.\,,
\end{align}
where $\lambda_1$ and $\lambda_2$ are constants, $\phi_\star$ is the position of the matching point and continuity of the potential requires
\begin{align}
\phi_c=\phi_\star \left(1-\frac{\lambda_1}{\lambda_2}\right)\,.
\end{align}
Note that we could set $\phi_\star=0$ via a redefinition of $V_\star$ without loss of generality. Nevertheless, we keep $\phi_\star$ as an explicit reminder of the matching point. 

Before getting into the details, let us make some clarifications. First, we will only be interested in the case when $|\lambda_1|,|\lambda_2|<1$ as it leads to cosmic inflation, as we show later. Second, throughout the paper, we do not assume any hierarchy between $\lambda_1$ and $\lambda_2$, although we advance that primordial fluctuations are enhanced when $\lambda_2< \lambda_1$. Then, for concreteness, we always set the initial conditions in the region where $\phi<\phi_\star$. Then, as the potential decreases for increasing $\phi$,  we impose $\lambda_1,\lambda_2>0$ so as to have uninterrupted inflation. Otherwise, if $\lambda_2<0$ the scalar field would eventually stop and roll back. While this may lead to an interesting situation like that of Ref.~\cite{Briaud:2023eae}, we leave the case when $\lambda_2<0$ for future work. We also show in App.~\ref{app:powerlawcmb} that in order to explain CMB observations \cite{Akrami:2018odb}, one needs $\lambda\sim 0.18$. When needed we will fix either $\lambda_1$ or $\lambda_2$ to such fiducial value. We refer the reader to App.~\ref{app:powerlawcmb} for an extended discussion of power-law inflation and CMB measurements.

The model \eqref{eq:action} has exact analytical solutions in both regions, that is $\phi<\phi_\star$ and $\phi>\phi_\star$, in a flat FLRW metric given by
\begin{align}\label{eq:FRLWbg}
ds^2=g_{\mu\nu}dx^\mu dx^\nu=-dt^2+a^2(t)d\mathbf{x}^2\,,
\end{align}
where $t$ is cosmic time and $a$ the scale factor. We first review the general solutions for a single exponential potential, say $V=V_\star e^{-\lambda\phi}$, and we later focus on matching the two solutions in the piecewise potential \eqref{eq:potential}. From Eqs.~\eqref{eq:action} and \eqref{eq:FRLWbg}, the equations of motions are given by the Klein-Gordon and Friedmann equations, namely
\begin{align}\label{eq:bgeq}
&\ddot\phi+3H\dot\phi+V_{,\phi}=0\,,\\
&3M_{\rm pl}^2H^2=\frac{1}{2}\dot\phi^2+V\,,
\end{align}
where $H=\dot a/a$ is the Hubble parameter, $\dot\,=d/dt$ and $V_{,\phi}=\partial V/\partial\phi$. Following \cite{Russo:2004ym,Andrianov:2011fg}, one finds that, 
after time and field redefinitions given by
\begin{align}\label{eq:redefinitions}
\frac{d\xi}{dt}=\sqrt{\frac{3}{4}\frac{V(\phi)}{M_{\rm pl}^2}}\quad,\quad
\frac{\phi}{M_{\rm pl}}=\sqrt{\frac{2}{3}}\left(v-u\right)\quad,\quad
{\ln a}=\frac{1}{3}\left(v+u\right)\,,
\end{align}
the exact solutions for arbitrary initial conditions which are given by
\begin{align}\label{eq:solutionsuandv}
u&=A_u+\sqrt{\frac{1-\alpha}{1+\alpha}}\,\xi+\frac{1}{1+\alpha}\ln\left(1+Be^{-2\omega \xi}\right)\,,\\
v&=A_v+\sqrt{\frac{1+\alpha}{1-\alpha}}\,\xi+\frac{1}{1-\alpha}\ln\left(1-Be^{-2\omega \xi}\right)\,,\label{eq:solutionsuandv2}
\end{align}
with $\alpha=\lambda/\sqrt{6}$ and $\omega^2=1-\alpha^2$. $A_u$, $A_v$ and $B$ are integration constants fixed by the initial conditions. The solutions \eqref{eq:solutionsuandv} and \eqref{eq:solutionsuandv2} are valid for both $\phi<\phi_\star$ and $\phi>\phi_\star$ by replacing $\lambda$ with the corresponding parameter, respectively $\lambda_1$ and $\lambda_2$. The same applies to $\alpha$ and $\omega$. Note that the solutions \eqref{eq:solutionsuandv} and \eqref{eq:solutionsuandv2} are valid as long as the arguments inside the logarithms are positive which sets a lower bound on $\xi$. The exact attractor solution of Ref.~\cite{Lucchin:1984yf} is given by $B=0$. We postpone a detailed discussion of the general inflationary trajectories in phase space in \S~\ref{sec:deltaN} as it is relevant for the $\delta N$ formalism, but see Refs.~\cite{Russo:2004ym,Andrianov:2011fg} for a broader discussion not restricted to inflation. From now on, we will set $M^2_{\rm pl}=1$ for simplicity and recover the units a posteriori.

For practical purposes, it is more convenient to recast the exact solutions for $\phi$ and $\ln a$, Eqs.~\eqref{eq:solutionsuandv} and \eqref{eq:solutionsuandv2}, through \eqref{eq:redefinitions} with a new time variable given by
\begin{align}\label{eq:Z}
\mathrm{Z}\equiv2\times{\rm arctanh}\left[Be^{-2\omega \xi}\right]\,.
\end{align}
We advance that in the regime of interest $\mathrm{Z}$ is directly related to the number of $e$-folds defined as $N=\ln a$.\footnote{With this new variable the relation with cosmic time is given by
\begin{align}
\frac{d\mathrm{Z}}{dt}=-2\omega \sinh \mathrm{Z} \sqrt{\frac{3}{4}\frac{V(\phi)}{M_{\rm pl}^2}}\,.
\end{align}
As a curiosity, we found that with $\mathrm{Z}$ one can find an exact expression for $t(\mathrm{Z})$ in terms of hypergeometric functions, as well as for the conformal time. However, they are not very informative so we omit them.} In fact, we can understand the physical range of $\mathrm{Z}$ by computing
\begin{align}\label{eq:phiN}
\phi_{,N}\equiv\frac{d\phi}{dN}=\frac{{d\phi}/{d\xi}}{d\ln a/d\xi}=\sqrt{6}\frac{\alpha+\tanh(\mathrm{Z})}{1+\alpha\tanh(\mathrm{Z})}=\sqrt{6}\tanh\left(\mathrm{Z}+{\rm arctanh}[\alpha]\right)\,,
\end{align}
where in the last step we used the properties of the $\tanh$ to simplify the formula. From Eq.~\eqref{eq:phiN}, we see that as soon as $\mathrm{|Z|}\ll 1$ the system is closing in into the attractor regime where
\begin{align}\label{eq:phiNatt}
\phi_{,N}^{\rm att}=\sqrt{6}\alpha=\lambda,
\end{align}
and superscript “att” refers to the attractor. Thus, $\mathrm{Z}$ always evolves towards zero with time. The sign of $\mathrm{Z}$ tells us whether we approach the attractor value from above ($\mathrm{Z}>0$) or from below ($\mathrm{Z}<0$).
In passing, we note that the first slow-roll parameter, which is given by
\begin{align}\label{eq:1stslowroll}
\epsilon\equiv -\frac{\dot H}{H^2}=\frac{1}{2}\phi_{,N}^2\,,
\end{align}
is exactly constant in the attractor regime and given by
\begin{align}\label{eq:epatt}
\epsilon^{\rm att}=\frac{1}{2}\lambda^2\,.
\end{align}
The second slow-roll parameter, defined by
\begin{align}\label{eq:eta}
\eta\equiv\frac{d\ln \epsilon}{dN}=2\frac{\phi_{,NN}}{\phi_{,N}}\,
\end{align}
vanishes exactly in the attractor. Eq.~\eqref{eq:epatt} also implies that the universe is expanding as a power-law of time, that is $a\propto t^p$ where $p={2}/\lambda^2$. So, in order to have an accelerated expansion and slow-roll inflation we need $p\gg 1$ and, hence, $\lambda\ll1$ \cite{Lucchin:1984yf}. This is the reason why we restrict our attention to $\lambda\ll1$.

The usefulness of $\mathrm{Z}$ \eqref{eq:Z} is that its value at a pivot scale, e.g. at $\mathrm{Z}_\star$, is determined only by the derivative of the field at that time, that is $\phi_{,N\star}$, which in our case will be given by the first phase of slow-roll inflation. After inverting \eqref{eq:phiN}, we have that
\begin{align}\label{eq:Zstar}
\mathrm{Z}_\star={\rm arctanh}\left[\frac{\phi_{,N\star}-\sqrt{6}\alpha}{\sqrt{6}-\alpha\phi_{,N\star}}\right]={\rm arctanh}\left[\frac{\phi_{,N\star}}{\sqrt{6}}\right]-{\rm arctanh}\left[\alpha\right]\,.
\end{align}
Lastly, we write $\phi$ and $N=\ln a$ \eqref{eq:redefinitions} as
\begin{align}\label{eq:phigeneral}
\phi&=\phi_\star+\frac{1}{\omega^2}\sqrt{\frac{2}{3}}\left({\mathrm{Z}_\star}(1-\mathrm{z}) + \alpha \ln \left(\frac{\sinh (  \mathrm{Z}_\star)}{\sinh (  \mathrm{Z}_\star \mathrm{z})}\right)\right)\,,\\
N&= N_\star+\frac{1}{3\omega^2} \left(\alpha \mathrm{Z}_\star\left(1-\mathrm{z}\right)+ \ln \left(\frac{\sinh ( \mathrm{Z}_\star)}{\sinh (   \mathrm{Z}_\star \mathrm{z})}\right)\right)\,,\label{eq:Ngeneral}
\end{align}
where we introduced the variable $\mathrm{z}$ given by
\begin{align}\label{eq:z}
\mathrm{z} =\mathrm{Z}/\mathrm{Z}_\star\,,
\end{align}
as it allows to take a smooth limit to the attractor solution by sending $\mathrm{Z}_\star \to 0$ but keeping $\mathrm{z} $ fixed. One can check that in the attractor we have a linear relation between $N$ and $\phi$, that is
\begin{align}\label{eq:generalNnomore}
N=N_\star+\frac{\phi-\phi_\star}{\sqrt{6}\alpha}\,,
\end{align}
which is independent of $\phi_{,N\star}$.
In the general case, once we have the boundary conditions for $\phi_\star$ and $\mathrm{Z}_{\star}$ (or alternatively $\phi_{,N\star}$) the evolution of the system is fixed. Note that although the value of $N_\star$ can be absorbed by a redefinition of time, we will keep it for later convenience.

\begin{figure}
\includegraphics[width=0.49\columnwidth]{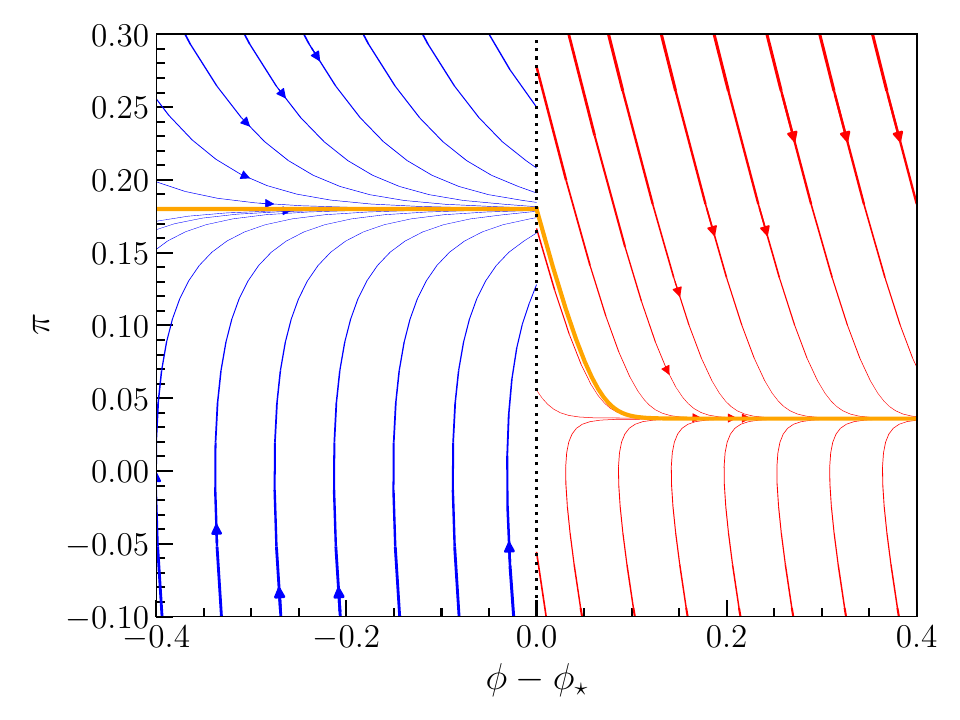}
\includegraphics[width=0.49\columnwidth]{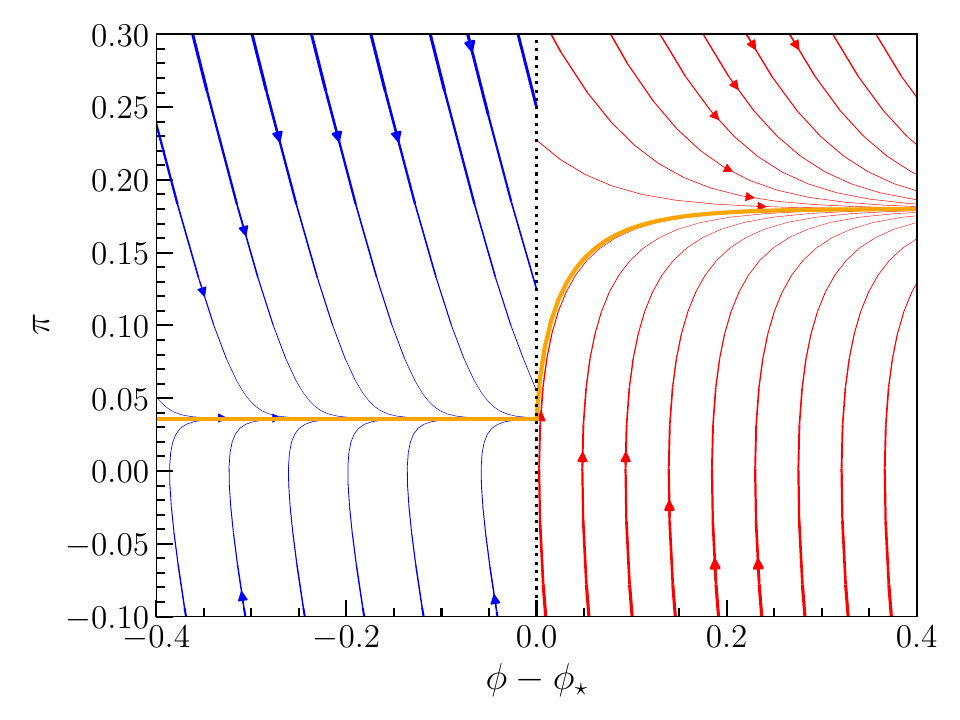}
\caption{Examples of phase space trajectories in the piecewise potential. Blue and red lines respectively show the trajectories in the first and second phases. The thickness of the lines indicates the magnitude of the second derivative ($\phi_{NN}$
 or $\pi_{N}$). The orange line shows the main trajectory of a transition from a slow-roll attractor to another slow-roll attractor. The vertical dashed lines show the matching point $\phi_\star$. On the left we show $\lambda_1=0.18$ and $\lambda_2=\lambda_1/5$. On the right, we show the opposite case, namely $\lambda_2=0.18$ and $\lambda_1=\lambda_2/5$.
\label{fig:ph}}
\end{figure}

In passing, we later found that the system of equations \eqref{eq:bgeq} can also be readily solved by using \textit{e}-folds as a time variable from the start. In that case, the Klein-Gordon equation reads
\begin{align}\label{eq:phiNNN}
\phi_{,NN}+\left(3-\frac{1}{2}\phi_{,N}^2\right)\left(\phi_{,N}+\frac{V_{,\phi}}{V}\right)=0\,.
\end{align}
This equation can be solved by splitting it into two first-order differential equations, namely
\begin{align}\label{eq:phasespace1}
\pi&=\phi_N\\
\pi_{,N}&=-\frac{1}{2}\left(6-\pi^2\right)\left(\pi-\sqrt{6}\alpha\right)\,.\label{eq:phasespace2}
\end{align}
Note that Eq.~\eqref{eq:phasespace2} shows that all phase space trajectories are independent of $\phi$, consistent with the shift symmetry of the exponential potential. Then, we integrate Eqs.~\eqref{eq:phasespace1} and \eqref{eq:phasespace2} once in terms of $\pi$, that is
\begin{align}
N-N_\star=\int^{\pi}_{\pi_\star} \frac{d\pi}{\pi_N}\quad{\rm and}\quad \phi-\phi_\star=\int_{\pi_\star}^\pi \frac{\pi}{\pi_N}d\pi\,,
\end{align}
which recovers Eqs.~\eqref{eq:phigeneral} and \eqref{eq:Ngeneral} after using Eq.~\eqref{eq:phiN}. We show some examples of trajectories in the phase space of ($\phi$, $\pi$) in Fig.~\ref{fig:ph}.

\subsection{Exact solutions in the piecewise potential}

\begin{figure}
\includegraphics[width=0.49\columnwidth]{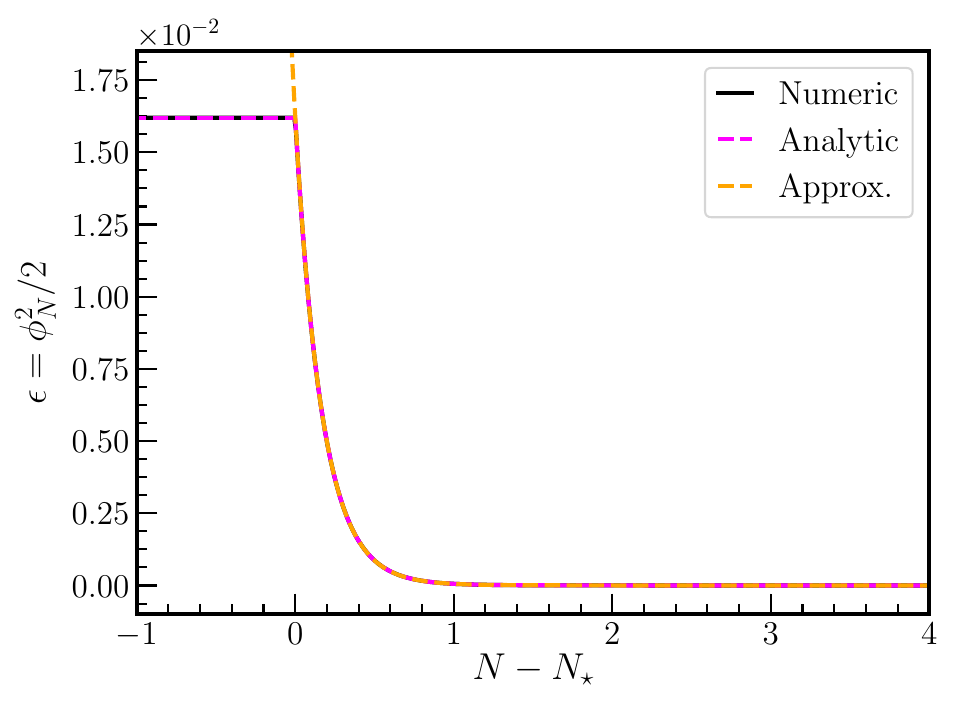}
\includegraphics[width=0.49\columnwidth]{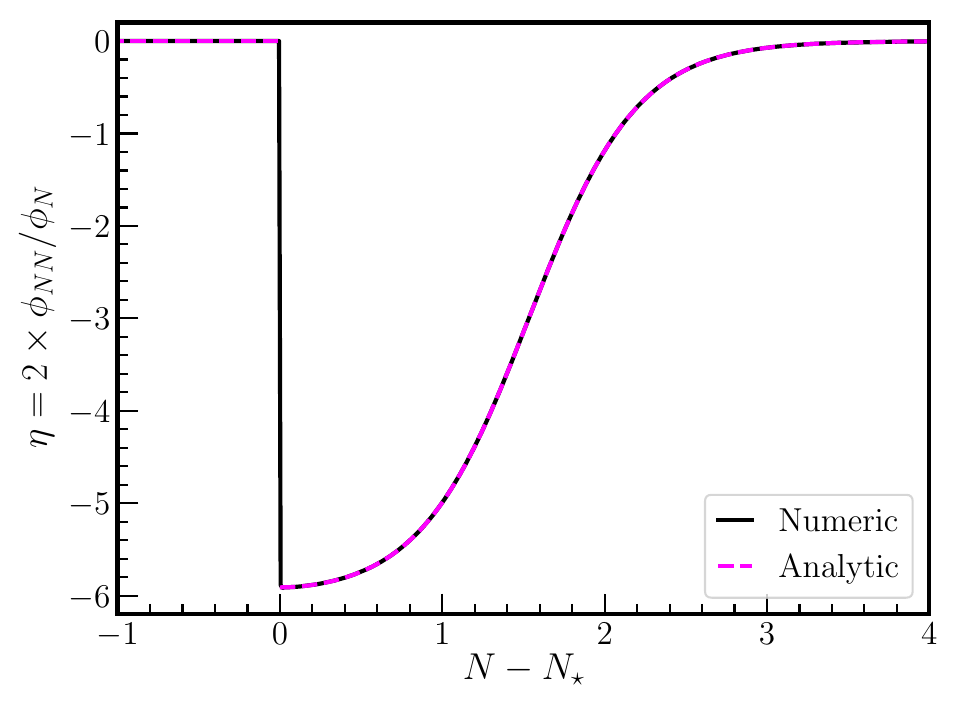}
\caption{First and second slow-roll parameters, Eqs.~\eqref{eq:1stslowroll} and \eqref{eq:eta}, respectively on the left and right figures, as a function of \textit{e}-folds $N$ for $\lambda_1=0.18$ and $\lambda_2=\lambda_1/100$. We show the numerical solutions in solid black lines, the exact solutions in magenta dashed lines, and the approximation for small $Z_\star$ in orange dashed lines. See how the approximation is very good after the matching point at $N=N_\star$. The $\eta$ parameter after the matching point is at most given by $\eta=-6$ as $\phi_{,NN}\approx -3 \phi_{,N}$.\label{fig:plotsbg}}
\end{figure}

\begin{figure}
\includegraphics[width=0.49\columnwidth]{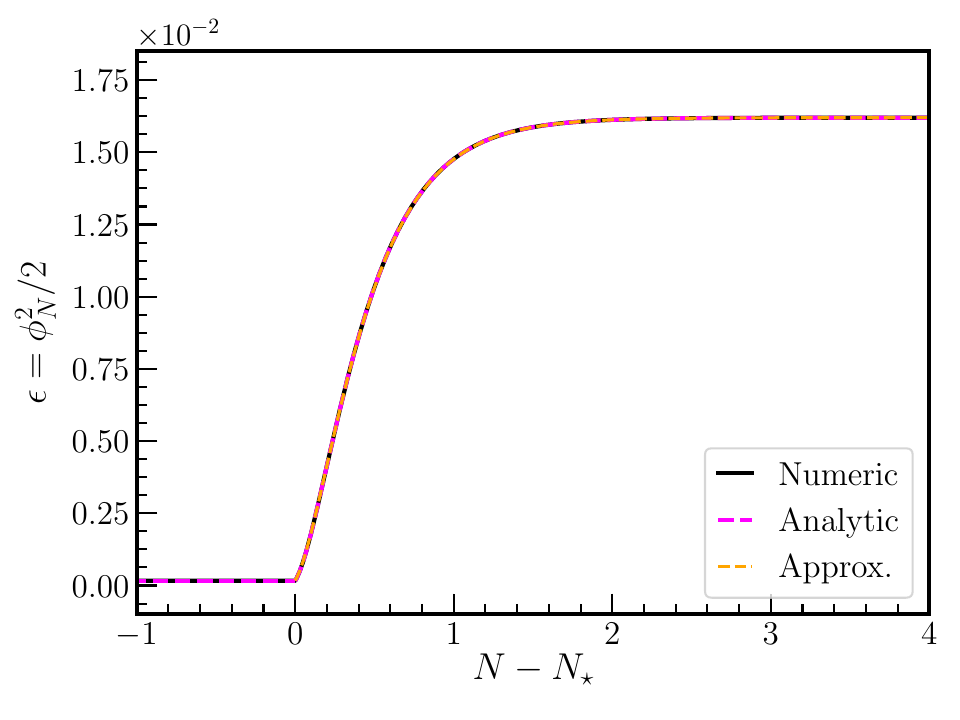}
\includegraphics[width=0.49\columnwidth]{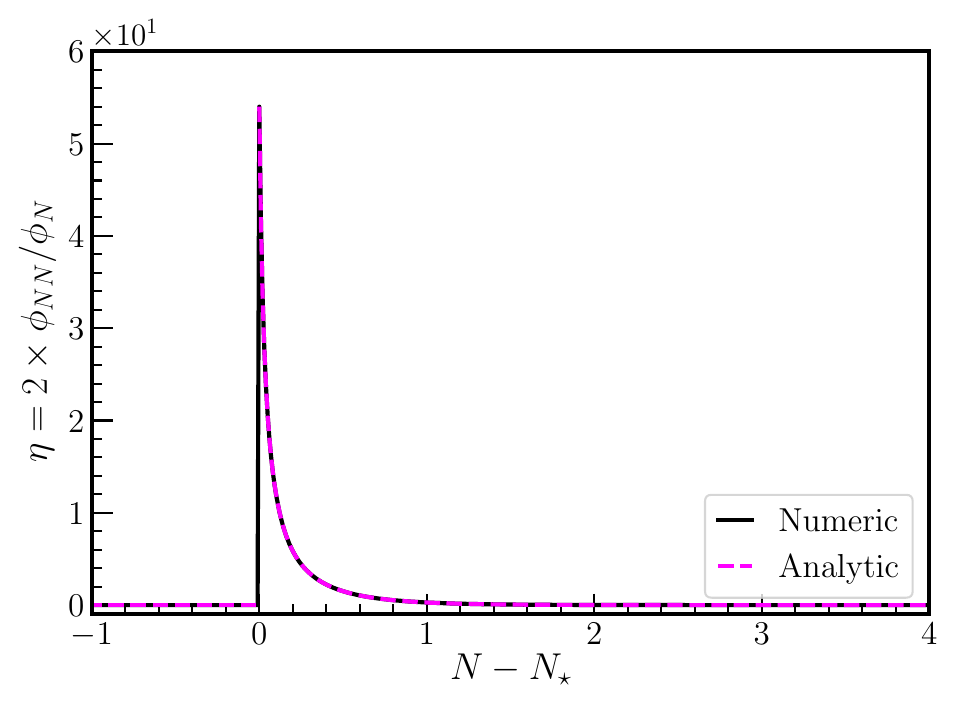}
\caption{First and second slow-roll parameters, Eqs.~\eqref{eq:1stslowroll} and \eqref{eq:eta}, respectively on the left and right figures, as a function of \textit{e}-folds $N$ for $\lambda_2=0.18$ and $\lambda_2=\lambda_1/10$. We show the numerical solutions in solid black lines, the exact solutions in magenta dashed lines, and the approximation for small $Z_\star$ in orange dashed lines. See how the approximation is very good after the matching point at $N=N_\star$. In this case, the $\eta$ parameter after the matching point can be arbitrarily large as $\phi_{,N}\ll \phi_{,NN}$.\label{fig:plotsbginv}}
\end{figure}

Now, let us come back to the piecewise potential \eqref{eq:potential}. Our inflationary model has two phases, one before and one after the matching point $\phi_\star$. We will respectively denote with subscripts $1$ and $2$ any variable before and after the matching point. Also, we assume that the first phase of inflation is a standard slow-roll inflation and the system is in the exact attractor regime. In this case we have that $\mathrm{Z}_{1\star}=0$,
\begin{align}\label{eq:phase1}
\phi_{1,N}=\lambda_1\quad ,\quad 
\mathrm{z}_1(N_1)= e^{-3\omega_1^2(N_1-N_\star)}\,,\quad {\rm and}\quad \phi_1=\lambda_1 N_1=\sqrt{6}\alpha_1 N_1\,,
\end{align}
where we fixed $N_\star=\lambda_1\phi_\star$. The first phase stops at $\phi_\star$ and we enter the second phase with the initial conditions $\phi_{2\star}=\phi_\star$ and $\phi_{2,N\star}=\lambda_1$. Continuity of $\phi_N$ at $\phi_\star$ implies from Eq.~\eqref{eq:Zstar} that in the second phase we have
\begin{align}
\mathrm{Z}_{2\star}={\rm arctanh}\left[\alpha_1\right]-{\rm arctanh}\left[\alpha_2\right]\,.
\end{align}
Interestingly, when $\alpha_2,\alpha_1\ll1$ we have that $|\mathrm{Z}_{2\star}|\sim{\cal O}({\rm min}(\alpha_1,\alpha_2))\ll 1$ and $\mathrm{z}_2$ is by definition bounded by $1>\mathrm{z}_2>0$. This conveniently justifies, in our set-up, a Taylor expansion for $|\mathrm{Z}_{2\star}|<1$, which turns out to give an accurate and useful approximation when dealing with perturbations. For later use, we present here the expressions for small $\mathrm{Z}_{2\star}$, namely
\begin{align}
\phi_2&\approx\phi_\star+\frac{1}{\omega_2^2}\sqrt{\frac{2}{3}}\left(\mathrm{Z}_{2\star}(1-\mathrm{z}_2) - \alpha_2 \ln \mathrm{z}_2\right)+{\cal O}(\mathrm{Z}_{2\star}^2)\,,\label{eq:approx} \\
N_2&\approx N_\star+\frac{1}{3\omega_2^2} \left(\alpha_2 \mathrm{Z}_{2\star}\left(1-\mathrm{z}_2\right)- \ln \mathrm{z}_2\right)+{\cal O}(\mathrm{Z}_{2\star}^2)\label{eq:approxN}\,.
\end{align}
In Figs.~\ref{fig:plotsbg} and \ref{fig:plotsbginv} we show the first and second slow-roll parameters for $\alpha_1>\alpha_2$ and $\alpha_1<\alpha_2$, respectively. See how the leading order approximations \eqref{eq:approx} and \eqref{eq:approxN} are in fact quite accurate in the regime of interest.

We can also explicitly write $\mathrm{z}_2$ in terms of \textit{e}-folds $N_2$ by inverting \eqref{eq:approx} by means of the Lambert function. However, as far as \textit{e}-folds are concerned, the term linear in $\mathrm{z}_2$ in \eqref{eq:approx} is always subdominant for $\mathrm{z}_2<1$. Thus, we have up to a very good approximation that
\begin{align}\label{eq:zN}
\mathrm{z}_2(N_2)\approx e^{-3\omega_2^2(N_2-N_\star)}\,,
\end{align}
which is exact in the attractor regime. Note that this simple relation \eqref{eq:zN} is only valid for $\mathrm{z}_2(N_2)$. The connection with cosmic and conformal time involves $V(\phi)$ by Eq.~\eqref{eq:redefinitions} for which the linear term in $\mathrm{z}$ of \eqref{eq:approx} cannot be initially neglected until close enough to the attractor. 

It is also useful to compute $\phi_{2,N}$ \eqref{eq:phiN} at leading order in $\mathrm{Z}_{2\star}$, which yields
\begin{align}\label{eq:phiNplus}
\phi_{2,N}\approx\sqrt{6}\left(\alpha_2+\omega_2^2 \mathrm{Z}_{2\star} \mathrm{z}_2\right)\approx\sqrt{6}\left(\alpha_2+\omega_2^2 \mathrm{Z}_{2\star} e^{-3\omega_2^2(N_2-N_\star)}\right)\,.
\end{align}
Using Eq.~\eqref{eq:phiNplus} we may estimate the duration of the transition from the slow-roll attractor to the next slow-roll attractor by finding the time it takes for the second slow-roll parameter \eqref{eq:eta} to fall below unity. This roughly leads us to
\begin{align}\label{eq:phiNpluse-fold}
\Delta N_{\rm tr.}=N_{\star\star}-N_\star\approx\frac{1}{3}\left(\ln\left|1-\frac{\alpha_1}{\alpha_2}\right|+\ln 5\right)\,,
\end{align}
where $N_{\star\star}$ is defined through $|\eta(N_{\star\star})|=1$, the subscript “tr.” refers to transition and we assumed that $\alpha_1,\alpha_2<1$ and so $\omega_2\sim 1$ and $\mathrm{Z}_{2\star}\sim \alpha_1-\alpha_2$.  For $\alpha_1<\alpha_2$ we have that $\Delta N_{\rm tr.}\approx 0.5$ and the transition is practically over after half an \textit{e}-fold. For $\alpha_1>\alpha_2$ though, we have that $\Delta N_{\rm tr.}\approx\frac{1}{3}\ln \frac{\alpha_1}{\alpha_2}$. For $\alpha_2\sim 10^{-3} \alpha_1$ this yields $\Delta N_{\rm tr.}\approx 2.8$. In fact, if we look closer to Eq.~\eqref{eq:phiNplus} we see that for $\alpha_2<\alpha_1$, i.e. the field decelerates after the matching point, the initial decay of $\phi_{2,N}$ is close to $\phi_{2,N}\sim a^{-3}$. This means that the system enters a period of quasi ultra-slow-roll \cite{Namjoo:2012aa} after the matching point. As expected, the larger the hierarchy between $\alpha_1$ and $\alpha_2$ the longer the duration of the ultra-slow-roll phase. We now proceed to study the linear perturbations.

\section{Fluctuations in a slow-roll to slow-roll transition \label{sec:perturbations}}

To study quantum fluctuations during inflation, we perturb the flat FLRW metric and focus only on scalar fluctuations (see e.g. Refs.~\cite{Kodama:1985bj,Mukhanov:1990me,Brandenberger:2003vk,Durrer:2004fx,Baumann:2009ds,Koyama:2010xj} for reviews). In general, we have fluctuations of the metric as well as the scalar field, say $\delta\phi$. For fluctuations during inflation, it is customary to work in the uniform-$\phi$ slicing, also referred to as comoving slicing, where perturbations of $\phi$ are set to zero, that is $
\delta\phi=0$. Then, one focuses on metric perturbations only, the so-called comoving curvature perturbation ${\cal R}$. The second order action in perturbation theory for the curvature perturbation is given in Fourier modes by
\begin{align}
S_2=\int d^3x d\tau\, a^2\epsilon \left({{\cal R}_k'}^2+k^2{\cal R}_k^2\right)\,,
\end{align}
where $\tau$ is conformal time, given by $dt=ad\tau$, $'\,=d/d\tau$, ${\cal R}_k$ are the mode functions and $k$ is the wavenumber of the fluctuation. The mode functions satisfy the Mukhanov-Sasaki equation, namely
\begin{align}
{\cal R}_k''+2aH{\cal R}'_k+k^2{\cal R}_k=0\,.
\end{align}
Canonical quantization then fixes the amplitude of quantum fluctuations of ${\cal R}_k$ by requiring
\begin{align}\label{eq:canonical}
{\cal R}_k^*{\cal R}_k'-{\cal R}_k{{\cal R}_k'}^*=-\frac{i}{2a^2\epsilon}\,.
\end{align}
For analytical and numerical purposes, we also present the Mukhanov-Sasaki equation in terms of \textit{e}-folds, which reads
\begin{align}\label{eq:RNN}
{\cal R}_{,NN}+\left(3-\frac{1}{2}\phi_{,N}^2+2\frac{\phi_{,NN}}{\phi_{,N}}\right){\cal R}_{,N}+\left(\frac{k}{aH}\right)^2{\cal R}=0\,,
\end{align}
where we omit the subscript $k$ hereon for simplicity. We may substitute $\phi_{,NN}$ using the Klein-Gordon equation in terms of \textit{e}-folds, that is Eq.~\eqref{eq:phiNNN}.

It is sometimes convenient to work in terms of $\delta\phi$ as well. The easiest way to do so is to relate the curvature perturbation with the scalar field fluctuations (in the spatially flat slicing where curvature perturbation vanishes) via a gauge transformation, which gives
\begin{align}\label{eq:Rtodeltaphi}
{\cal R}=-\frac{H}{\dot \phi}\delta\phi=-\frac{\delta\phi}{\phi_{,N}}\,.
\end{align}
In the attractor, we have that ${\cal R}$ and $\delta\phi$ are just related by a constant, explicitly 
\begin{align}\label{eq:relationattractor}
{\cal R}^{\rm att}=-\frac{1}{\sqrt{6}\alpha}\delta\phi^{\rm att}\,,
\end{align}
where we used Eq.~\eqref{eq:phiNatt}.
Thus, in the attractor, the dynamics of ${\cal R}$ and $\delta\phi$ are exactly the same.\footnote{In fact, they are also the same for tensor modes. The only practical difference is the overall normalization.} We obtain the equations of motion for $\delta\phi$ after plugging \eqref{eq:Rtodeltaphi} into \eqref{eq:RNN}, which yields
\begin{align}\label{eq:deltaphiNN}
{\delta \phi}_{,NN}+\left(3-\frac{1}{2}\phi_{,N}^2\right){\delta \phi}_{,N}+\left(\frac{k}{aH}\right)^2{\delta \phi}+\frac{m^2_{\delta\phi}}{H^2}{\delta \phi}=0\,,
\end{align}
where we defined
\begin{align}\label{eq:massterm}
\frac{m^2_{\delta\phi}}{H^2}=\left(3-\frac{1}{2}\phi_{,N}^2\right)\left(\phi_{,N}^2+2\phi_{,N}\frac{V_{,\phi}}{V}+\frac{V_{,\phi\phi}}{V}\right)\,,
\end{align}
and we used Eq.~\eqref{eq:phiNNN} to simplify the form of $m^2_{\delta\phi}$.
In passing, we also write explicitly for later use that 
\begin{align}\label{eq:aH}
a^2H^2=\frac{e^{2N}V(\phi)}{3-\frac{1}{2}\phi_{,N}^2}\,.
\end{align}
Since the relation $\mathrm{z}(N)$ \eqref{eq:zN} is exact in the attractor and a very good approximation in the non-attractor phase, it will be convenient to use $\mathrm{z}$ instead of $N$. In terms of $\mathrm{z}$ the equations for $\delta\phi$, Eq.~\eqref{eq:deltaphiNN}, read
\begin{align}\label{eq:deltaphizz}
\frac{d^2}{d\mathrm{z}^2}{\delta \phi}+\left(\frac{1}{6}\phi_{,N}^2-\alpha^2\right)\frac{1}{\omega^2\mathrm{z}}\frac{d}{d\mathrm{z}}{\delta \phi}+\left(\frac{k}{3\omega^2 \mathrm{z} aH}\right)^2{\delta \phi}+\frac{m^2_{\delta\phi}}{H^2}\frac{1}{9\omega^4\mathrm{z}^2}{\delta \phi}=0\,.
\end{align}

Before studying the behavior of the perturbations in the piecewise potential, let us first review the exact solutions for the mode functions in the attractor regime. In the attractor, we have that $\phi_{N}=\sqrt{6}\alpha$ and, hence, Eq.~\eqref{eq:deltaphizz} exactly becomes
\begin{align}\label{eq:deltaphizzatt}
\frac{d^2}{d\mathrm{z}^2}{\delta \phi}+\kappa^2\mathrm{z}^{1/\mu-2}{\delta \phi}=0\,.
\end{align}
where we defined
\begin{align}
\kappa=\frac{k}{3\omega^2k_\star}\quad{,}\quad k_\star=a_\star H_\star
\quad{\rm and}\quad
 \mu=\frac{3}{2}\frac{\omega^2}{3\omega^2-2}\,.
 \end{align}
Eq.~\eqref{eq:deltaphizzatt} also holds for ${\cal R}$ in the attractor. The general solutions to \eqref{eq:deltaphizzatt} are given by \cite{Lyth:1991bc}
\begin{align}\label{eq:exactsolpert}
\delta\phi=2^{-\mu}\frac{i\pi}{\Gamma[\mu]}\left(C x^\mu H^{(1)}_\mu(x)-D x^\mu H^{(2)}_\mu(x)\right)\,,
\end{align}
where $C$ and $D$ are constants, $H_\mu^{(1)}(x)$ and $H_\mu^{(2)}(x)$ respectively are the Hankel functions of the first and second kind and we defined
\begin{align}\label{eq:xeq}
x=2\mu\kappa \mathrm{z}^{\frac{1}{2\mu}}\,.
\end{align}
We note that in the attractor $x$ actually coincides with $-k\tau$ where $\tau$ is the conformal time.
Also note that in Eq.~\eqref{eq:exactsolpert} we have chosen the prefactor  such that
\begin{align}
\delta\phi(x\to 0)= C+D\,.
\end{align}
Bunch-Davies initial conditions further sets $D=0$ \cite{Lyth:1991bc} and canonical quantization \eqref{eq:canonical} yields
\begin{align}\label{eq:C1}
C=\frac{i  4^{\mu -1} (2 \mu -1)^{1/2-\mu }  \Gamma [\mu]}{\sqrt{\pi } }\frac{{\kappa}^{-\mu }}{a_\star\sqrt{k_\star}}\,.
\end{align}

If we only had a single stage of power-law inflation, then the dimensionless primordial spectrum of curvature fluctuations at the end of inflation would be given by
\begin{align}
P_{\delta\phi}(x\to0)=\frac{k^3}{2\pi^2}|\delta\phi|^2=\frac{k^3}{2\pi^2}|C|^2=\frac{  2^{4 \mu -5} (2 \mu -1)^{1- 2\mu } 
   \Gamma^2[\mu ]}{\pi ^3 }\kappa ^{3-2 \mu }\frac{H_\star^2}{M_{\rm pl}^2}\,.
\end{align}
Using Eq.~\eqref{eq:relationattractor} we recover the well-known result for the curvature perturbation at leading order in $\lambda$ ($\mu \sim 3/2$), namely
\begin{align}\label{eq:slowroll}
{\cal P}_{\cal R}(k)=\frac{1}{8\pi^2}\frac{H_\star^2}{\epsilon^{\rm att}M_{\rm pl}^2}\left(\frac{k}{k_\star}\right)^{-\lambda^2}\,.
\end{align}
The main difference with the standard approach is that we need not evaluate the mode functions at Hubble horizon crossing, that is when $k=aH$, as we have exact solutions.

\subsection{Perturbations in the piecewise potential}

In the piecewise power-law inflation case, we have that the initial conditions for the mode functions are given by the attractor solution Eqs.~\eqref{eq:exactsolpert} and \eqref{eq:C1} by fixing the parameters of the first phase, e.g. $\mu\to\mu_1$. To avoid later confusions, they are explicitly given by
\begin{align}\label{eq:exactsolpertminus}
\delta\phi_1=C_1\,2^{-\mu_1}\frac{i\pi}{\Gamma[\mu_1]}x_1^{\mu_1} H^{(1)}_{\mu_1}(x_1)\,,
\end{align}
where
\begin{align}
 \kappa_1=\frac{k}{3\omega_1^2k_\star}\quad,\quad\mu_1=\frac{3}{2}\frac{\omega_1^2}{3\omega_1^2-2} \quad {\rm and} \quad x_1=2\mu_1\kappa_1 z_1^{\frac{1}{2\mu_1}}
\end{align}
and
\begin{align}\label{eq:C11}
C_1=\frac{i  4^{\mu_1 -1} (2 \mu_1 -1)^{1/2-\mu_1 }  \Gamma [\mu_1]}{\sqrt{\pi } }\frac{{\kappa_1}^{-\mu_1 }}{a_\star\sqrt{k_\star}}\,.
\end{align}
Then, the equations of motion for $\delta\phi$ \eqref{eq:deltaphizz} after the transition in general read
\begin{align}\label{eq:deltaphizzplus}
\frac{d^2}{d\mathrm{z}_2^2}{\delta \phi_2}+\left(\frac{1}{6}\phi_{2,N}^2-\alpha_2^2\right)&\frac{1}{\omega_2^2 \mathrm{z}_2}\frac{d}{d\mathrm{z}_2}{\delta \phi_2}\nonumber\\&+\frac{1-{\phi_{2,N}^2}/{6}}{\omega_2^2}\left(\frac{V(\phi_2)}{V(\phi_2^{\rm att})}\kappa_2^2\mathrm{z}_2^{1/\mu-2}+\frac{(\phi_{2,N}-\sqrt{6}\alpha_2)^2}{3\omega_2^4\mathrm{z}_2^2}\right){\delta \phi_2}=0\,,
\end{align}
where to keep the same notation for $\kappa$ we defined 
\begin{align}
\kappa_2=\frac{\omega_1^2}{\omega_2^2}\kappa_1\,.
\end{align}

We now proceed as follows in order to investigate the dynamics of $\delta\phi$ after the transition. To simplify the equations and remove the friction term, we first define
\begin{align}\label{eq:relationvarphi}
\delta\phi_2=e^{f(\mathrm{z}_2)}\varphi\,,
\end{align}
where 
\begin{align}
f(\mathrm{z}_2)=\int_{1}^{\mathrm{z}_2} d\tilde{\mathrm{z}}_2 \left(\alpha_2^2-\frac{1}{6}\phi_{,N}^2\right)\frac{1}{2\omega_2^2\tilde{\mathrm{z}}_2}\approx \mathrm{Z}_{2\star} (1-\mathrm{z}_2)\left(\alpha_2+\frac{1}{2}\mathrm{Z}_{2\star}\omega_2^2(1+\mathrm{z}_2)\right)\,,
\end{align}
where in the last step we used Eq.~\eqref{eq:phiNplus}.
We note that since $f\ll1$ the redefinition will not have much effect when relating $\varphi$ and $\delta\phi_2$ via Eq.~\eqref{eq:relationvarphi}. Then we find at next to leading order in $\mathrm{Z}_{2\star}$ that
\begin{align}\label{eq:deltaphizzplus2}
\frac{d^2\varphi}{d\mathrm{z}_2^2}+(1-2 \mathrm{Z}_{2\star} \mathrm{z}_2 \alpha_2)e^{\frac{2\alpha_2}{\omega_2^2} \mathrm{Z}_{2\star}(1-\mathrm{z}_2)}\kappa_2^2\mathrm{z}_2^{1/\mu_2-2}{\varphi}+ \frac{1}{2} \mathrm{Z}_{2\star}^2 ( 5 \omega_2^2-2)\varphi+O(\mathrm{Z}_{2\star}^3)\varphi=0\,.
\end{align}
The mass term in \eqref{eq:deltaphizzplus2} is $O(\mathrm{Z}_{2\star}^2)\sim O({\rm min}(\alpha_1^2,\alpha_2^2))$ and since we have $0<\mathrm{z}_2<1$ the suppression due to the constant mass is negligible. For the same reason, we neglect the leading order terms in $\mathrm{Z}_{2\star}$ and we arrive at
\begin{align}\label{eq:deltaphizzplus3}
\frac{d^2\varphi}{d\mathrm{z}_2^2}+\kappa_2^2\mathrm{z}_2^{1/\mu_2-2}{\varphi}+O(\mathrm{Z}_{2\star}^2)=0\,,
\end{align}
which is the equation for the mode function in the attractor. We note that this result is rather general. It shows that, in the exponential potential, scalar field fluctuations follow very closely the attractor solution even though the background trajectory is off-attractor, as long as the system is in the slow-roll regime, i.e. $\phi_N<1$.

The approximate solutions for $\delta\phi_2$ are then given by
\begin{align}\label{eq:deltaphi2}
\delta\phi_2=2^{-\mu_2}\frac{i\pi}{\Gamma[\mu_2]}\left(C_2x_2^{\mu_2} H^{(1)}_{\mu_2}(x_2)-D_2 x_2^{\mu_2} H^{(2)}_{\mu_2}(x_2)\right)\,,
\end{align}
where
\begin{align}
 \mu_2=\frac{3}{2}\frac{\omega_2^2}{3\omega_2^2-2} \quad {\rm and} \quad x_2=2\mu_2\kappa_2 \mathrm{z}_2^{\frac{1}{2\mu_2}}\,.
\end{align}
The reason that fluctuations of $\delta\phi$ are well approximated by the attractor solution is that, although the attractor of the second phase is reached for $\mathrm{z}_2>\mathrm{z}(N_{\star\star})$ (see Eq.~\eqref{eq:phiNpluse-fold}), the change of vacuum is effectively instantaneous. In fact, the coefficients $C_2$ and $D_2$ in Eq.~\eqref{eq:deltaphi2} are the Bogolyubov coefficients due to the sudden change of vacuum.

\begin{figure}
\includegraphics[width=0.49\columnwidth]{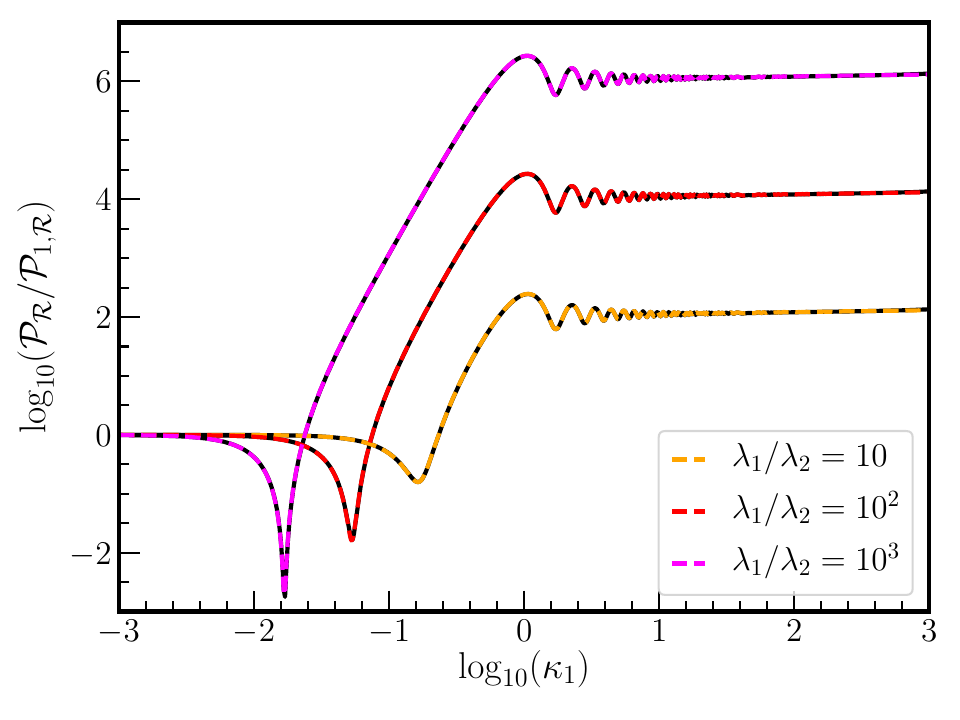}
\includegraphics[width=0.49\columnwidth]{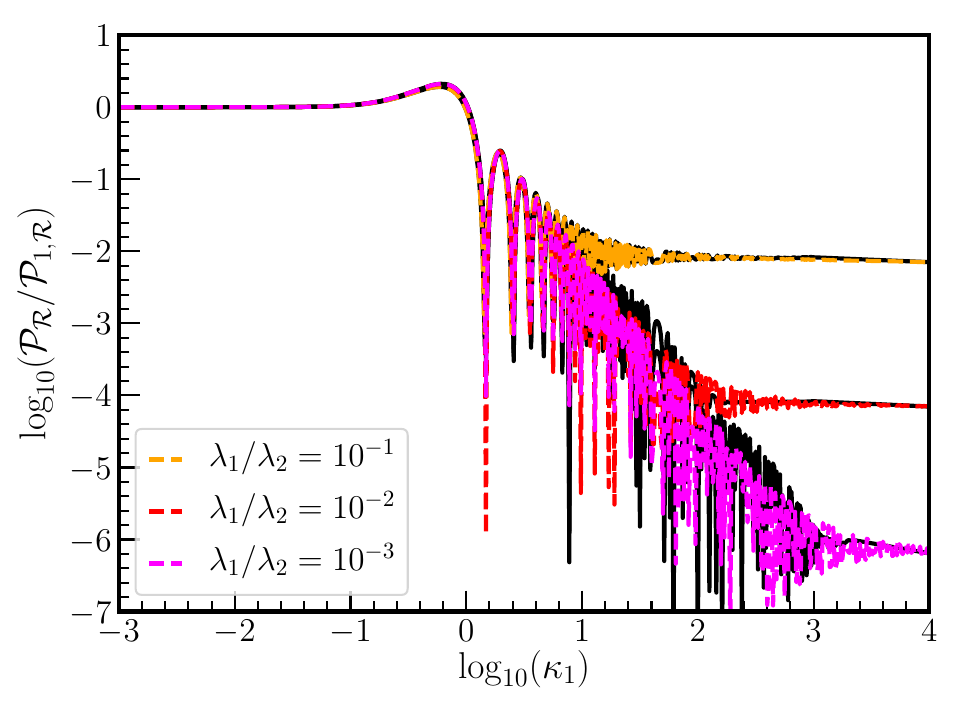}
\caption{Power spectrum of curvature fluctuations at the end of inflation in the piecewise exponential potential. Black solid lines show the result of numerical integration. Dashed lines show the analytical result \eqref{eq:resultdeltaphi} after using Eq.~\eqref{eq:relationattractor} to relate $\delta\phi$ and ${\cal R}$. On the left figure, we show the case when $\lambda_1>\lambda_2$ ($\alpha_1>\alpha_2$) with $\lambda_1=0.18$. Orange, red and magenta dashed lines respectively show $\lambda_1/\lambda_2=\{10,10^2,10^3\}$. On the right figure, we show the opposite case, namely $\lambda_2>\lambda_1$ with $\lambda_2=0.18$. In this figure orange, red and magenta dashed lines respectively show $\lambda_2/\lambda_1=\{10,10^2,10^3\}$. Deviations between the analytical and numerical solutions during the oscillations are due to numerical errors.\label{fig:numvsana}}
\end{figure}

The derivation above tells us that the main effect of the transition is the matching between attractor solutions. If we consider the full piecewise potential \eqref{eq:potential}, the mass term \eqref{eq:massterm} in the full equation for $\delta\phi$, that is \eqref{eq:deltaphiNN}, has a Dirac delta-like feature in the second derivative of the potential, concretely given by
\begin{align}\label{eq:vphiphidelta}
V_{,\phi\phi}^{\delta}=\left(V_{1,\phi}-V_{2,\phi}\right)\delta(\phi-\phi_\star)\,.
\end{align}
Inserting Eq.~\eqref{eq:vphiphidelta} into \eqref{eq:deltaphiNN}, and integrating once, yields the matching conditions,\footnote{We used that
\begin{align}
\frac{d\phi_2}{d\mathrm{z}}\Big|_{\mathrm{z}=\mathrm{z}_\star}=\frac{d\phi_1}{d\mathrm{z}}\Big|_{\mathrm{z}=\mathrm{z}_\star}=-\frac{\phi_{1,N}}{3\omega_1^2\mathrm{z}_\star}=-\frac{\sqrt{6}\alpha_1}{3\omega_1^2\mathrm{z}_\star}\,.
\end{align}
} namely
\begin{align}\label{eq:matchingconditions}
\delta\phi_2(\mathrm{z}_\star)&=\delta\phi_1(\mathrm{z}_\star)\,,\\
\frac{d}{d\mathrm{z}_2}\delta\phi_2(\mathrm{z}_\star)&=\frac{d}{d\mathrm{z}_1}\delta\phi_1(\mathrm{z}_\star)+\left(1-\frac{\alpha_2}{\alpha_1}\right)\delta\phi_1(\mathrm{z}_\star)\,,
\end{align}
where by definition $\mathrm{z}_\star=1$. By performing the matching we obtain
\begin{align}\label{eq:C2}
C_2/{\cal C}&=\frac{1}{\kappa_1}\left(1-\frac{\alpha_2}{\alpha_1}\right)H_{\mu_1}^{(1)}(x_{1\star})
   H_{\mu_2}^{(2)}\left(x_{2\star}\right)-\frac{\omega_1^2}{\omega_2^2}H_{\mu_1}^{(1)}(x_{1\star}) H_{\mu_2-1}^{(2)}\left(x_{2\star}\right)+H_{\mu_1-1}^{(1)}(x_{1\star}) H_{\mu_2}^{(2)}\left(x_{2\star}\right)\,,\\
D_2/{\cal C}&=\frac{1}{\kappa_1}\left(1-\frac{\alpha_2}{\alpha_1}\right)H_{\mu_1}^{(1)}(x_{1\star})
   H_{\mu_2}^{(1)}\left(x_{2\star}\right)-\frac{\omega_1^2}{\omega_2^2}H_{\mu_1}^{(1)}(x_{1\star}) H_{\mu_2-1}^{(1)}\left(x_{2\star}\right)+H_{\mu_1-1}^{(1)}(x_{1\star}) H_{\mu_2}^{(1)}\left(x_{2\star}\right)\,,\label{eq:D2}
\end{align}
where we defined $x_{1\star}=2\kappa_1\mu_1$, $x_{2\star}=2\kappa_2\mu_2$ and
\begin{align}\label{eq:calC}
{\cal C}=-i \pi  \mu_1^{\mu_1} \mu_2^{1-\mu_2}
   \omega_1^{-2\mu_2} \omega_2^{2\mu_2}\kappa_1^{\mu_1-\mu_2+1}\frac{ \Gamma
   (\mu_2) }{2 \Gamma
   (\mu_1)}\times C_1\,.
\end{align}
In Eq.~\eqref{eq:calC} $C_1$ is given by the canonical normalization in the first phase, that is Eq.~\eqref{eq:C11}. Then, after some simplifications, the power spectrum of $\delta\phi$ at the end of inflation is given by
\begin{align}\label{eq:resultdeltaphi}
{\cal P}_{\delta\phi}&(x_2\to0)=\frac{k^3}{2\pi^2}|\delta\phi|^2=\frac{k^3}{2\pi^2}|C_2+D_2|^2\nonumber\\&
=\frac{H_\star^2}{4\pi}\kappa_1^{3-2 \mu_2}\times\left(\frac{3 \omega_1^2}{2}\right)^3{ \mu_1^{2 (\mu_1-\mu_2)}\Gamma^2[\mu_2+1]
    \frac{ 2^{4 \mu_1}(2 \mu_1-1)^{1+2(\mu_2-\mu_1)}}{(2 \mu_2-1)^{2
   \mu_2}} }\nonumber\\&\,\,\,\,\,\,\times
   \left|H_{\mu_1}^{(1)}(x_{1\star})
   \left({\left(1-\frac{\alpha_2}{\alpha_1}\right)
   J_{\mu_2}(x_{2\star})}-\kappa_2{ J_{\mu_2-1}(x_{2\star}})\right)+ \kappa_1
   H_{\mu_1-1}^{(1)}(x_{1\star})
   J_{\mu_2}(x_{2\star})\right|^2\,.
\end{align}
As we show in Fig.~\ref{fig:numvsana} this provides a very good approximation to the numerical solution at all scales for both the cases when $\alpha_2<\alpha_1$ (enhancement) and $\alpha_2>\alpha_1$ (suppression). In our numerical calculations, we solved for ${\cal R}$ using \textit{e}-folds, that is Eq.~\eqref{eq:RNN}, with initial conditions given by the exact attractor solutions few \textit{e}-folds before the transition. We now investigate the two cases, enhancement and suppression, separately below.

\subsection{Enhancement of fluctuations \label{sec:enhancement}}

We turn our attention to the case when primordial fluctuations are enhanced due to the transition. Since $\delta\phi$ on small scales maintains the amplitude, this case corresponds to $\lambda_2<\lambda_1$ from Eq.~\eqref{eq:Rtodeltaphi}. With our close to exact analytical solutions for the perturbations, we can study two limits: large scales $\kappa_1\ll1$ and small scales $\kappa_1\gg1$. On one hand, for $\kappa_1\ll1$ we find that
\begin{align}\label{eq:lowklimit}
(C_2+D_2)&(\kappa_1\ll1)\approx C_1\nonumber\\&\times\left\{\frac{\alpha_2}{\alpha_1}+\frac{\kappa_1^2
   }{\alpha_1 }\left(\frac{\mu_1 }{\mu_1-1} 
   (\alpha_1-\alpha_2 \mu_1)-\frac{\mu_2}{\mu_2+1} \frac{\omega_1^4}{\omega_2^4} (\alpha_1+\alpha_2\mu_2)\right)\right\}\,.
\end{align}
From Eq.~\eqref{eq:lowklimit} we first see that for very long wavelengths we have
\begin{align}
\lim_{\kappa_1\to 0}|C_2+D_2|^2=|C_1|^2\times\frac{\lambda_2^2}{\lambda_1^2}\,,
\end{align}
which means that there is a suppression of $\lambda_2^2/\lambda_1^2$ between the large and small scale power spectrum since the overall amplitude of the small scale power spectrum does not change. Second, we note that the power spectrum vanishes at
\begin{align}\label{eq:kdip}
\kappa_{1,{\rm dip}}^2=\frac{\alpha_1 (\mu_1-1) (\mu_2+1) \omega_2^4}{(\mu_1-1) \mu_2 \omega_1^4 (\alpha_1+\alpha_2 \mu_2)+\mu_1 (\mu_2+1) \omega_2^4 (\alpha_1-\alpha_2 \mu_1)}\approx \frac{5\alpha_2}{18\alpha_1}=\frac{5\lambda_2}{18\lambda_1}\,,
\end{align}
which is the position of the dip. In the second step in Eq.~\eqref{eq:kdip} we only took the leading order value in $\alpha_2<\alpha_1\ll1$ for simplicity. Thus, the higher the enhancement of the power spectrum, the lower the position of the dip. Lastly, we find that there is a phase of $\kappa_1^4$ in the power spectrum 
\begin{align}\label{eq:k4}
|C_2+D_2|^2(\kappa_{1,\rm dip}\ll \kappa_1\ll 1)=\frac{ \kappa_1^4}{\alpha_1^2}\left(\frac{\mu_1 (\alpha_1-\alpha_1\mu_1)}{\mu_1-1}+\frac{\mu_2 \omega_1^4 (\alpha_1+\alpha_2 \mu_2)}{\omega_2^4(\mu_2+1)}\right)^2\approx\frac{324}{25} \kappa_1^4\,,
\end{align}
where the second step is again at leading order in $\alpha_2<\alpha_1\ll1$. In that approximation, we note that the amplitude of the $\kappa_1^4$ piece is independent of the ratio of $\alpha_2/\alpha_1$. The $\kappa_1^4$ scaling is the general expectation for sharp transitions and, in most situations, it is the steepest growth in single field inflation \cite{Byrnes:2018txb,Cole:2022xqc}. The position of the dip is also consistent with the general analysis of Refs.~\cite{Leach:2001zf,Byrnes:2018txb,Cole:2022xqc}.

\begin{figure}
\includegraphics[width=0.49\columnwidth]{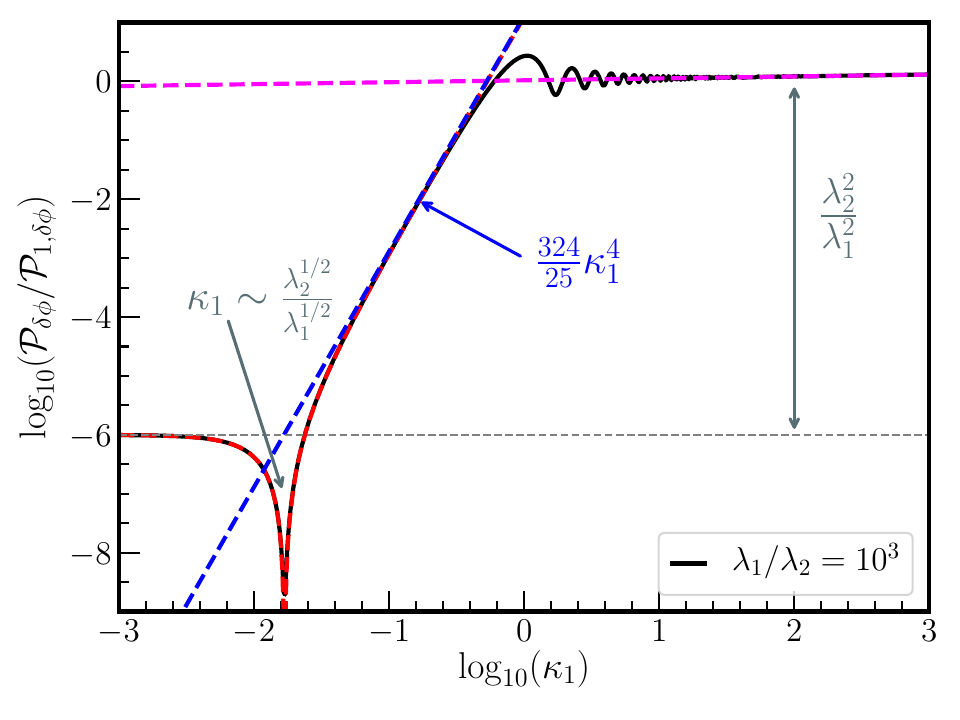}
\includegraphics[width=0.49\columnwidth]{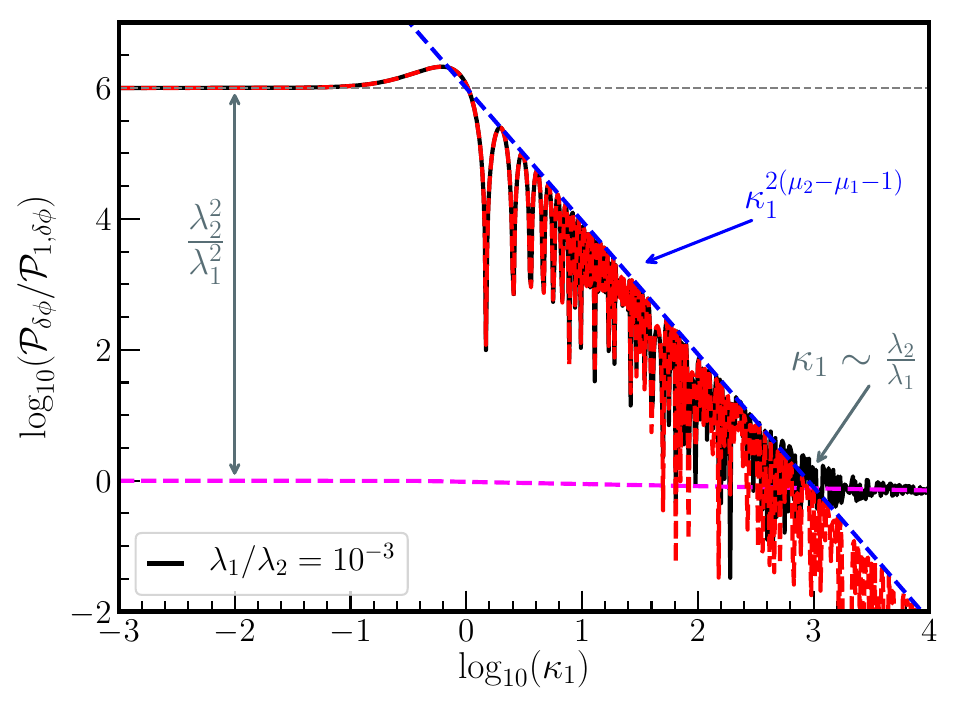}
\caption{Analytical power spectrum for $\delta\phi$ fluctuations \eqref{eq:resultdeltaphi} as a function of $\kappa_1$. In black solid lines, we show the exact analytical result from Eq.~\eqref{eq:resultdeltaphi}. On the left we respectively show in red, blue and magenta dashed lines the limiting behaviors for $\alpha_2<\alpha_1$ for low $\kappa_1$, Eqs.~\eqref{eq:lowklimit} and \eqref{eq:k4}, and large $\kappa_1$, Eq.~\eqref{eq:largek}. On the right the red, blue and magenta dashed lines the limiting behaviors for $\alpha_2>\alpha_1$ for low $\kappa_1$, Eqs.~\eqref{eq:lowklimit}, and large $\kappa_1$, Eq.~\eqref{eq:largeksupp}.\label{fig:ana}}
\end{figure}

In the other limit, that is for $\kappa_1\gg1$, we have that
\begin{align}\label{eq:largek}
|C_2+D_2|^2(\kappa_1\gg1)\approx |C_1|^2\times&\frac{\kappa_1^{2 (\mu_1-\mu_2)}\Gamma^2[\mu_2]}{2 \Gamma^2[\mu_1]}\mu_1^{2 \mu_1-1} \mu_2^{1-2 \mu_2}
   \omega_1^{-4 \mu_2-2} \omega_2^{4 \mu_2-2}\nonumber\\&\times
   \left(\omega_1^4+\omega_2^2+\left(\omega_1^4-\omega_2^4\right) \sin\left[
   \pi\mu_2 -4 \kappa_2\mu_2\right]\right)\,.
\end{align}
Although the oscillations in Eq.~\eqref{eq:largek} are subdominant, they share the same frequency of the larger oscillations, which is about $4\mu_2\frac{\omega^2_1}{\omega_2^2}$. This frequency is twice the argument of the Hankel functions at leading order in $\alpha_2<\alpha_1\ll1$, that is $x_{\star1}\approx x_{\star 2}\approx 3\kappa_1$. Besides the oscillations, we checked that 
\begin{align}
\lim_{\kappa_1\to \infty}|C_2+D_2|^2\approx|C_1(1\to2)|^2\,,
\end{align}
where $C_1(1\to2)$ is $C_1$ \eqref{eq:C11} but replacing $\mu_1\to\mu_2$ and $\kappa_1\to\kappa_2$,
as if the very short wavelength modes were in the Bunch-Davies vacuum since the start. Thus, our analytical solution recovers both the expectations for long and short wavelengths. Lastly, although we have not found any simple formula for the largest oscillations, we find that the maximum lies at $\kappa_1\approx 1$ with amplitude of the order
\begin{align}
|C_2+D_2|^2(\kappa_1\approx 1)\approx 0.45\times|{\cal C}(\kappa_1=1)|^2\,,
\end{align}
and that the relative amplitude with the second peak is about $1.5$. We obtained the number $\sim 0.45$ by taking the limit $\alpha_1,\alpha_2\to0$ of $\kappa_1\times|C_2+D_2|^2/|{\cal C}|^2$, which is independent of $\alpha_1$ and $\alpha_2$, and then we found the maximum value numerically. We show the accuracy of the approximations and explain the features in the primordial spectrum on the left plot of Fig.~\ref{fig:ana}.

From Fig.~\ref{fig:ana} we also see that our results are qualitatively similar to Starobinsky’s piecewise linear potential shown in Fig.~2 of Ref.~\cite{Pi:2022zxs}. The reason for the similarity is the fact that around the matching point, the exponential potential is well approximated by the linear one. The plateaus at low and high $\kappa$ are similar in slope due to the slow-roll approximation but its precise value is different. The main differences with \cite{Pi:2022zxs} are that we do not assume $H$ to be a constant and that the exponential potential allows us to solve directly $\delta\phi$ instead of ${\cal R}$.

\subsection{Suppression of fluctuations \label{sec:suppression}}

Our formulas are also valid in the case where $\lambda_2>\lambda_1$, which suppresses fluctuations on small scales. This case might also be relevant for PBHs and induced GWs. First, a suppression of the spectrum can be useful if after the enhancement inflation returns to the original slow-roll phase. This model would include a second transition point and a two-piecewise potential. Second, we later find in \S~\ref{sec:deltaN} using the $\delta N$ formalism that an abrupt end from the ultra-slow-roll transition phase due to, e.g., a new stage with $\lambda_2>\lambda_1$ yields larger non-Gaussianity and an exponential tail of the PDF. For these reasons, we use our exact formulas to study the behavior of the decay in the primordial spectrum and understand the main features. This would be a good approximation if the second transition takes place a few \textit{e}-folds after the first matching point.

We start with the low $\kappa_1$ limit. In this case, i.e. when $\kappa_1\ll1$, we recover Eq.~\eqref{eq:lowklimit}. However, we find no cancellations when $\alpha_2>\alpha_1$. In the limit when $\kappa_1\to 0$ we have that $P_{\delta\phi}\propto{\lambda_2^2}/{\lambda_1^2}$, which is enhanced with respect to the small scale spectrum. The large $\kappa_1$ limit now corresponds to $\kappa_1\gg \alpha_2/\alpha_1$, so that we can drop the first term in the right hand side of Eqs.~\eqref{eq:C2} and \eqref{eq:D2}. This limit is also given by Eq.~\eqref{eq:largek}, which recovers the amplitude of fluctuations in the second phase as if no transition happened. The different feature with respect to the $\lambda_2<\lambda_1$ case is that the decay is approximated by
\begin{align}\label{eq:largeksupp}
|C_2+D_2|^2(1\ll \kappa_1&\ll \alpha_2/\alpha_1)=|C_1|^2\times\frac{\kappa_1^{2(
   \mu_1-\mu_2-1)} \Gamma[\mu_2]^2 }{2 \alpha_1^2 \Gamma[\mu_1]^2}\nonumber\\&\times(\alpha_1-\alpha_2)^2{\mu_1}^{2 \mu_1-1} \mu_2^{1-2
   \mu_2} 
   (\omega_2/\omega_1)^{4
   \mu_2+2}\left(1-\sin \left[
   \mu_2\pi -4\mu_2 \kappa_2\right]\right)\,.
\end{align}
We see that while the enhancement of fluctuations led to a $\kappa_1^4$ growth, the decay instead follows a milder power-law given by $\kappa_1^{2(\mu_1-\mu_2-1)}\sim \kappa_1^{-2}$. We also note that during the decay there are ${\cal O}(1)$ oscillations linear in $\kappa_1$ with frequency given by $4\mu_2\frac{ \omega_1^2}{\omega_2^2}$. We show the analytical solution and the different features in the power spectrum on the right plot of Fig.~\ref{fig:ana}. This possibility was briefly considered by Ref.~\cite{Pi:2022zxs} in Starobinsky’s piecewise linear potential but the change in slopes was taken to be much milder. However, we also expect the main features to be similar.

\subsection{The induced GW signal \label{sec:GWs}}

\begin{figure}
\includegraphics[width=0.49\columnwidth]{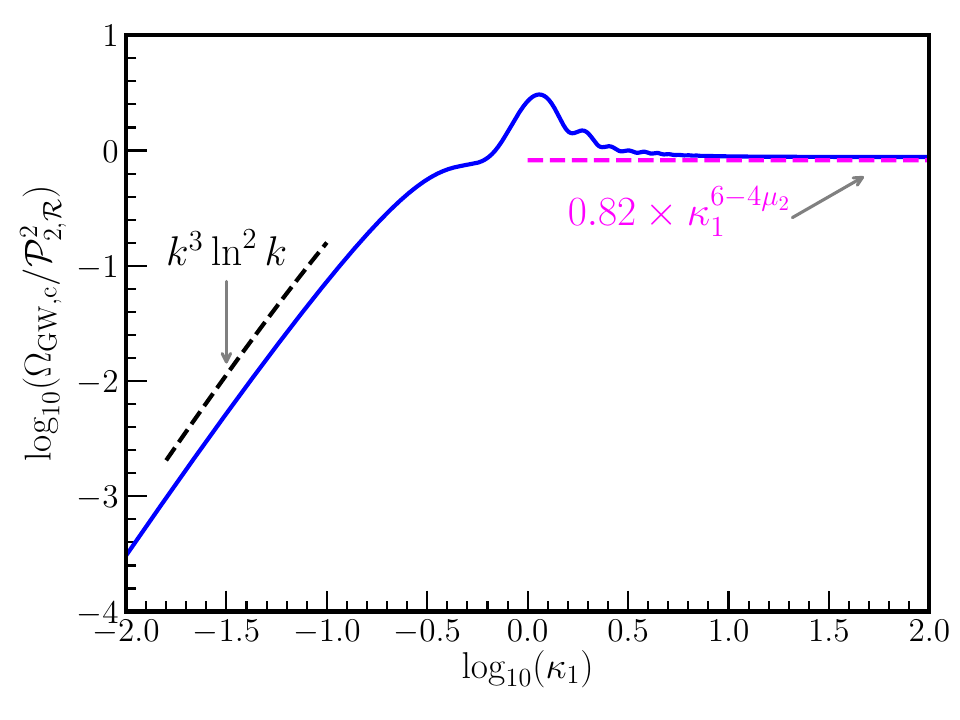}
\includegraphics[width=0.49\columnwidth]{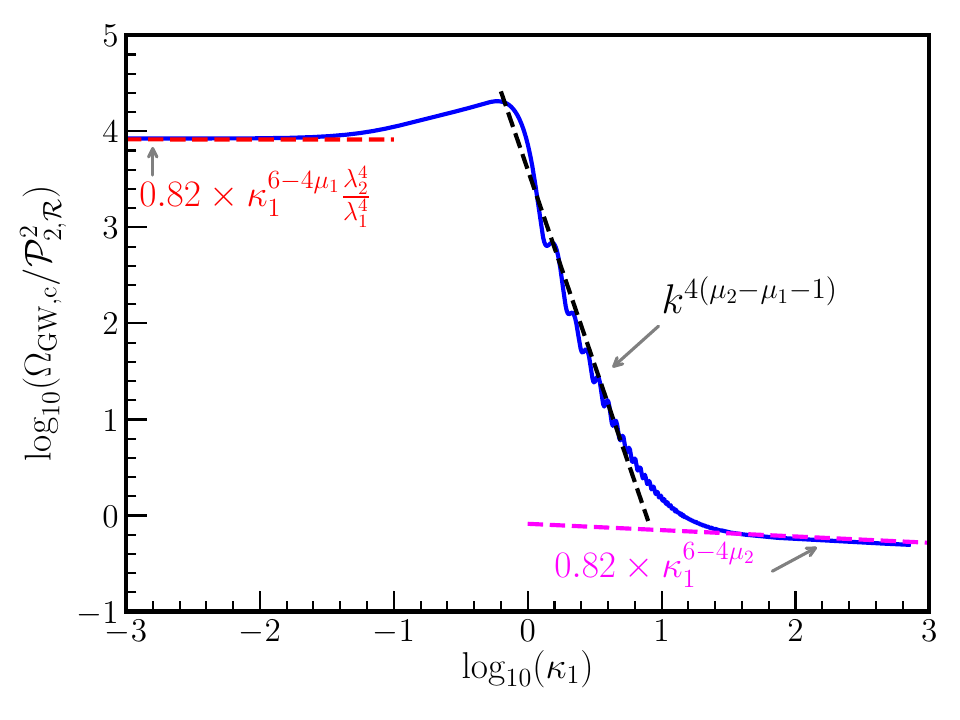}
\caption{Induced GW spectrum with amplitude normalized to ${\cal P}_{2,{\cal R}}^2$ (Eq.~\eqref{eq:slowroll} with the parameters of the second phase) as a function of $\kappa_1$. In solid blue lines we show the results of numerical integration with \href{https://github.com/Lukas-T-W/SIGWfast/releases}{\textsc{SIGWfast}} \cite{2022arXiv220905296W} and in dashed lines the behaviors described in \S~\ref{sec:GWs}. On the left, we show the case when $\alpha_1>\alpha_2$ with $\alpha_2=10^{-3}\,\alpha_1$. See how the low-frequency tail follows the universal scaling of $k^3$ with a logarithmic correction \cite{Cai:2018dig,Yuan:2019wwo}. To be precise, in our model, the induced GW spectrum also present a plateau for very low frequencies but it is $3\times 4=12$ orders of magnitude below the high plateau. Therefore, we ignore it. On the right figure, we show the case $\alpha_1<\alpha_2$ with $\alpha_2=10\,\alpha_1$. In this case, the two plateaus are $4$ orders of magnitude apart. \label{fig:GWs}}
\end{figure}

An important signal to test the enhancement of primordial fluctuations is the associated induced gravitational waves after inflation. With an analytical expression for the primordial spectrum in the piecewise exponential potential, we can also compute the predicted induced GW spectrum for different parameters. We use \href{https://github.com/Lukas-T-W/SIGWfast/releases}{\textsc{SIGWfast}} \cite{2022arXiv220905296W} to compute the resulting induced GW spectrum, which uses the analytical expressions of Refs.~\cite{Espinosa_2018,Kohri:2018awv,Domenech:2019quo,Domenech:2020kqm}, assuming Gaussian primordial fluctuations. We show our results in Fig.~\ref{fig:GWs}. Although we show the spectrum for specific choices of the parameters, we provide explanations for the different slopes in the induced GW spectrum below. We neglect induced GW during inflation as they are expected to be much smaller within single field inflation \cite{Inomata:2021zel}. Note that this might not be the case in general, specially when several fields are involved \cite{Fumagalli:2021mpc}.

We first approximate the primordial spectrum by two broken power-laws as
\begin{align}
{\cal P}_{\cal R}(\kappa)\propto A_{\cal R}\times\left\{
\begin{aligned}
 &\kappa^{3-2\mu_1}& (\kappa <\kappa_a)\\
 &\kappa^{n}&(\kappa_a<\kappa <\kappa_b)\\
 &\kappa^{3-2\mu_2}&(\kappa >\kappa_b)\\
\end{aligned}
\right.\,,
\end{align}
where $n=4$, $\kappa_a\sim \sqrt{\alpha_2/\alpha_1}$, $\kappa_b\sim 1$ for $\alpha_1>\alpha_2$ and $n=2(\mu_2-\mu_1-1)$, $\kappa_a\sim 1$ and $\kappa_b\sim \alpha_2/\alpha_1$ for $\alpha_1<\alpha_2$. $A_{\cal R}$ is an arbitrary amplitude of the power spectrum to be determined from Eq.~\eqref{eq:resultdeltaphi}. Then we use the results of Refs.~\cite{Cai:2019cdl,Yuan:2019wwo,Liu:2020oqe,Atal:2021jyo,Balaji:2022dbi} to give the asymptotic behavior of the induced GW spectrum in the three different regions. For the almost flat plateaus we obtain
\begin{align}
\Omega_{\rm GW,c}\approx 0.82 A_{\cal R}^2\times \kappa^{6-4\mu_{i}}\propto {\cal P}^2_{\cal R}(\kappa)\,,
\end{align}
where $i=\{1,2\}$ depending on which plateau we consider and the numerical coefficient is computed for the exact flat case ($\mu\sim 3/2$). On one hand, for $\alpha_1>\alpha_2$ where the spectrum grows as $\kappa^4$ we find that
\begin{align}
\Omega_{\rm GW,c}(\kappa_a<\kappa <\kappa_b)\propto A_{\cal R}^2\times \kappa^3\ln^2\kappa\,,
\end{align}
which follows the expected universal low-frequency scaling for induced GWs \cite{Cai:2018dig,Cai:2019cdl,Yuan:2019wwo}.
On the other hand, for $\alpha_1<\alpha_2$ where there is a suppression of the spectrum as $\kappa^{-2}$ we have that
\begin{align}
\Omega_{\rm GW,c}(\kappa_a<\kappa <\kappa_b)\propto A_{\cal R}^2\times \kappa^{4(\mu_2-\mu_1-1)}\propto {\cal P}^2_{\cal R}(\kappa)\,.
\end{align}
In a nutshell, if the spectrum is not steep enough, which corresponds to $n<3/2$ for the growth or $n>-4$ for the suppression, the induced GW spectrum is proportional to the squared of the primordial spectrum. We show in Fig.~\ref{fig:GWs} that the approximate power-law behaviors described above fit well with the numerically computed induced GW spectrum. The bump in the left plot of Fig.~\ref{fig:GWs} is due to the typical resonance of induced GWs during the radiation of the largest peak in the primordial spectrum, see Fig.~\ref{fig:ana}. It is also interesting to note that the induced GW spectrum is not very sensitive to the oscillations in the primordial spectrum. This is because, as explained in Refs.~\cite{Fumagalli:2020nvq,Fumagalli:2021cel,Witkowski:2021raz}, the induced GWs are a secondary effect and oscillations tend to smear out. That being said, ${\cal O}(1)$ oscillations in the primordial spectrum do become visible in the induced GW spectrum \cite{Fumagalli:2020nvq,Braglia:2020taf,Zhou:2020kkf}.

The mass function of the PBH counterpart is studied in Ref.~\cite{Pi:2022zxs} for Starobinsky’s piecewise linear potential but due to the similarities of the resulting primordial spectrum (see \S~\ref{sec:enhancement}), it will be very similar for the piecewise exponential potential. In addition to the similarities, the high amplitude plateau for large $\kappa_1$ might be producing too many small PBHs, as commented in \cite{Pi:2022zxs}. To avoid this, we should invoke the suppression of fluctuations as in \S~\ref{sec:suppression}. Furthermore, the main point of this paper is the analytical treatment in an exact model and not the phenomenology of PBHs. For these reasons, we omit the calculation of the PBH mass function and refer the interested reader to Ref.~\cite{Pi:2022zxs} for further details.

\section{\texorpdfstring{$\delta N$}{} formalism and the tail of the distribution\label{sec:deltaN}}

In \S~\ref{sec:perturbations} we have studied the enhancement/suppression of linear primordial fluctuations during inflation and their induced GW signal after inflation. For the calculation of the induced GWs, we assumed that linear fluctuations are mostly Gaussian, which is expected to be a good approximation since the GWs are induced by typical fluctuations. However, PBHs form from large and very rare fluctuations, which highly depend on the tail of the PDF of primordial fluctuations. The $\delta N$ formalism provides a way to study the probability distribution of large primordial curvature fluctuations, under the separate universe approach \cite{Starobinsky:1985ibc,Salopek:1990jq,Sasaki:1995aw,Wands:2000dp,Lyth:2004gb}.

Intuitively speaking, the separate universe approach tells us that each Hubble patch evolves according to the background equations of motion, including the contribution of any superhorizon fluctuation, say $\delta\phi$, to the energy density inside each patch. The difference in the “local” expansion of a given Hubble patch is then determined by the fluctuation $\delta\phi$ in that patch, the size of which is randomly distributed according to the PDF. At the same time, the difference in the expansion measured in terms of \textit{e}-folds from the end of inflation, say $\delta N$, is related to the curvature perturbation ${\cal R}$ at the end of inflation, i.e. $\delta N={\cal R}$. In this way, the $\delta N$ formalism gives a non-linear relation between $\delta\phi$ (evaluated in the spatially flat slices) and  ${\cal R}$. 

In our model, we have exact solutions for the background equations of motion and for the fluctuations $\delta\phi$, both of which are important for the $\delta N$ formalism. Since we have the exact solutions, we study each step relevant to $\delta N$ formalism in some detail. First, we will have a closer look at the general phase space trajectories. Second, we will study the solutions of superhorizon perturbations from the perturbed background equations and the equations for the mode functions. In this way, we will understand from which moment the $\delta N$ formalism is applicable. Lastly, we will use $\delta N$ formalism to investigate non-Gaussianities and the tail of the PDF. Recent references in this direction are \cite{Biagetti:2018pjj,Atal:2019cdz,Ezquiaga:2019ftu,Atal:2019erb,Pi:2021dft,Cai:2022erk,Abe:2022xur,Pi:2022ysn}.

\subsection{Phase space trajectories and number of \textit{e}-folds}

To use the $\delta N$ formalism, we need the total number of \textit{e}-folds from the end of inflation to an initial time with general initial conditions, given by quantum fluctuations exiting the Hubble horizon. In other words, we have to compute how a given off-attractor trajectory modifies the total number of \textit{e}-folds. In section \S~\ref{sec:PLreview} we presented the general solutions in a parametric form, namely we have $\phi(\mathrm{Z})$ and $N(\mathrm{Z})$ in Eqs.~\eqref{eq:phigeneral} and \eqref{eq:Ngeneral}. 
To study trajectories in the phase space though, it is more useful to express $\mathrm{Z}$ in terms of $\phi_N$ via Eq.~\eqref{eq:phiN}. To be consistent with the recent literature, we use the notation $\pi\equiv \phi_N$ as in Eq.~\eqref{eq:phasespace1}. Then, we arrive at
\begin{align}\label{eq:phigeneralphiN}
\phi&=\phi_\star+\frac{1}{\omega^2}\sqrt{\frac{2}{3}}\left({\rm arctanh}\left[\frac{\pi_\star}{\sqrt{6}}\right]-{\rm arctanh}\left[\frac{\pi}{\sqrt{6}}\right] + \alpha \ln \left(\frac{\pi_\star-\sqrt{6}\alpha}{\pi-\sqrt{6}\alpha}\sqrt{\frac{\pi^2-6}{\pi_\star^2-6}}\right)\right)\,,\\
N&= N_\star+\frac{1}{3\omega^2} \left(\alpha\, {\rm arctanh}\left[\frac{\pi_\star}{\sqrt{6}}\right]-\alpha\,{\rm arctanh}\left[\frac{\pi}{\sqrt{6}}\right]+ \ln \left(\frac{\pi_\star-\sqrt{6}\alpha}{\pi-\sqrt{6}\alpha}\sqrt{\frac{\pi^2-6}{\pi_\star^2-6}}\right)\right)\,.\label{eq:NgeneralphiN}
\end{align}
From Eq.~\eqref{eq:phigeneralphiN} we find that any trajectory in phase space satisfies
\begin{align}\label{eq:generalphasespace}
F(\phi,\pi)\equiv\frac{\left|\pi-\sqrt{6}\alpha\right|^{2\alpha}}{\left(\sqrt{6}-\pi\right)^{\alpha+1}\left(\sqrt{6}+\pi\right)^{\alpha-1}}\times e^{{\sqrt{6}\omega^2}\phi}={\rm constant}\,.
\end{align}
The case where the constant is exactly zero corresponds to the exact attractor solution.
For the number of \textit{e}-folds it is more convenient to use Eq.~\eqref{eq:phigeneralphiN} to replace the logarithm in Eq.~\eqref{eq:NgeneralphiN} for $\phi$, which leads us to
\begin{align}\label{eq:generalN2}
N-N_\star=\frac{\phi-\phi_\star}{\sqrt{6}\alpha}+\frac{1}{3\alpha}\left({\rm arctanh}\left[\frac{\pi}{\sqrt{6}}\right]-{\rm arctanh}\left[\frac{\pi_\star}{\sqrt{6}}\right]\right)\,.
\end{align}
Constant $N$ slices in the phase space are then given by
\begin{align}
%\pi=\sqrt{6}\tanh\left(3\alpha (N-N_\star)-\sqrt{\frac{3}{2}}({\phi-\phi_\star})+{\rm arctanh}\left[\frac{\pi_\star}{\sqrt{6}}\right]\right)
%\pi=-\sqrt{6}\tanh\left(\sqrt{\frac{3}{2}}{\phi}+{\rm constant}\right)
\phi+\sqrt{\frac{2}{3}}{\rm arctanh}\left[\frac{\pi}{\sqrt{6}}\right]={\rm constant}\,.
\end{align}

Although we can work with the general solution, we proceed with some simplifications for clarity. We will limit ourselves in the regime where $\pi\ll\sqrt{6}$ so that $3H^2\approx V$, which corresponds to the slow-roll regime. In the slow-roll regime, Eqs.~\eqref{eq:phigeneralphiN} and \eqref{eq:NgeneralphiN} are approximately given by
\begin{align}\label{eq:phigeneralsmallphiN}
\phi-\phi_\star&\approx\frac{1}{3\omega^2}\left({\tilde\pi_\star}-{\tilde\pi}+\sqrt{6}\alpha\,\ln \frac{\tilde\pi_\star}{\tilde\pi}\right)\,,\\
N-N_\star&\approx \frac{1}{3\omega^2}\left(\frac{\alpha}{\sqrt{6}} \left({\tilde\pi_\star}-{\tilde\pi}\right)+ \ln \frac{\tilde\pi_\star}{\tilde\pi}\right)=\frac{1}{\sqrt{6}\alpha}\left(\phi-\phi_\star+\frac{1}{3}\left({\tilde\pi}-{\tilde\pi_\star}\right)\right)\,,\label{eq:NgeneralsmallphiN}
\end{align}
where for compactness we have introduced
\begin{align}
\tilde\pi=\pi-\sqrt{6}\alpha.
\end{align}
Note that the variable $\tilde\pi$ vanishes in the attractor trajectory.
These simplifications also allow us to compare our results with the case of ultra-slow-roll, which corresponds to the limit $\alpha\to0$. In the ultra-slow-roll limit we respectively have from Eqs.~\eqref{eq:phigeneralsmallphiN} and \eqref{eq:NgeneralsmallphiN} that $N-N_\star=\frac{1}{3}\ln(\pi_\star/\pi)$ and $\pi-\pi_\star=3(\phi-\phi_\star)$, as in Ref.~\cite{Cai:2018dkf}. Now, in contrast with the implicit relations given by the exact solutions, we may isolate $\tilde\pi$ in terms of $\phi$ by inverting Eq.~\eqref{eq:phigeneralsmallphiN}, namely
\begin{align}\label{eq:lambert}
\frac{\tilde\pi}{\sqrt{6}\alpha}=W\left[\frac{\tilde\pi_\star}{\sqrt{6}\alpha}e^{\frac{\tilde\pi_\star-3\omega^2(\phi-\phi_\star)}{\sqrt{6}\alpha}}\right]\,,
\end{align}
where $W$ is the Lambert function. Note that for $\alpha_2<\alpha_1$ the argument of $W$ is always positive but for $\alpha_1<\alpha_2$ we have that $\tilde\pi_\star$ is negative in the second phase and so one must carefully use the appropriate branch of the Lambert function. With Eq.~\eqref{eq:lambert} the number of \textit{e}-folds is a function of $\phi$, $\phi_\star$ and $\pi_\star$ only.

\begin{figure}
\includegraphics[width=0.49\columnwidth]{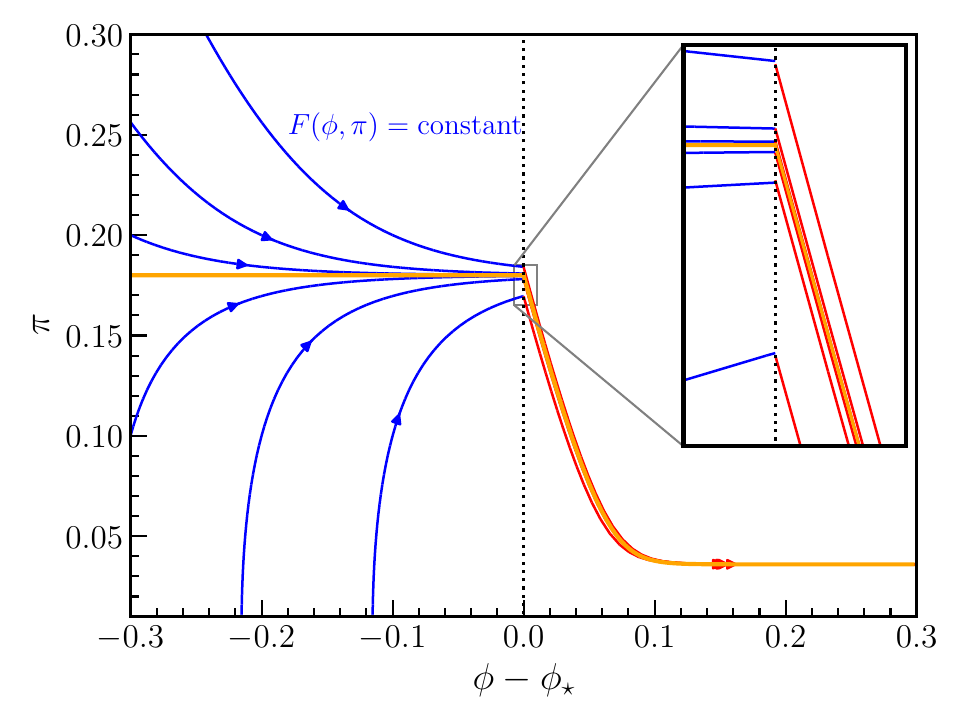}
\includegraphics[width=0.49\columnwidth]{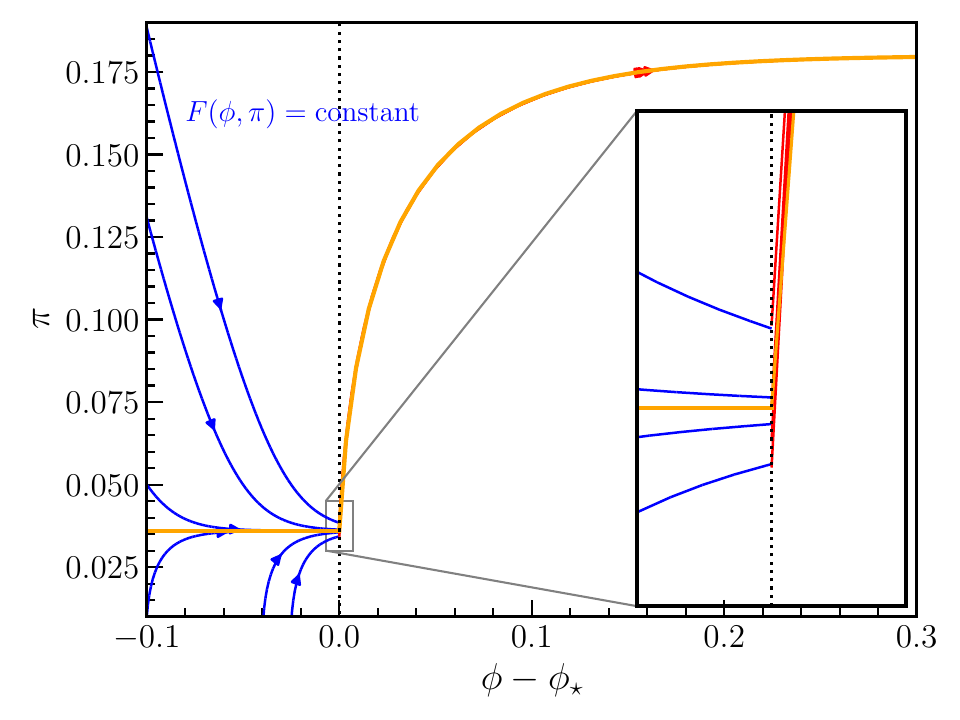}
\caption{Illustration of the trajectories involved in the $\delta N$ formalism: a quantum fluctuation $\delta\phi$ exits the horizon and takes the trajectory out of the attractor background solution (in orange) to a non-attractor one (in blue for the first phase and in red for the second). The new trajectories approach the attractor exponentially fast and by the end of inflation, all of them end up in the second attractor. On the left and right figures, we respectively show the cases when $\alpha_2<\alpha_1$ and $\alpha_2>\alpha_1$. \label{fig:plotsphasespacedeltaN}}
\end{figure}

Before proceeding, let us study two limiting cases of Eq.~\eqref{eq:lambert}: close to the attractor and far off the attractor. First, we note from Eq.~\eqref{eq:lambert} that whenever $\tilde\pi_\star\sim0$, i.e. we are close to the attractor, we have
\begin{align}\label{eq:lambertnearatt}
\tilde\pi\approx \tilde\pi_\star\times e^{-\frac{3\omega^2}{\sqrt{6}\alpha}(\phi-\phi_\star)}\,,
\end{align}
where we used that $W[x\ll1]\approx x$. We thus see that the system approaches the attractor exponentially fast. However, if we are very far from the attractor, e.g. when $\sqrt{6}\alpha\ll\tilde\pi_\star$ or $\alpha\ll\phi_\star-\phi$, we have that $W[x\gg1]\approx \ln x$ and then
\begin{align}\label{eq:lambertUSR}
{\tilde\pi}\approx{\tilde\pi_\star-3\omega^2(\phi-\phi_\star)}\,,
\end{align}
almost like in ultra-slow-roll. We show examples of trajectories in phase space in the piecewise potential in Fig.~\ref{fig:plotsphasespacedeltaN}. We note that Eqs.~\eqref{eq:phigeneralphiN} and \eqref{eq:NgeneralphiN}, and the simplified version Eqs.~\eqref{eq:phigeneralsmallphiN} and \eqref{eq:NgeneralsmallphiN}, are valid for both phases with the appropriate parameters of each phase. 

\subsection{Evolution of superhorizon scalar fluctuations \label{subsec:evolution}}

The $\delta N$ formalism assumes that $\delta\phi$ is constant on superhorizon scales and that $\delta\pi$ quickly decays and becomes unimportant. However, in a sudden transition, this might not be the case and as we proceed to show, for some scales, $\delta\pi$ does not have enough time to decay sufficiently for the standard $\delta N$ approach to be valid.

Let us start by comparing the equations of motion for the perturbations, Eq.~\eqref{eq:deltaphiNN} in the exact $k=0$ limit, and the perturbed background equations, Eq.~\eqref{eq:phasespace2}, in a given phase. They are respectively given by
\begin{align}\label{eq:perturbationsk0}
\delta\phi_{NN}+\frac{1}{2}(6-\pi^2)\delta\phi_N+\frac{1}{2}(6-\pi^2)(\pi-\sqrt{6}\alpha)^2\delta\phi=0\,,
\end{align}
and
\begin{align}\label{eq:perturbed}
\delta\pi_N+\frac{1}{2}(6-\pi^2)\delta\pi-\pi(\pi-\sqrt{6}\alpha)\delta\pi=0\,.
\end{align}
These two equations are equivalent when
\begin{align}\label{eq:deltapideltaphi}
\delta\pi=-\frac{1}{2\pi}(6-\pi^2)(\pi-\sqrt{6}\alpha)\delta\phi=-\frac{\pi_N}{\pi}\delta\phi\,,
\end{align}
which is the relation between the perturbations of $\delta\pi$ and $\delta\phi$ along a given background trajectory by, e.g., perturbing Eq.~\eqref{eq:generalphasespace}. % Most importantly, Eq.~\eqref{eq:deltapideltaphi} shows that for non-attractor trajectories, for which the coefficients in Eq.~\eqref{eq:deltapideltaphi} vanish, there is a direct relation between $\delta\pi$ and $\delta\phi$. 
For trajectories close to the attractor, we have that the perturbed background equations \eqref{eq:perturbed} yield
\begin{align}\label{eq:deltapisuperhorizon}
\delta\pi\propto e^{-3\omega^2N}\approx e^{-3N}\,.
\end{align}

However, superhorizon fluctuations may behave differently, as we proceed to show. During the first phase, $\delta\phi_k$ on superhorizon scales is given by 
\begin{align}\label{eq:deltaphisuperhorizon1}
\delta\phi_{1k} &\approx C_1\times \left(1+\frac{x^2}{4 (\mu_1
   -1)}+\frac{\pi  4^{-\mu_1 } (i-\cot (\pi  \mu_1
   )) }{\Gamma[\mu_1]
   \Gamma[1+\mu_1]}x^{2 \mu_1 }\right)\nonumber\\&
   \approx C_1\times \left(1+\frac{(\mu_1\kappa_1)^{2}}{\mu_1
   -1}e^{-2(1-3\alpha_1^2)(N_1-N_\star)}+\frac{\pi  (i-\cot (\pi  \mu_1
   )) }{\Gamma[\mu_1]
   \Gamma[1+\mu_1]}(\mu_1\kappa_1)^{2\mu_1}e^{-3\omega_1^2(N_1-N_\star)}\right)\,,
\end{align}
where we expanded Eq.~\eqref{eq:exactsolpertminus} for small arguments. We recover here the subscript “$k$” to denote quantum fluctuations and no subscript for the perturbed background solutions. As clear from Eq.~\eqref{eq:deltaphisuperhorizon1}, the leading order term in $\delta\pi_{1k}=\delta\phi_{1k,N}$ is $\delta\pi_{1k}\propto e^{-2N}$, which decays slower than the perturbed background Eq.~\eqref{eq:deltapisuperhorizon}, namely $\delta\pi\propto  e^{-3N}$. The slower decay per se is not a problem but it may become important at the transition, as we shall see later.

For superhorizon fluctuations after the transition, we have from Eq.~\eqref{eq:deltaphi2} that
\begin{align}\label{eq:superhorizondeltaphi2}
\delta\phi_{2k} \approx (C_2+D_2)& \left(1+\frac{(\mu_2\kappa_2)^{2}}{\mu_2
   -1}e^{-2(1-3\alpha_2^2)(N_2-N_\star)}\right)\nonumber\\&+(C_2-D_2)\frac{\pi  (i-\cot (\pi  \mu_2
   )) }{\Gamma[\mu_2]
   \Gamma[1+\mu_2]}(\mu_2\kappa_2)^{2\mu_2}e^{-3\omega_2^2(N_2-N_\star)}\,.
\end{align}
We see that the coefficient of the constant and the $e^{-2N}$ terms are the same and, therefore, once a mode is superhorizon the $e^{-2N}$ contribution quickly becomes negligible. This time, however, depending on the value of $k$ the coefficient of the third term in the Taylor series \eqref{eq:superhorizondeltaphi2}, the one with $e^{-3N}$, dominates. By equating the constant term with the latter we find that $\delta\phi_{2k}$ can only be regarded as a constant for
\begin{align}\label{eq:negligible}
N_2>N_{2,\rm min}= N_\star&+\frac{1}{3\omega_2^2}\ln\left|\frac{C_2-D_2}{|C_2+D_2|}\frac{\pi  (i-\cot (\pi  \mu_2
   )) }{\Gamma[\mu_2]
   \Gamma[1+\mu_2]}(\mu_2\kappa_2)^{2\mu_2}\right|\,.
\end{align}
Note that since $\mu\sim 3/2$ one may drop the $\cot (\pi  \mu_2)$ term.
Eq.~\eqref{eq:negligible} in the $\kappa_1\ll1$ limit gives
\begin{align}
N_{2,\rm min}= N_\star&+\frac{1}{3\omega_2^2}\ln\left|1-\frac{\alpha_2}{\alpha_1}\right|\nonumber\\&-\frac{1}{3\omega_2^2}\ln\left|\frac{\alpha_2}{\alpha_1}+\frac{\kappa_1^2
   }{\alpha_1 }\left(\frac{\mu_1 }{\mu_1-1} 
   (\alpha_1-\alpha_2 \mu_1)-\frac{\mu_2}{\mu_2+1} \frac{\omega_1^4}{\omega_2^4} (\alpha_1+\alpha_2\mu_2)\right)\right|\,.
\end{align}
In particular, we see that for $\kappa_1^2\ll\alpha_2/\alpha_1 $ we have $N_{2,\rm min}-N_\star\approx\tfrac{1}{3}\ln(\alpha_1/\alpha_2) $ which corresponds to the end of the ultra-slow-roll phase in the case where $\alpha_1>\alpha_2$ (see the discussion around Eq.~\eqref{eq:phiNpluse-fold}). For $\alpha_1<\alpha_2$, we find that the $e^{-3N}$ term is barely important, and $\delta\phi$ can be regarded as constant soon after horizon crossing.

\begin{figure}
\includegraphics[width=0.49\columnwidth]{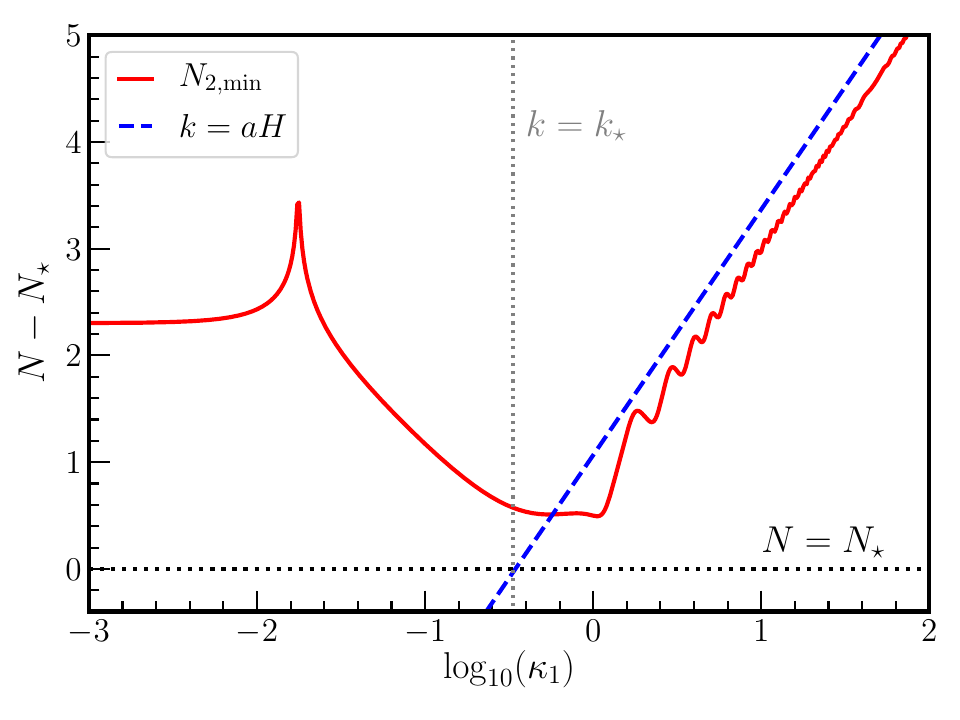}
\includegraphics[width=0.49\columnwidth]{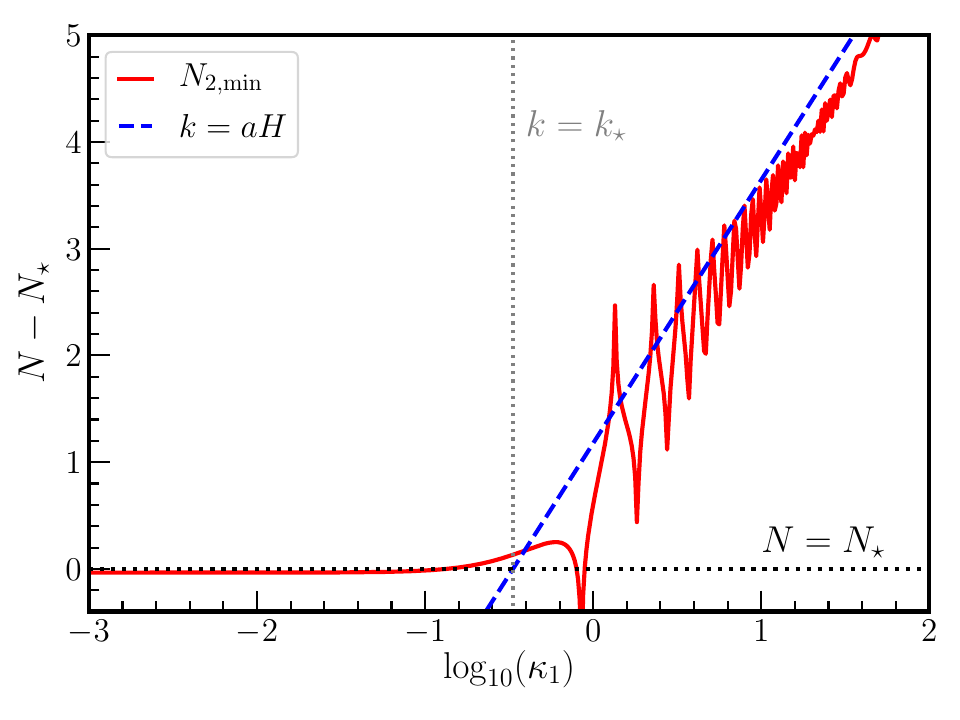}
\caption{Number of \textit{e}-folds versus wavenumber $\kappa_1$. Blue dashed lines show the relation at horizon crossing. The black dotted line shows the matching point $N=N_\star$. The red solid line is the minimum number of \textit{e}-folds, Eq.~\eqref{eq:negligible}, after which a mode $\delta\phi_{2k}$ may be regarded as constant. Thus, the $\delta N$ formalism where $\delta\phi$ is assumed to be constant is only valid for $N$ above the red line. On the left we show the case where $\alpha_2=10^{-3}\,\alpha_1$ and on the right when $\alpha_2=10^3\,\alpha_1$. See how only for $\alpha_2<\alpha_1$ this effect is important for $\kappa_1<1$. \label{fig:plotscross}}
\end{figure}

In Fig.~\ref{fig:plotscross} we show the value of $N_{2,\rm min}$ for $\alpha_1>\alpha_2$ on the left and $\alpha_1<\alpha_2$ on the right. We also show the \textit{e}-fold corresponding to horizon crossing, which using Eqs.~\eqref{eq:aH}, \eqref{eq:phase1} and \eqref{eq:zN} is given by
\begin{align}\label{eq:Nhor}
N-N_\star=\left(\mu-\frac{1}{2}\right)\ln\left(\frac{k}{k_\star}\right)\,.
\end{align}
From Fig.~\ref{fig:plotscross} it is clear that for $\alpha_1>\alpha_2$ and $\kappa_1<1$, the fluctuation $\delta\phi$ is not constant until few \textit{e}-folds after the matching point. This has implications for the standard $\delta N$ approach where $\delta\phi$ is assumed to be constant. For $\alpha_2>\alpha_1$ this effect is only important for scales close to the matching scale $k_\star$. We also note that Eq.~\eqref{eq:negligible} for the scale corresponding to the dip in the primordial spectrum yields a divergent $N_{2,\rm min}$. This is because the constant term vanishes exactly. This is artificial as quantum one-loop corrections are expected to yield a non-vanishing dip \cite{Franciolini:2023lgy,Fumagalli:2023loc}.\footnote{For an interesting discussion on the effects of one-loop corrections on large scales see Refs.~\cite{Kristiano:2022maq,Riotto:2023hoz,Kristiano:2023scm,Riotto:2023gpm,Firouzjahi:2023aum,Firouzjahi:2023ahg,Franciolini:2023lgy,Tasinato:2023ukp,Cheng:2023ikq,Fumagalli:2023hpa}. Also see Ref.~\cite{Inomata:2022yte} for the one-loop calculation of models with enhancements due to resonances in oscillating potential, which shows that the resonant models might be out of perturbative control.}

\subsection{Linear \texorpdfstring{$\delta N$}{}  in the piecewise exponential potential}

In \S~\ref{subsec:evolution} we have found that $\delta\phi$ can only be regarded as constant some time after the matching point. This means that strictly speaking, we can only use the standard $\delta N$ approach for $N>N_\star$. Nevertheless, one may wonder if $\delta N$ still gives meaningful results for $N<N_\star$. Let us show that, if that is the case, the momentum perturbation, which is often neglected, of modes which left the horizon at $\phi<\phi_\star$ plays a very important role. 

The total number of $e$-folds from the end of inflation back to an initial time for $\phi<\phi_\star$ reads
\begin{align}
N_{\rm tot}(\phi_i<\phi_\star)=N_{2e}-N_{2\star}+N_{1\star}-N_{1i}\,,
\end{align}
where the subscripts $1$ and $2$ respectively refer to Eq.~\eqref{eq:generalN2}, or Eq.~\eqref{eq:phigeneralsmallphiN}, evaluated during the first and second phase and $i$ and $e$ respectively refer to the initial and end points. The difference in the number of \textit{e}-folds from a perturbed trajectory then reads
\begin{align}\label{eq:deltaNNNN}
\delta N=N_{\rm tot}(\phi_i+\delta\phi,\pi_i+\delta\pi)-N_{\rm tot}(\phi_i,\pi_i)\,,
\end{align}
which applied to the current case yields
\begin{align}\label{eq:deltaNphi1less}
\delta N(\phi_i<\phi_\star)\approx-\frac{1}{\sqrt{6}\alpha_1}\left(\delta\phi+\frac{1}{3}\left({\delta\pi}-{\delta\pi}_{1\star}\right)\right)-\frac{1}{3\sqrt{6}\alpha_2}\left({\delta\pi_{2\star}}-{\delta\pi_e}\right)\,.
\end{align}
Interestingly, we see that while for $\phi_i\ll\phi_\star$ the $\delta N$ gets the right normalization, i.e. ${\cal R}=-\frac{1}{\sqrt{6}\alpha_1}\delta\phi$, this is not enough to explain the growth of fluctuations for $\kappa_1<1$. In that respect, we see that the coefficient in front of $\delta\pi_{2\star}$ in Eq.~\eqref{eq:deltaNphi1less} is enhanced by a factor $\alpha_1/\alpha_2$ with respect to $\delta\phi$. Thus, for modes close to the matching point, $\delta\pi$ did not have enough time to decay and dominates the $\delta N$ calculation. However, the results now depend on the time when $\delta\pi$ is evaluated and whether one uses the results from the perturbed background equations or the solutions for perturbations, which decay differently as we showed in \S~\ref{subsec:evolution}. We found no straightforward way to generalize the $\delta N$ calculation to include the $\delta\pi$ dependence and recover the exact results of \S~\ref{sec:perturbations} for the power spectrum of curvature fluctuations. As this is out of the scope of this paper, we leave this issue for future work.

Doing the same calculation for $\phi_i>\phi_\star$ we obtain
\begin{align}\label{eq:deltaNphi2less}
\delta N(\phi_i>\phi_\star)\approx-\frac{1}{\sqrt{6}\alpha_2}\left(\delta\phi+\frac{1}{3}\left({\delta\pi}-{\delta\pi}_{e}\right)\right)\,.
\end{align}
If the initial time is sufficiently far after the transition, we may neglect the momentum contribution entirely and find that ${\cal R}=-\frac{1}{\sqrt{6}\alpha_2}\delta\phi$ which recovers the exact solution. Here we used that if $\phi_e\gg\phi_i$ the momentum at the end of inflation is exponentially suppressed.

\subsection{Local non-Gaussianity and probability distribution}

Now that the range of applicability of the $\delta N$ is clear, let us use the $\delta N$ formalism to explore the non-Gaussianity of the model and the tail of the PDF. We shall focus on $\phi_i>\phi_\star$, at times when $\delta\phi$ is constant according to the results of \S~\ref{subsec:evolution}. In that case, we have from Eq.~\eqref{eq:deltaNNNN} that 
\begin{align}\label{eq:deltaNphi2full}
\delta N(\phi_i>\phi_\star)&\approx\frac{1}{3\omega_2^2}\left(\frac{\alpha}{\sqrt{6}}\left(\delta\pi+\tilde\pi_e(\phi_i,\pi_i)-\tilde\pi_e(\phi_i+\delta\phi,\pi_i+\delta\pi)\right)+\ln\frac{\tilde\pi_e(\phi_i,\pi_i)}{\tilde\pi_e(\phi_i+\delta\phi,\pi_i+\delta\pi)}\right)\,,
\end{align}
where $\tilde\pi_e(\phi_i,\pi_i)$ is given by Eq.~\eqref{eq:lambert}, explicitly
\begin{align}
\tilde\pi_e(\phi_i,\pi_i)=\sqrt{6}\alpha_2\,W\left[\frac{\tilde\pi_i}{\sqrt{6}\alpha_2}e^{\frac{\tilde\pi_i+3\omega_2^2(\phi_i-\phi_e)}{\sqrt{6}\alpha_2}}\right]\,.
\end{align}
For the moment, let us consider that $\phi_e$ is an arbitrary point where the off-attractor dynamics ends. It may be the end of inflation or an abrupt change due to another piecewise exponential potential. We anticipate that the results of the non-Gaussianity and the PDF will depend on how close $\pi_e$ is to the attractor solution of the second phase.

Since we are evaluating the field perturbation $\delta\phi$ in the regime when it is practically a constant, we may neglect the initial momentum perturbation $\delta\pi$ as long as the system does not transition to another ultra-slow-roll like phase. In this case, we may find the “Gaussian” curvature perturbation ${\cal R}_g$ by expanding Eq.~\eqref{eq:deltaNphi2full} at linear order, which yields
\begin{align}\label{eq:deltaNphi2linear}
\delta N(\phi_i>\phi_\star)&\approx-\frac{1+\frac{\alpha_2}{\sqrt{6}}\tilde\pi_e}{\sqrt{6}\alpha_2+\tilde\pi_e}\delta\phi+{\cal O}(\delta\phi^2)\,,
\end{align}
and so we identify
\begin{align}\label{eq:Rg}
{\cal R}_g\approx-\frac{\delta\phi}{\sqrt{6}\alpha_2+\tilde\pi_e}=-\frac{\delta\phi}{\pi_e}\,,
\end{align}
where to be consistent with our approximations, namely that $\pi\ll\sqrt{6}$, we drop terms proportional to $\alpha \tilde \pi_e$.
${\cal R}_g$ is called the Gaussian variable since $\delta\phi$ is mostly Gaussian if it has a canonical kinetic term as in our case \eqref{eq:action}.

We find the value of local non-Gaussianity by the second-order expansion of Eq.~\eqref{eq:deltaNphi2full}, which gives
\begin{align}\label{eq:deltaNphi2full222}
{\cal R}\approx &\,{\cal R}_g+\frac{3}{5}f_{\rm NL}{\cal R}_g^2\,,
\end{align}
where
\begin{align}\label{eq:fNL}
f_{\rm NL}\approx\frac{5}{2}\frac{\tilde\pi_e}{\tilde\pi_e+\sqrt{6}\alpha_2}=\frac{5}{2}\frac{\pi_e-\sqrt{6}\alpha_2}{\pi_e}\approx -\frac{5}{12}\eta\,.
\end{align}
From Eq.~\eqref{eq:fNL} we see that if the endpoint is very close to the attractor, where $\tilde\pi_e\ll1$, then it follows that $f_{\rm NL}\ll1$. In the exact limit where $\tilde\pi_e=0$ then $f_{\rm NL}=0$.\footnote{Strictly speaking $f_{\rm NL}\neq 0$ because of the intrinsic non-Gaussianity of $\delta\phi$ \cite{Domenech:2016zxn}, which satisfies Maldacena’s consistency relation \cite{Maldacena:2002vr} (see also Ref.~\cite{Suyama:2021adn}). However, intrinsic non-Gaussianity of $\delta\phi$ is proportional to $\epsilon$ and, therefore, negligible.} We also see that for $\alpha_2<\alpha_1$ (enhancement), $\tilde\pi_e$ is at most close to $\alpha_1$, which yields $f_{\rm NL}\sim 5/2$ as in ultra-slow-roll \cite{Cai:2018dkf}. Interestingly, for $\alpha_1<\alpha_2$ (suppression), we have that $\tilde\pi_e$ can be of the order of $\alpha_1$, which leads to $f_{\rm NL}<0$ and $|f_{\rm NL}|\propto \alpha_1/\alpha_2\gg1$. This result is consistent with the fact that $\eta$ can become very large in this case (see e.g. Fig~\ref{fig:plotsbginv}).\footnote{A large $\eta$ might be in trouble with perturbativity but this issue can be solved by considering a sharp but smooth transition. See the discussion in Refs.~\cite{Adshead:2014sga,Cannone:2014qna,Achucarro:2014msa}.} This means that ending the non-attractor transition in the suppression case ($\alpha_1<\alpha_2$) can lead to very large and negative non-Gaussianities.

To investigate the tail of the PDF, we need the full non-linear relation between ${\cal R}_g$ and ${\cal R}$. To do so, we start by expressing $\tilde\pi_e(\phi_i+\delta\phi)$ in terms of $\delta N={\cal R}$ from Eq.~\eqref{eq:deltaNphi2full}, which results in
\begin{align}\label{eq:Rpie}
\tilde\pi_e(\phi_i+\delta\phi)=\frac{\sqrt{6}}{\alpha_2}\,W\left[\frac{\alpha_2\tilde\pi_e(\phi_i)}{\sqrt{6}}e^{\frac{\alpha_2\tilde\pi_e(\phi_i)}{\sqrt{6}}}\times e^{-3\omega_2^2{\cal R}}\right]\approx\tilde\pi_e(\phi_i)\times e^{-3\omega_2^2{\cal R}} \,,
\end{align}
and in the last step we used that $\alpha_2\tilde\pi_e(\phi_i)\ll1$. We may also write $\delta\phi$ in terms of $\tilde\pi_e(\phi_i+\delta\phi)$ by Eq.~\eqref{eq:phigeneralsmallphiN}, namely
\begin{align}\label{eq:deltaphipie}
\delta\phi \approx \frac{1}{3\omega_2^2}\left(\tilde\pi_e(\phi_i+\delta\phi)-\tilde\pi_e(\phi_i)+\sqrt{6}\alpha_2 \ln \frac{\tilde\pi_e(\phi_i+\delta\phi)}{\tilde\pi_e(\phi_i)}\right)\,.
\end{align}
Combining Eqs.~\eqref{eq:Rpie}, \eqref{eq:deltaphipie} and \eqref{eq:Rg} we arrive at
\begin{align}\label{eq:nonlinearrelation}
{\cal R}_g\approx \frac{1}{\tilde\pi_e(\phi_i)+\sqrt{6}\alpha_2}\left(\sqrt{6}\alpha_2 {\cal R}+\frac{\tilde \pi_e(\phi_i)}{3\omega_2^2}\left(1-e^{-3\omega_2^2{\cal R}}\right)\right)\,.
\end{align}
This is the general, approximate, relation between the Gaussian curvature fluctuation ${\cal R}_g$ and the non-linear curvature fluctuation ${\cal R}$. 

With Eq.~\eqref{eq:nonlinearrelation} we can compute the PDF for ${\cal R}$, which is given by
\begin{align}
P[{\cal R}]=\left|\frac{\partial {\cal R}_g}{\partial {\cal R}}\right|P[{\cal R}_g]\quad{\rm where}\quad
P[{\cal R}_g]=\frac{1}{\sqrt{2\pi}\,\sigma_{\cal R}}e^{-\frac{{\cal R}^2_g}{2\sigma_{\cal R}^2}}\,,
\end{align}
and $\sigma_{\cal R}^2=\int d\ln k\, {\cal P}_{{\cal R}_g}(k)$ is the variance of the Gaussian curvature fluctuation. And, after some simplifications, we arrive at
%\begin{align}
%\left|\frac{\partial{\cal R}_g}{\partial{\cal R}}\right|&
%\approx\frac{1}{\tilde\pi_e(\phi_i)+\sqrt{6}\alpha_2}\left({\sqrt{6}\alpha_2}+\tilde\pi_e(\phi_i)\,e^{-3\omega_2^2{\cal R}}\right)
%\end{align}
\begin{align}\label{eq:P[R]}
P[{\cal R}]\approx&\frac{1+\frac{\tilde\pi_e(\phi_i)}{{\sqrt{6}\alpha_2}}\,e^{-3\omega_2^2{\cal R}}}{\sqrt{2\pi}\,\sigma_{\cal R}\left(1+\frac{\tilde\pi_e(\phi_i)}{\sqrt{6}\alpha_2}\right)}\times\exp\left[-\left(\frac{ {\cal R}+\frac{\tilde \pi_e(\phi_i)}{3\omega_2^2\sqrt{6}\alpha_2}\left(1-e^{-3\omega_2^2{\cal R}}\right)}{\sqrt{2}\sigma_{\cal R}\left(1+\frac{\tilde\pi_e(\phi_i)}{\sqrt{6}\alpha_2}\right)}\right)^2\,\right]\,.
\end{align}
Let us discuss some interesting features of Eq.~\eqref{eq:P[R]}, which is valid for any value of $\alpha_1,\alpha_2<1$. First, we note that if the endpoint is in the attractor regime where $\tilde\pi_e(\phi_i)\sim 0$, the PDF of ${\cal R}$ is basically Gaussian. This is consistent with the results of Ref.~\cite{Cai:2018dkf} for smooth transitions using the in-in formalism and Ref.~\cite{Pattison:2021oen} for Starobinsky’s piecewise linear potential using the stochastic-$\delta N$ formalism that takes into account quantum effects. Second, when there is enhancement of fluctuations, i.e. $\alpha_1>\alpha_2$, we may have that $\tilde\pi_e\gg \sqrt{6}\alpha_2$. In that case, we find the exponential tail $e^{-3\omega_2^2{\cal R}}$ typical of ultra-slow-roll \cite{Biagetti:2018pjj,Atal:2019cdz,Ezquiaga:2019ftu,Atal:2019erb} and PBH formation is also enhanced. Lastly, when there is suppression of fluctuations, i.e. $\alpha_1<\alpha_2$, it is possible that $\tilde\pi_e\sim -\alpha_2$. This implies that the exponential tail is never important for this case but depending on how close $\tilde\pi_e$ is to $-\sqrt{6}\alpha_2$ we have a large prefactor in the exponential of \eqref{eq:P[R]} which suppresses PBH formation, consistent with the large negative non-Gaussianity we found earlier. Let us end by remarking that it is quite interesting that using the $\delta N$ formalism we arrived at Eq.~\eqref{eq:P[R]} which is valid in various different situations.

\section{Conclusions\label{sec:conclusions}}

We investigated the enhancement/suppression of primordial fluctuations during inflation in the piecewise exponential potential. This model has exact background solutions for arbitrary initial conditions as well as exact solutions for perturbations in the attractor. We found that the same solutions are also valid for scalar field perturbations $\delta\phi$ in non-attractor trajectories as long as the scalar field is slowly rolling. This result allowed us to find close to exact solutions for $\delta\phi$ in a general slow-roll to slow-roll transition (Eqs.~\eqref{eq:deltaphi2}, \eqref{eq:C2} and \eqref{eq:D2}). The power spectrum of $\delta\phi$ evaluated at the end of inflation is given by Eq.~\eqref{eq:resultdeltaphi}, also see Fig.~\ref{fig:ana}. General features of the primordial spectrum only depend on the ratio of the exponents of the potential and are consistent with the general analysis of Refs.~\cite{Leach:2001zf,Byrnes:2018txb,Carrilho:2019oqg,Ozsoy:2019lyy,Cole:2022xqc,Tasinato:2023ukp}. We showed how these features are imprinted in the induced GW spectrum in \S~\ref{sec:GWs} and summarized them in Fig.~\ref{fig:GWs}. We also found that the final spectrum of curvature fluctuations is similar to that of Starobinsky’s piecewise linear potential \cite{Pi:2022zxs}, which is consistent with the fact that the enhancement/suppression is mostly due to a sudden change of vacuum at the matching point. One interesting characteristic of the piecewise exponential potential though is that we are able to find exact solutions for $\delta\phi$.

Most interestingly, exact background solutions for arbitrary initial conditions enabled us to make use of the $\delta N$ formalism to study the tail of the PDF of curvature fluctuations. With the exact solutions for $\delta\phi$ fluctuations we estimated from which moment on the standard $\delta N$ formalism is valid, i.e. by requiring that $\delta\phi$ is constant and the momentum perturbation $\delta\pi$ is irrelevant. This time is roughly a few \textit{e}-folds after the matching point for the case of enhancement and roughly soon after the matching point for the case of suppression (see Fig.~\ref{fig:plotscross} for a summary). We found that if we try to use the $\delta N$ formalism earlier than that time, given by Eq.~\eqref{eq:negligible}, the momentum perturbation $\delta\pi$ becomes very important at the matching point. This hints at a possible generalization of the $\delta N$ formalism by also taking into account $\delta\pi$. However, as this is out of the scope of this paper, we left it for future work.

Nevertheless, the $\delta N$ formalism is particularly powerful to study non-Gaussianities of a model. We derived the general PDF for the non-linear curvature perturbation in the piecewise exponential potential in Eq.~\eqref{eq:P[R]} for modes which exit the horizon during the slow-roll to slow-roll transition. In the case of a single piecewise potential the transition is very smooth and the resulting non-Gaussianity is negligible, consistent with the analysis of Ref.~\cite{Cai:2018dkf}. However, our analysis is also valid for multiple piecewise exponential potentials. Therefore, we considered that the transition may end abruptly due to an additional matching point. In that case, non-Gaussianity and the PDF are sensitive to the end of the transition \cite{Cai:2018dkf}. For the ultra-slow-roll like transition, i.e. enhancement of the spectrum, non-Gaussianity is positive and at most ${\cal O}(1)$ as expected \cite{Cai:2018dkf} and the exponential tail of the PDF \cite{Biagetti:2018pjj,Atal:2019cdz,Ezquiaga:2019ftu,Atal:2019erb} may arise, depending on when the transition ends. For the suppression of fluctuations, we find that non-Gaussianity is negative and can be very large. Furthermore, we find that the PDF does not have a relevant exponential tail but that the main effect is a large prefactor in the “Gaussian” exponential. It is remarkable that the PDF $P[{\cal R}]$ \eqref{eq:P[R]} derived using the $\delta N$ is applicable in various different situations.

Our work can be extended in several ways. First, it would be interesting to generalize our results for the primordial spectrum of a single exponential potential to a succession of piecewise potentials. It is plausible that depending on the different matching points the growth of fluctuations exceeds the $k^4$ bound \cite{Byrnes:2018txb}, as in Ref.~\cite{Ozsoy:2019lyy}. It would also be interesting to repeat the analysis in the case when the slope in the second phase changes signs and eventually the system reaches $\phi_N=0$ before rolling backward, as in Ref.~\cite{Briaud:2023eae}. This case has been shown to abundantly produce tiny PBHs \cite{Briaud:2023eae}. We note that the analysis of this paper might not be straightforwardly applicable to this case since the scalar field spends most of the time in non-attractor trajectories, as it must first reach $\phi_N=0$ before approaching the attractor solution. With the analytical background solutions, one may also be able to make use of the $\delta N$ formalism to investigate the PDF of curvature fluctuations. Lastly, it would be intriguing to do the one-loop calculations in our exact model.

\section*{Acknowledgments} 

We would like to thank G.~Franciolini, S.~Pi and M.~Sasaki for useful discussions and N.~Bartolo, S.~Matarrese, A.~Ricciardone and G.~Tasinato for useful comments on the draft. G.D. is supported by the DFG under the Emmy-Noether program grant no. DO 2574/1-1, project number 496592360. T. V. is supported by the Vice-rectorate for Research and Postgraduate B20131101 and CONCYTEC through the Research Academic grant.  Calculations of the SIGW spectrum have been performed with  \href{https://github.com/Lukas-T-W/SIGWfast/releases}{\textsc{SIGWfast}} \cite{2022arXiv220905296W}.
	
\appendix

\section{Power-law inflation and CMB scales\label{app:powerlawcmb}}

Here we briefly compare power-law inflation with CMB results \cite{Akrami:2018odb}. From CMB measurements we have that at the CMB pivot scale ($k_{\rm pivot}\approx0.05{\rm Mpc}^{-1}$)
\begin{align}
{\cal P}_{\cal R}\approx 2\times 10^{-9}\,.
\quad{\rm and}\quad 
n_s-1\approx 0.035\,.
\end{align}
Applied to the spectrum from power-law inflation \eqref{eq:slowroll} gives
\begin{align}
p\approx 58 \quad{\rm and}\quad H_\star \approx 3\times 10^{-5}M_{\rm pl}\,.
\end{align}
In the attractor regime, this corresponds to
\begin{align}
\lambda\approx0.18\quad{ \rm and}\quad V_\star e^{-\lambda\phi_{\rm pivot} }=\frac{9}{64}H_\star ^2M_{\rm pl}^2(6-\lambda_1^2)\approx 7\times 10^{-10}M_{\rm pl}^4\,.
\end{align}

The bound on the tensor to scalar ratio is \cite{Akrami:2018odb}
\begin{align}
r=\frac{{\cal P}_{t}}{{\cal P}_{\cal R}}=\frac{16}{p}<0.06\,.
\end{align}
But this requires $p>266$. As it is well-known, power-law inflation predicts too many primordial tensor modes. Nevertheless, the tensor-to-scalar ratio can be suppressed by the curvaton mechanism \cite{Lyth:2001nq,Enqvist:2001zp,Moroi:2002rd}. Nevertheless, we stress that the main point of our paper is the analytical treatment of the enhancement/suppression and not the exact fit to CMB data.

\bibliography{refgwscalar.bib} 

\begin{thebibliography}{198}%
\makeatletter
\providecommand \@ifxundefined [1]{%
 \@ifx{#1\undefined}
}%
\providecommand \@ifnum [1]{%
 \ifnum #1\expandafter \@firstoftwo
 \else \expandafter \@secondoftwo
 \fi
}%
\providecommand \@ifx [1]{%
 \ifx #1\expandafter \@firstoftwo
 \else \expandafter \@secondoftwo
 \fi
}%
\providecommand \natexlab [1]{#1}%
\providecommand \enquote  [1]{``#1''}%
\providecommand \bibnamefont  [1]{#1}%
\providecommand \bibfnamefont [1]{#1}%
\providecommand \citenamefont [1]{#1}%
\providecommand \href@noop [0]{\@secondoftwo}%
\providecommand \href [0]{\begingroup \@sanitize@url \@href}%
\providecommand \@href[1]{\@@startlink{#1}\@@href}%
\providecommand \@@href[1]{\endgroup#1\@@endlink}%
\providecommand \@sanitize@url [0]{\catcode `\\12\catcode `\$12\catcode
  `\&12\catcode `\#12\catcode `\^12\catcode `\_12\catcode `\%12\relax}%
\providecommand \@@startlink[1]{}%
\providecommand \@@endlink[0]{}%
\providecommand \url  [0]{\begingroup\@sanitize@url \@url }%
\providecommand \@url [1]{\endgroup\@href {#1}{\urlprefix }}%
\providecommand \urlprefix  [0]{URL }%
\providecommand \Eprint [0]{\href }%
\providecommand \doibase [0]{https://doi.org/}%
\providecommand \selectlanguage [0]{\@gobble}%
\providecommand \bibinfo  [0]{\@secondoftwo}%
\providecommand \bibfield  [0]{\@secondoftwo}%
\providecommand \translation [1]{[#1]}%
\providecommand \BibitemOpen [0]{}%
\providecommand \bibitemStop [0]{}%
\providecommand \bibitemNoStop [0]{.\EOS\space}%
\providecommand \EOS [0]{\spacefactor3000\relax}%
\providecommand \BibitemShut  [1]{\csname bibitem#1\endcsname}%
\let\auto@bib@innerbib\@empty
%</preamble>
\bibitem [{\citenamefont {Aghanim}\ \emph {et~al.}(2020)\citenamefont {Aghanim}
  \emph {et~al.}}]{Planck:2018vyg}%
  \BibitemOpen
  \bibfield  {author} {\bibinfo {author} {\bibfnamefont {N.}~\bibnamefont
  {Aghanim}} \emph {et~al.} (\bibinfo {collaboration} {Planck}),\ }\bibfield
  {title} {\bibinfo {title} {{Planck 2018 results. VI. Cosmological
  parameters}},\ }\href {https://doi.org/10.1051/0004-6361/201833910}
  {\bibfield  {journal} {\bibinfo  {journal} {Astron. Astrophys.}\ }\textbf
  {\bibinfo {volume} {641}},\ \bibinfo {pages} {A6} (\bibinfo {year} {2020})},\
  \bibinfo {note} {[Erratum: Astron.Astrophys. 652, C4 (2021)]},\ \Eprint
  {https://arxiv.org/abs/1807.06209} {arXiv:1807.06209 [astro-ph.CO]}
  \BibitemShut {NoStop}%
\bibitem [{\citenamefont {Akrami}\ \emph {et~al.}(2020)\citenamefont {Akrami}
  \emph {et~al.}}]{Akrami:2018odb}%
  \BibitemOpen
  \bibfield  {author} {\bibinfo {author} {\bibfnamefont {Y.}~\bibnamefont
  {Akrami}} \emph {et~al.} (\bibinfo {collaboration} {Planck}),\ }\bibfield
  {title} {\bibinfo {title} {{Planck 2018 results. X. Constraints on
  inflation}},\ }\href {https://doi.org/10.1051/0004-6361/201833887} {\bibfield
   {journal} {\bibinfo  {journal} {Astron. Astrophys.}\ }\textbf {\bibinfo
  {volume} {641}},\ \bibinfo {pages} {A10} (\bibinfo {year} {2020})},\ \Eprint
  {https://arxiv.org/abs/1807.06211} {arXiv:1807.06211 [astro-ph.CO]}
  \BibitemShut {NoStop}%
\bibitem [{\citenamefont {Hawking}(1971)}]{Hawking:1971ei}%
  \BibitemOpen
  \bibfield  {author} {\bibinfo {author} {\bibfnamefont {S.}~\bibnamefont
  {Hawking}},\ }\bibfield  {title} {\bibinfo {title} {{Gravitationally
  collapsed objects of very low mass}},\ }\href@noop {} {\bibfield  {journal}
  {\bibinfo  {journal} {Mon. Not. Roy. Astron. Soc.}\ }\textbf {\bibinfo
  {volume} {152}},\ \bibinfo {pages} {75} (\bibinfo {year} {1971})}\BibitemShut
  {NoStop}%
\bibitem [{\citenamefont {Carr}\ and\ \citenamefont
  {Hawking}(1974)}]{Carr:1974nx}%
  \BibitemOpen
  \bibfield  {author} {\bibinfo {author} {\bibfnamefont {B.~J.}\ \bibnamefont
  {Carr}}\ and\ \bibinfo {author} {\bibfnamefont {S.}~\bibnamefont {Hawking}},\
  }\bibfield  {title} {\bibinfo {title} {{Black holes in the early Universe}},\
  }\href@noop {} {\bibfield  {journal} {\bibinfo  {journal} {Mon. Not. Roy.
  Astron. Soc.}\ }\textbf {\bibinfo {volume} {168}},\ \bibinfo {pages} {399}
  (\bibinfo {year} {1974})}\BibitemShut {NoStop}%
\bibitem [{\citenamefont {Carr}(1975)}]{Carr:1975qj}%
  \BibitemOpen
  \bibfield  {author} {\bibinfo {author} {\bibfnamefont {B.~J.}\ \bibnamefont
  {Carr}},\ }\bibfield  {title} {\bibinfo {title} {{The Primordial black hole
  mass spectrum}},\ }\href {https://doi.org/10.1086/153853} {\bibfield
  {journal} {\bibinfo  {journal} {Astrophys. J.}\ }\textbf {\bibinfo {volume}
  {201}},\ \bibinfo {pages} {1} (\bibinfo {year} {1975})}\BibitemShut {NoStop}%
\bibitem [{\citenamefont {Khlopov}\ \emph {et~al.}(1985)\citenamefont
  {Khlopov}, \citenamefont {Malomed},\ and\ \citenamefont
  {Zeldovich}}]{Khlopov:1985jw}%
  \BibitemOpen
  \bibfield  {author} {\bibinfo {author} {\bibfnamefont {M.}~\bibnamefont
  {Khlopov}}, \bibinfo {author} {\bibfnamefont {B.~A.}\ \bibnamefont
  {Malomed}},\ and\ \bibinfo {author} {\bibfnamefont {I.~B.}\ \bibnamefont
  {Zeldovich}},\ }\bibfield  {title} {\bibinfo {title} {{Gravitational
  instability of scalar fields and formation of primordial black holes}},\
  }\href@noop {} {\bibfield  {journal} {\bibinfo  {journal} {Mon. Not. Roy.
  Astron. Soc.}\ }\textbf {\bibinfo {volume} {215}},\ \bibinfo {pages} {575}
  (\bibinfo {year} {1985})}\BibitemShut {NoStop}%
\bibitem [{\citenamefont {Tomita}(1967)}]{Tomita}%
  \BibitemOpen
  \bibfield  {author} {\bibinfo {author} {\bibfnamefont {K.}~\bibnamefont
  {Tomita}},\ }\bibfield  {title} {\bibinfo {title} {{Non-Linear Theory of
  Gravitational Instability in the Expanding Universe}},\ }\href
  {https://doi.org/10.1143/PTP.37.831} {\bibfield  {journal} {\bibinfo
  {journal} {Progress of Theoretical Physics}\ }\textbf {\bibinfo {volume}
  {37}},\ \bibinfo {pages} {831} (\bibinfo {year} {1967})},\ \Eprint
  {https://arxiv.org/abs/https://academic.oup.com/ptp/article-pdf/37/5/831/5234391/37-5-831.pdf}
  {https://academic.oup.com/ptp/article-pdf/37/5/831/5234391/37-5-831.pdf}
  \BibitemShut {NoStop}%
\bibitem [{\citenamefont {Matarrese}\ \emph {et~al.}(1993)\citenamefont
  {Matarrese}, \citenamefont {Pantano},\ and\ \citenamefont
  {Saez}}]{Matarrese:1992rp}%
  \BibitemOpen
  \bibfield  {author} {\bibinfo {author} {\bibfnamefont {S.}~\bibnamefont
  {Matarrese}}, \bibinfo {author} {\bibfnamefont {O.}~\bibnamefont {Pantano}},\
  and\ \bibinfo {author} {\bibfnamefont {D.}~\bibnamefont {Saez}},\ }\bibfield
  {title} {\bibinfo {title} {{A General relativistic approach to the nonlinear
  evolution of collisionless matter}},\ }\href
  {https://doi.org/10.1103/PhysRevD.47.1311} {\bibfield  {journal} {\bibinfo
  {journal} {Phys. Rev. D}\ }\textbf {\bibinfo {volume} {47}},\ \bibinfo
  {pages} {1311} (\bibinfo {year} {1993})}\BibitemShut {NoStop}%
\bibitem [{\citenamefont {Matarrese}\ \emph {et~al.}(1994)\citenamefont
  {Matarrese}, \citenamefont {Pantano},\ and\ \citenamefont
  {Saez}}]{Matarrese:1993zf}%
  \BibitemOpen
  \bibfield  {author} {\bibinfo {author} {\bibfnamefont {S.}~\bibnamefont
  {Matarrese}}, \bibinfo {author} {\bibfnamefont {O.}~\bibnamefont {Pantano}},\
  and\ \bibinfo {author} {\bibfnamefont {D.}~\bibnamefont {Saez}},\ }\bibfield
  {title} {\bibinfo {title} {{General relativistic dynamics of irrotational
  dust: Cosmological implications}},\ }\href
  {https://doi.org/10.1103/PhysRevLett.72.320} {\bibfield  {journal} {\bibinfo
  {journal} {Phys. Rev. Lett.}\ }\textbf {\bibinfo {volume} {72}},\ \bibinfo
  {pages} {320} (\bibinfo {year} {1994})},\ \Eprint
  {https://arxiv.org/abs/astro-ph/9310036} {arXiv:astro-ph/9310036}
  \BibitemShut {NoStop}%
\bibitem [{\citenamefont {Ananda}\ \emph {et~al.}(2007)\citenamefont {Ananda},
  \citenamefont {Clarkson},\ and\ \citenamefont {Wands}}]{Ananda:2006af}%
  \BibitemOpen
  \bibfield  {author} {\bibinfo {author} {\bibfnamefont {K.~N.}\ \bibnamefont
  {Ananda}}, \bibinfo {author} {\bibfnamefont {C.}~\bibnamefont {Clarkson}},\
  and\ \bibinfo {author} {\bibfnamefont {D.}~\bibnamefont {Wands}},\ }\bibfield
   {title} {\bibinfo {title} {{The Cosmological gravitational wave background
  from primordial density perturbations}},\ }\href
  {https://doi.org/10.1103/PhysRevD.75.123518} {\bibfield  {journal} {\bibinfo
  {journal} {Phys. Rev. D}\ }\textbf {\bibinfo {volume} {75}},\ \bibinfo
  {pages} {123518} (\bibinfo {year} {2007})},\ \Eprint
  {https://arxiv.org/abs/gr-qc/0612013} {arXiv:gr-qc/0612013} \BibitemShut
  {NoStop}%
\bibitem [{\citenamefont {Baumann}\ \emph {et~al.}(2007)\citenamefont
  {Baumann}, \citenamefont {Steinhardt}, \citenamefont {Takahashi},\ and\
  \citenamefont {Ichiki}}]{Baumann:2007zm}%
  \BibitemOpen
  \bibfield  {author} {\bibinfo {author} {\bibfnamefont {D.}~\bibnamefont
  {Baumann}}, \bibinfo {author} {\bibfnamefont {P.~J.}\ \bibnamefont
  {Steinhardt}}, \bibinfo {author} {\bibfnamefont {K.}~\bibnamefont
  {Takahashi}},\ and\ \bibinfo {author} {\bibfnamefont {K.}~\bibnamefont
  {Ichiki}},\ }\bibfield  {title} {\bibinfo {title} {{Gravitational Wave
  Spectrum Induced by Primordial Scalar Perturbations}},\ }\href
  {https://doi.org/10.1103/PhysRevD.76.084019} {\bibfield  {journal} {\bibinfo
  {journal} {Phys. Rev. D}\ }\textbf {\bibinfo {volume} {76}},\ \bibinfo
  {pages} {084019} (\bibinfo {year} {2007})},\ \Eprint
  {https://arxiv.org/abs/hep-th/0703290} {arXiv:hep-th/0703290} \BibitemShut
  {NoStop}%
\bibitem [{\citenamefont {Saito}\ and\ \citenamefont
  {Yokoyama}(2009)}]{Saito:2008jc}%
  \BibitemOpen
  \bibfield  {author} {\bibinfo {author} {\bibfnamefont {R.}~\bibnamefont
  {Saito}}\ and\ \bibinfo {author} {\bibfnamefont {J.}~\bibnamefont
  {Yokoyama}},\ }\bibfield  {title} {\bibinfo {title} {{Gravitational wave
  background as a probe of the primordial black hole abundance}},\ }\href
  {https://doi.org/10.1103/PhysRevLett.102.161101} {\bibfield  {journal}
  {\bibinfo  {journal} {Phys. Rev. Lett.}\ }\textbf {\bibinfo {volume} {102}},\
  \bibinfo {pages} {161101} (\bibinfo {year} {2009})},\ \bibinfo {note}
  {[Erratum: Phys.Rev.Lett. 107, 069901 (2011)]},\ \Eprint
  {https://arxiv.org/abs/0812.4339} {arXiv:0812.4339 [astro-ph]} \BibitemShut
  {NoStop}%
\bibitem [{\citenamefont {Saito}\ and\ \citenamefont
  {Yokoyama}(2010)}]{Saito:2009jt}%
  \BibitemOpen
  \bibfield  {author} {\bibinfo {author} {\bibfnamefont {R.}~\bibnamefont
  {Saito}}\ and\ \bibinfo {author} {\bibfnamefont {J.}~\bibnamefont
  {Yokoyama}},\ }\bibfield  {title} {\bibinfo {title} {{Gravitational-Wave
  Constraints on the Abundance of Primordial Black Holes}},\ }\href
  {https://doi.org/10.1143/PTP.126.351} {\bibfield  {journal} {\bibinfo
  {journal} {Prog. Theor. Phys.}\ }\textbf {\bibinfo {volume} {123}},\ \bibinfo
  {pages} {867} (\bibinfo {year} {2010})},\ \bibinfo {note} {[Erratum:
  Prog.Theor.Phys. 126, 351--352 (2011)]},\ \Eprint
  {https://arxiv.org/abs/0912.5317} {arXiv:0912.5317 [astro-ph.CO]}
  \BibitemShut {NoStop}%
\bibitem [{\citenamefont {Assadullahi}\ and\ \citenamefont
  {Wands}(2010)}]{Assadullahi:2009jc}%
  \BibitemOpen
  \bibfield  {author} {\bibinfo {author} {\bibfnamefont {H.}~\bibnamefont
  {Assadullahi}}\ and\ \bibinfo {author} {\bibfnamefont {D.}~\bibnamefont
  {Wands}},\ }\bibfield  {title} {\bibinfo {title} {{Constraints on primordial
  density perturbations from induced gravitational waves}},\ }\href
  {https://doi.org/10.1103/PhysRevD.81.023527} {\bibfield  {journal} {\bibinfo
  {journal} {Phys. Rev. D}\ }\textbf {\bibinfo {volume} {81}},\ \bibinfo
  {pages} {023527} (\bibinfo {year} {2010})},\ \Eprint
  {https://arxiv.org/abs/0907.4073} {arXiv:0907.4073 [astro-ph.CO]}
  \BibitemShut {NoStop}%
\bibitem [{\citenamefont {Bugaev}\ and\ \citenamefont
  {Klimai}(2010{\natexlab{a}})}]{Bugaev:2009zh}%
  \BibitemOpen
  \bibfield  {author} {\bibinfo {author} {\bibfnamefont {E.}~\bibnamefont
  {Bugaev}}\ and\ \bibinfo {author} {\bibfnamefont {P.}~\bibnamefont
  {Klimai}},\ }\bibfield  {title} {\bibinfo {title} {{Induced gravitational
  wave background and primordial black holes}},\ }\href
  {https://doi.org/10.1103/PhysRevD.81.023517} {\bibfield  {journal} {\bibinfo
  {journal} {Phys. Rev. D}\ }\textbf {\bibinfo {volume} {81}},\ \bibinfo
  {pages} {023517} (\bibinfo {year} {2010}{\natexlab{a}})},\ \Eprint
  {https://arxiv.org/abs/0908.0664} {arXiv:0908.0664 [astro-ph.CO]}
  \BibitemShut {NoStop}%
\bibitem [{\citenamefont {Bugaev}\ and\ \citenamefont
  {Klimai}(2010{\natexlab{b}})}]{Bugaev:2009kq}%
  \BibitemOpen
  \bibfield  {author} {\bibinfo {author} {\bibfnamefont {E.~V.}\ \bibnamefont
  {Bugaev}}\ and\ \bibinfo {author} {\bibfnamefont {P.~A.}\ \bibnamefont
  {Klimai}},\ }\bibfield  {title} {\bibinfo {title} {{Bound on induced
  gravitational wave background from primordial black holes}},\ }\href
  {https://doi.org/10.1134/S0021364010010017} {\bibfield  {journal} {\bibinfo
  {journal} {JETP Lett.}\ }\textbf {\bibinfo {volume} {91}},\ \bibinfo {pages}
  {1} (\bibinfo {year} {2010}{\natexlab{b}})},\ \Eprint
  {https://arxiv.org/abs/0911.0611} {arXiv:0911.0611 [astro-ph.CO]}
  \BibitemShut {NoStop}%
\bibitem [{\citenamefont {Bugaev}\ and\ \citenamefont
  {Klimai}(2011)}]{Bugaev:2010bb}%
  \BibitemOpen
  \bibfield  {author} {\bibinfo {author} {\bibfnamefont {E.}~\bibnamefont
  {Bugaev}}\ and\ \bibinfo {author} {\bibfnamefont {P.}~\bibnamefont
  {Klimai}},\ }\bibfield  {title} {\bibinfo {title} {{Constraints on the
  induced gravitational wave background from primordial black holes}},\ }\href
  {https://doi.org/10.1103/PhysRevD.83.083521} {\bibfield  {journal} {\bibinfo
  {journal} {Phys. Rev. D}\ }\textbf {\bibinfo {volume} {83}},\ \bibinfo
  {pages} {083521} (\bibinfo {year} {2011})},\ \Eprint
  {https://arxiv.org/abs/1012.4697} {arXiv:1012.4697 [astro-ph.CO]}
  \BibitemShut {NoStop}%
\bibitem [{\citenamefont {Inomata}\ and\ \citenamefont
  {Nakama}(2019)}]{Inomata:2018epa}%
  \BibitemOpen
  \bibfield  {author} {\bibinfo {author} {\bibfnamefont {K.}~\bibnamefont
  {Inomata}}\ and\ \bibinfo {author} {\bibfnamefont {T.}~\bibnamefont
  {Nakama}},\ }\bibfield  {title} {\bibinfo {title} {{Gravitational waves
  induced by scalar perturbations as probes of the small-scale primordial
  spectrum}},\ }\href {https://doi.org/10.1103/PhysRevD.99.043511} {\bibfield
  {journal} {\bibinfo  {journal} {Phys. Rev. D}\ }\textbf {\bibinfo {volume}
  {99}},\ \bibinfo {pages} {043511} (\bibinfo {year} {2019})},\ \Eprint
  {https://arxiv.org/abs/1812.00674} {arXiv:1812.00674 [astro-ph.CO]}
  \BibitemShut {NoStop}%
\bibitem [{\citenamefont {Sato-Polito}\ \emph {et~al.}(2019)\citenamefont
  {Sato-Polito}, \citenamefont {Kovetz},\ and\ \citenamefont
  {Kamionkowski}}]{Sato-Polito:2019hws}%
  \BibitemOpen
  \bibfield  {author} {\bibinfo {author} {\bibfnamefont {G.}~\bibnamefont
  {Sato-Polito}}, \bibinfo {author} {\bibfnamefont {E.~D.}\ \bibnamefont
  {Kovetz}},\ and\ \bibinfo {author} {\bibfnamefont {M.}~\bibnamefont
  {Kamionkowski}},\ }\bibfield  {title} {\bibinfo {title} {{Constraints on the
  primordial curvature power spectrum from primordial black holes}},\ }\href
  {https://doi.org/10.1103/PhysRevD.100.063521} {\bibfield  {journal} {\bibinfo
   {journal} {Phys. Rev. D}\ }\textbf {\bibinfo {volume} {100}},\ \bibinfo
  {pages} {063521} (\bibinfo {year} {2019})},\ \Eprint
  {https://arxiv.org/abs/1904.10971} {arXiv:1904.10971 [astro-ph.CO]}
  \BibitemShut {NoStop}%
\bibitem [{\citenamefont {Kalaja}\ \emph {et~al.}(2019)\citenamefont {Kalaja},
  \citenamefont {Bellomo}, \citenamefont {Bartolo}, \citenamefont {Bertacca},
  \citenamefont {Matarrese}, \citenamefont {Musco}, \citenamefont
  {Raccanelli},\ and\ \citenamefont {Verde}}]{Kalaja:2019uju}%
  \BibitemOpen
  \bibfield  {author} {\bibinfo {author} {\bibfnamefont {A.}~\bibnamefont
  {Kalaja}}, \bibinfo {author} {\bibfnamefont {N.}~\bibnamefont {Bellomo}},
  \bibinfo {author} {\bibfnamefont {N.}~\bibnamefont {Bartolo}}, \bibinfo
  {author} {\bibfnamefont {D.}~\bibnamefont {Bertacca}}, \bibinfo {author}
  {\bibfnamefont {S.}~\bibnamefont {Matarrese}}, \bibinfo {author}
  {\bibfnamefont {I.}~\bibnamefont {Musco}}, \bibinfo {author} {\bibfnamefont
  {A.}~\bibnamefont {Raccanelli}},\ and\ \bibinfo {author} {\bibfnamefont
  {L.}~\bibnamefont {Verde}},\ }\bibfield  {title} {\bibinfo {title} {{From
  Primordial Black Holes Abundance to Primordial Curvature Power Spectrum (and
  back)}},\ }\href {https://doi.org/10.1088/1475-7516/2019/10/031} {\bibfield
  {journal} {\bibinfo  {journal} {JCAP}\ }\textbf {\bibinfo {volume} {10}},\
  \bibinfo {pages} {031}},\ \Eprint {https://arxiv.org/abs/1908.03596}
  {arXiv:1908.03596 [astro-ph.CO]} \BibitemShut {NoStop}%
\bibitem [{\citenamefont {Gow}\ \emph {et~al.}(2021)\citenamefont {Gow},
  \citenamefont {Byrnes}, \citenamefont {Cole},\ and\ \citenamefont
  {Young}}]{Gow:2020bzo}%
  \BibitemOpen
  \bibfield  {author} {\bibinfo {author} {\bibfnamefont {A.~D.}\ \bibnamefont
  {Gow}}, \bibinfo {author} {\bibfnamefont {C.~T.}\ \bibnamefont {Byrnes}},
  \bibinfo {author} {\bibfnamefont {P.~S.}\ \bibnamefont {Cole}},\ and\
  \bibinfo {author} {\bibfnamefont {S.}~\bibnamefont {Young}},\ }\bibfield
  {title} {\bibinfo {title} {{The power spectrum on small scales: Robust
  constraints and comparing PBH methodologies}},\ }\href
  {https://doi.org/10.1088/1475-7516/2021/02/002} {\bibfield  {journal}
  {\bibinfo  {journal} {JCAP}\ }\textbf {\bibinfo {volume} {02}},\ \bibinfo
  {pages} {002}},\ \Eprint {https://arxiv.org/abs/2008.03289} {arXiv:2008.03289
  [astro-ph.CO]} \BibitemShut {NoStop}%
\bibitem [{\citenamefont {Kimura}\ \emph {et~al.}(2021)\citenamefont {Kimura},
  \citenamefont {Suyama}, \citenamefont {Yamaguchi},\ and\ \citenamefont
  {Zhang}}]{Kimura:2021sqz}%
  \BibitemOpen
  \bibfield  {author} {\bibinfo {author} {\bibfnamefont {R.}~\bibnamefont
  {Kimura}}, \bibinfo {author} {\bibfnamefont {T.}~\bibnamefont {Suyama}},
  \bibinfo {author} {\bibfnamefont {M.}~\bibnamefont {Yamaguchi}},\ and\
  \bibinfo {author} {\bibfnamefont {Y.-L.}\ \bibnamefont {Zhang}},\ }\bibfield
  {title} {\bibinfo {title} {{Reconstruction of Primordial Power Spectrum of
  curvature perturbation from the merger rate of Primordial Black Hole
  Binaries}},\ }\href {https://doi.org/10.1088/1475-7516/2021/04/031}
  {\bibfield  {journal} {\bibinfo  {journal} {JCAP}\ }\textbf {\bibinfo
  {volume} {04}},\ \bibinfo {pages} {031}},\ \Eprint
  {https://arxiv.org/abs/2102.05280} {arXiv:2102.05280 [astro-ph.CO]}
  \BibitemShut {NoStop}%
\bibitem [{\citenamefont {Wang}\ \emph
  {et~al.}(2023{\natexlab{a}})\citenamefont {Wang}, \citenamefont {Zhang},
  \citenamefont {Kimura},\ and\ \citenamefont {Yamaguchi}}]{Wang:2022nml}%
  \BibitemOpen
  \bibfield  {author} {\bibinfo {author} {\bibfnamefont {X.}~\bibnamefont
  {Wang}}, \bibinfo {author} {\bibfnamefont {Y.-l.}\ \bibnamefont {Zhang}},
  \bibinfo {author} {\bibfnamefont {R.}~\bibnamefont {Kimura}},\ and\ \bibinfo
  {author} {\bibfnamefont {M.}~\bibnamefont {Yamaguchi}},\ }\bibfield  {title}
  {\bibinfo {title} {{Reconstruction of power spectrum of primordial curvature
  perturbations on small scales from primordial black hole binaries scenario of
  LIGO/VIRGO detection}},\ }\href {https://doi.org/10.1007/s11433-023-2091-x}
  {\bibfield  {journal} {\bibinfo  {journal} {Sci. China Phys. Mech. Astron.}\
  }\textbf {\bibinfo {volume} {66}},\ \bibinfo {pages} {260462} (\bibinfo
  {year} {2023}{\natexlab{a}})},\ \Eprint {https://arxiv.org/abs/2209.12911}
  {arXiv:2209.12911 [astro-ph.CO]} \BibitemShut {NoStop}%
\bibitem [{\citenamefont {Dandoy}\ \emph {et~al.}(2023)\citenamefont {Dandoy},
  \citenamefont {Domcke},\ and\ \citenamefont {Rompineve}}]{Dandoy:2023jot}%
  \BibitemOpen
  \bibfield  {author} {\bibinfo {author} {\bibfnamefont {V.}~\bibnamefont
  {Dandoy}}, \bibinfo {author} {\bibfnamefont {V.}~\bibnamefont {Domcke}},\
  and\ \bibinfo {author} {\bibfnamefont {F.}~\bibnamefont {Rompineve}},\
  }\bibfield  {title} {\bibinfo {title} {{Search for scalar induced
  gravitational waves in the International Pulsar Timing Array Data Release 2
  and NANOgrav 12.5 years datasets}},\ }\href@noop {} {\  (\bibinfo {year}
  {2023})},\ \Eprint {https://arxiv.org/abs/2302.07901} {arXiv:2302.07901
  [astro-ph.CO]} \BibitemShut {NoStop}%
\bibitem [{\citenamefont {Kawasaki}\ \emph {et~al.}(1998)\citenamefont
  {Kawasaki}, \citenamefont {Sugiyama},\ and\ \citenamefont
  {Yanagida}}]{Kawasaki:1997ju}%
  \BibitemOpen
  \bibfield  {author} {\bibinfo {author} {\bibfnamefont {M.}~\bibnamefont
  {Kawasaki}}, \bibinfo {author} {\bibfnamefont {N.}~\bibnamefont {Sugiyama}},\
  and\ \bibinfo {author} {\bibfnamefont {T.}~\bibnamefont {Yanagida}},\
  }\bibfield  {title} {\bibinfo {title} {{Primordial black hole formation in a
  double inflation model in supergravity}},\ }\href
  {https://doi.org/10.1103/PhysRevD.57.6050} {\bibfield  {journal} {\bibinfo
  {journal} {Phys. Rev. D}\ }\textbf {\bibinfo {volume} {57}},\ \bibinfo
  {pages} {6050} (\bibinfo {year} {1998})},\ \Eprint
  {https://arxiv.org/abs/hep-ph/9710259} {arXiv:hep-ph/9710259} \BibitemShut
  {NoStop}%
\bibitem [{\citenamefont {Frampton}\ \emph {et~al.}(2010)\citenamefont
  {Frampton}, \citenamefont {Kawasaki}, \citenamefont {Takahashi},\ and\
  \citenamefont {Yanagida}}]{Frampton:2010sw}%
  \BibitemOpen
  \bibfield  {author} {\bibinfo {author} {\bibfnamefont {P.~H.}\ \bibnamefont
  {Frampton}}, \bibinfo {author} {\bibfnamefont {M.}~\bibnamefont {Kawasaki}},
  \bibinfo {author} {\bibfnamefont {F.}~\bibnamefont {Takahashi}},\ and\
  \bibinfo {author} {\bibfnamefont {T.~T.}\ \bibnamefont {Yanagida}},\
  }\bibfield  {title} {\bibinfo {title} {{Primordial Black Holes as All Dark
  Matter}},\ }\href {https://doi.org/10.1088/1475-7516/2010/04/023} {\bibfield
  {journal} {\bibinfo  {journal} {JCAP}\ }\textbf {\bibinfo {volume} {04}},\
  \bibinfo {pages} {023}},\ \Eprint {https://arxiv.org/abs/1001.2308}
  {arXiv:1001.2308 [hep-ph]} \BibitemShut {NoStop}%
\bibitem [{\citenamefont {Kawasaki}\ \emph {et~al.}(2013)\citenamefont
  {Kawasaki}, \citenamefont {Kitajima},\ and\ \citenamefont
  {Yanagida}}]{Kawasaki:2012wr}%
  \BibitemOpen
  \bibfield  {author} {\bibinfo {author} {\bibfnamefont {M.}~\bibnamefont
  {Kawasaki}}, \bibinfo {author} {\bibfnamefont {N.}~\bibnamefont {Kitajima}},\
  and\ \bibinfo {author} {\bibfnamefont {T.~T.}\ \bibnamefont {Yanagida}},\
  }\bibfield  {title} {\bibinfo {title} {{Primordial black hole formation from
  an axionlike curvaton model}},\ }\href
  {https://doi.org/10.1103/PhysRevD.87.063519} {\bibfield  {journal} {\bibinfo
  {journal} {Phys. Rev. D}\ }\textbf {\bibinfo {volume} {87}},\ \bibinfo
  {pages} {063519} (\bibinfo {year} {2013})},\ \Eprint
  {https://arxiv.org/abs/1207.2550} {arXiv:1207.2550 [hep-ph]} \BibitemShut
  {NoStop}%
\bibitem [{\citenamefont {Namjoo}\ \emph {et~al.}(2013)\citenamefont {Namjoo},
  \citenamefont {Firouzjahi},\ and\ \citenamefont {Sasaki}}]{Namjoo:2012aa}%
  \BibitemOpen
  \bibfield  {author} {\bibinfo {author} {\bibfnamefont {M.~H.}\ \bibnamefont
  {Namjoo}}, \bibinfo {author} {\bibfnamefont {H.}~\bibnamefont {Firouzjahi}},\
  and\ \bibinfo {author} {\bibfnamefont {M.}~\bibnamefont {Sasaki}},\
  }\bibfield  {title} {\bibinfo {title} {{Violation of non-Gaussianity
  consistency relation in a single field inflationary model}},\ }\href
  {https://doi.org/10.1209/0295-5075/101/39001} {\bibfield  {journal} {\bibinfo
   {journal} {EPL}\ }\textbf {\bibinfo {volume} {101}},\ \bibinfo {pages}
  {39001} (\bibinfo {year} {2013})},\ \Eprint {https://arxiv.org/abs/1210.3692}
  {arXiv:1210.3692 [astro-ph.CO]} \BibitemShut {NoStop}%
\bibitem [{\citenamefont {Inomata}\ \emph {et~al.}(2017)\citenamefont
  {Inomata}, \citenamefont {Kawasaki}, \citenamefont {Mukaida}, \citenamefont
  {Tada},\ and\ \citenamefont {Yanagida}}]{Inomata:2017okj}%
  \BibitemOpen
  \bibfield  {author} {\bibinfo {author} {\bibfnamefont {K.}~\bibnamefont
  {Inomata}}, \bibinfo {author} {\bibfnamefont {M.}~\bibnamefont {Kawasaki}},
  \bibinfo {author} {\bibfnamefont {K.}~\bibnamefont {Mukaida}}, \bibinfo
  {author} {\bibfnamefont {Y.}~\bibnamefont {Tada}},\ and\ \bibinfo {author}
  {\bibfnamefont {T.~T.}\ \bibnamefont {Yanagida}},\ }\bibfield  {title}
  {\bibinfo {title} {{Inflationary Primordial Black Holes as All Dark
  Matter}},\ }\href {https://doi.org/10.1103/PhysRevD.96.043504} {\bibfield
  {journal} {\bibinfo  {journal} {Phys. Rev. D}\ }\textbf {\bibinfo {volume}
  {96}},\ \bibinfo {pages} {043504} (\bibinfo {year} {2017})},\ \Eprint
  {https://arxiv.org/abs/1701.02544} {arXiv:1701.02544 [astro-ph.CO]}
  \BibitemShut {NoStop}%
\bibitem [{\citenamefont {Pi}\ \emph {et~al.}(2018)\citenamefont {Pi},
  \citenamefont {Zhang}, \citenamefont {Huang},\ and\ \citenamefont
  {Sasaki}}]{Pi:2017gih}%
  \BibitemOpen
  \bibfield  {author} {\bibinfo {author} {\bibfnamefont {S.}~\bibnamefont
  {Pi}}, \bibinfo {author} {\bibfnamefont {Y.-l.}\ \bibnamefont {Zhang}},
  \bibinfo {author} {\bibfnamefont {Q.-G.}\ \bibnamefont {Huang}},\ and\
  \bibinfo {author} {\bibfnamefont {M.}~\bibnamefont {Sasaki}},\ }\bibfield
  {title} {\bibinfo {title} {{Scalaron from $R^2$-gravity as a heavy field}},\
  }\href {https://doi.org/10.1088/1475-7516/2018/05/042} {\bibfield  {journal}
  {\bibinfo  {journal} {JCAP}\ }\textbf {\bibinfo {volume} {05}},\ \bibinfo
  {pages} {042}},\ \Eprint {https://arxiv.org/abs/1712.09896} {arXiv:1712.09896
  [astro-ph.CO]} \BibitemShut {NoStop}%
\bibitem [{\citenamefont {Cai}\ \emph {et~al.}(2018{\natexlab{a}})\citenamefont
  {Cai}, \citenamefont {Chen}, \citenamefont {Namjoo}, \citenamefont {Sasaki},
  \citenamefont {Wang},\ and\ \citenamefont {Wang}}]{Cai:2018dkf}%
  \BibitemOpen
  \bibfield  {author} {\bibinfo {author} {\bibfnamefont {Y.-F.}\ \bibnamefont
  {Cai}}, \bibinfo {author} {\bibfnamefont {X.}~\bibnamefont {Chen}}, \bibinfo
  {author} {\bibfnamefont {M.~H.}\ \bibnamefont {Namjoo}}, \bibinfo {author}
  {\bibfnamefont {M.}~\bibnamefont {Sasaki}}, \bibinfo {author} {\bibfnamefont
  {D.-G.}\ \bibnamefont {Wang}},\ and\ \bibinfo {author} {\bibfnamefont
  {Z.}~\bibnamefont {Wang}},\ }\bibfield  {title} {\bibinfo {title}
  {{Revisiting non-Gaussianity from non-attractor inflation models}},\ }\href
  {https://doi.org/10.1088/1475-7516/2018/05/012} {\bibfield  {journal}
  {\bibinfo  {journal} {JCAP}\ }\textbf {\bibinfo {volume} {05}},\ \bibinfo
  {pages} {012}},\ \Eprint {https://arxiv.org/abs/1712.09998} {arXiv:1712.09998
  [astro-ph.CO]} \BibitemShut {NoStop}%
\bibitem [{\citenamefont {Cai}\ \emph {et~al.}(2018{\natexlab{b}})\citenamefont
  {Cai}, \citenamefont {Tong}, \citenamefont {Wang},\ and\ \citenamefont
  {Yan}}]{Cai:2018tuh}%
  \BibitemOpen
  \bibfield  {author} {\bibinfo {author} {\bibfnamefont {Y.-F.}\ \bibnamefont
  {Cai}}, \bibinfo {author} {\bibfnamefont {X.}~\bibnamefont {Tong}}, \bibinfo
  {author} {\bibfnamefont {D.-G.}\ \bibnamefont {Wang}},\ and\ \bibinfo
  {author} {\bibfnamefont {S.-F.}\ \bibnamefont {Yan}},\ }\bibfield  {title}
  {\bibinfo {title} {{Primordial Black Holes from Sound Speed Resonance during
  Inflation}},\ }\href {https://doi.org/10.1103/PhysRevLett.121.081306}
  {\bibfield  {journal} {\bibinfo  {journal} {Phys. Rev. Lett.}\ }\textbf
  {\bibinfo {volume} {121}},\ \bibinfo {pages} {081306} (\bibinfo {year}
  {2018}{\natexlab{b}})},\ \Eprint {https://arxiv.org/abs/1805.03639}
  {arXiv:1805.03639 [astro-ph.CO]} \BibitemShut {NoStop}%
\bibitem [{\citenamefont {Cai}\ \emph {et~al.}(2019{\natexlab{a}})\citenamefont
  {Cai}, \citenamefont {Chen}, \citenamefont {Tong}, \citenamefont {Wang},\
  and\ \citenamefont {Yan}}]{Cai:2019jah}%
  \BibitemOpen
  \bibfield  {author} {\bibinfo {author} {\bibfnamefont {Y.-F.}\ \bibnamefont
  {Cai}}, \bibinfo {author} {\bibfnamefont {C.}~\bibnamefont {Chen}}, \bibinfo
  {author} {\bibfnamefont {X.}~\bibnamefont {Tong}}, \bibinfo {author}
  {\bibfnamefont {D.-G.}\ \bibnamefont {Wang}},\ and\ \bibinfo {author}
  {\bibfnamefont {S.-F.}\ \bibnamefont {Yan}},\ }\bibfield  {title} {\bibinfo
  {title} {{When Primordial Black Holes from Sound Speed Resonance Meet a
  Stochastic Background of Gravitational Waves}},\ }\href
  {https://doi.org/10.1103/PhysRevD.100.043518} {\bibfield  {journal} {\bibinfo
   {journal} {Phys. Rev. D}\ }\textbf {\bibinfo {volume} {100}},\ \bibinfo
  {pages} {043518} (\bibinfo {year} {2019}{\natexlab{a}})},\ \Eprint
  {https://arxiv.org/abs/1902.08187} {arXiv:1902.08187 [astro-ph.CO]}
  \BibitemShut {NoStop}%
\bibitem [{\citenamefont {Chen}\ and\ \citenamefont
  {Cai}(2019)}]{Chen:2019zza}%
  \BibitemOpen
  \bibfield  {author} {\bibinfo {author} {\bibfnamefont {C.}~\bibnamefont
  {Chen}}\ and\ \bibinfo {author} {\bibfnamefont {Y.-F.}\ \bibnamefont {Cai}},\
  }\bibfield  {title} {\bibinfo {title} {{Primordial black holes from sound
  speed resonance in the inflaton-curvaton mixed scenario}},\ }\href
  {https://doi.org/10.1088/1475-7516/2019/10/068} {\bibfield  {journal}
  {\bibinfo  {journal} {JCAP}\ }\textbf {\bibinfo {volume} {10}},\ \bibinfo
  {pages} {068}},\ \Eprint {https://arxiv.org/abs/1908.03942} {arXiv:1908.03942
  [astro-ph.CO]} \BibitemShut {NoStop}%
\bibitem [{\citenamefont {Ashoorioon}\ \emph {et~al.}(2021)\citenamefont
  {Ashoorioon}, \citenamefont {Rostami},\ and\ \citenamefont
  {Firouzjaee}}]{Ashoorioon:2019xqc}%
  \BibitemOpen
  \bibfield  {author} {\bibinfo {author} {\bibfnamefont {A.}~\bibnamefont
  {Ashoorioon}}, \bibinfo {author} {\bibfnamefont {A.}~\bibnamefont
  {Rostami}},\ and\ \bibinfo {author} {\bibfnamefont {J.~T.}\ \bibnamefont
  {Firouzjaee}},\ }\bibfield  {title} {\bibinfo {title} {{EFT compatible PBHs:
  effective spawning of the seeds for primordial black holes during
  inflation}},\ }\href {https://doi.org/10.1007/JHEP07(2021)087} {\bibfield
  {journal} {\bibinfo  {journal} {JHEP}\ }\textbf {\bibinfo {volume} {07}},\
  \bibinfo {pages} {087}},\ \Eprint {https://arxiv.org/abs/1912.13326}
  {arXiv:1912.13326 [astro-ph.CO]} \BibitemShut {NoStop}%
\bibitem [{\citenamefont {Chen}\ \emph {et~al.}(2020)\citenamefont {Chen},
  \citenamefont {Ma},\ and\ \citenamefont {Cai}}]{Chen:2020uhe}%
  \BibitemOpen
  \bibfield  {author} {\bibinfo {author} {\bibfnamefont {C.}~\bibnamefont
  {Chen}}, \bibinfo {author} {\bibfnamefont {X.-H.}\ \bibnamefont {Ma}},\ and\
  \bibinfo {author} {\bibfnamefont {Y.-F.}\ \bibnamefont {Cai}},\ }\bibfield
  {title} {\bibinfo {title} {{Dirac-Born-Infeld realization of sound speed
  resonance mechanism for primordial black holes}},\ }\href
  {https://doi.org/10.1103/PhysRevD.102.063526} {\bibfield  {journal} {\bibinfo
   {journal} {Phys. Rev. D}\ }\textbf {\bibinfo {volume} {102}},\ \bibinfo
  {pages} {063526} (\bibinfo {year} {2020})},\ \Eprint
  {https://arxiv.org/abs/2003.03821} {arXiv:2003.03821 [astro-ph.CO]}
  \BibitemShut {NoStop}%
\bibitem [{\citenamefont {Garcia-Bellido}\ \emph {et~al.}(1996)\citenamefont
  {Garcia-Bellido}, \citenamefont {Linde},\ and\ \citenamefont
  {Wands}}]{Garcia-Bellido:1996mdl}%
  \BibitemOpen
  \bibfield  {author} {\bibinfo {author} {\bibfnamefont {J.}~\bibnamefont
  {Garcia-Bellido}}, \bibinfo {author} {\bibfnamefont {A.~D.}\ \bibnamefont
  {Linde}},\ and\ \bibinfo {author} {\bibfnamefont {D.}~\bibnamefont {Wands}},\
  }\bibfield  {title} {\bibinfo {title} {{Density perturbations and black hole
  formation in hybrid inflation}},\ }\href
  {https://doi.org/10.1103/PhysRevD.54.6040} {\bibfield  {journal} {\bibinfo
  {journal} {Phys. Rev. D}\ }\textbf {\bibinfo {volume} {54}},\ \bibinfo
  {pages} {6040} (\bibinfo {year} {1996})},\ \Eprint
  {https://arxiv.org/abs/astro-ph/9605094} {arXiv:astro-ph/9605094}
  \BibitemShut {NoStop}%
\bibitem [{\citenamefont {Yokoyama}(1998)}]{Yokoyama:1998pt}%
  \BibitemOpen
  \bibfield  {author} {\bibinfo {author} {\bibfnamefont {J.}~\bibnamefont
  {Yokoyama}},\ }\bibfield  {title} {\bibinfo {title} {{Chaotic new inflation
  and formation of primordial black holes}},\ }\href
  {https://doi.org/10.1103/PhysRevD.58.083510} {\bibfield  {journal} {\bibinfo
  {journal} {Phys. Rev. D}\ }\textbf {\bibinfo {volume} {58}},\ \bibinfo
  {pages} {083510} (\bibinfo {year} {1998})},\ \Eprint
  {https://arxiv.org/abs/astro-ph/9802357} {arXiv:astro-ph/9802357}
  \BibitemShut {NoStop}%
\bibitem [{\citenamefont {Kohri}\ \emph {et~al.}(2013)\citenamefont {Kohri},
  \citenamefont {Lin},\ and\ \citenamefont {Matsuda}}]{Kohri:2012yw}%
  \BibitemOpen
  \bibfield  {author} {\bibinfo {author} {\bibfnamefont {K.}~\bibnamefont
  {Kohri}}, \bibinfo {author} {\bibfnamefont {C.-M.}\ \bibnamefont {Lin}},\
  and\ \bibinfo {author} {\bibfnamefont {T.}~\bibnamefont {Matsuda}},\
  }\bibfield  {title} {\bibinfo {title} {{Primordial black holes from the
  inflating curvaton}},\ }\href {https://doi.org/10.1103/PhysRevD.87.103527}
  {\bibfield  {journal} {\bibinfo  {journal} {Phys. Rev. D}\ }\textbf {\bibinfo
  {volume} {87}},\ \bibinfo {pages} {103527} (\bibinfo {year} {2013})},\
  \Eprint {https://arxiv.org/abs/1211.2371} {arXiv:1211.2371 [hep-ph]}
  \BibitemShut {NoStop}%
\bibitem [{\citenamefont {Clesse}\ and\ \citenamefont
  {Garc\'\i{}a-Bellido}(2015)}]{Clesse:2015wea}%
  \BibitemOpen
  \bibfield  {author} {\bibinfo {author} {\bibfnamefont {S.}~\bibnamefont
  {Clesse}}\ and\ \bibinfo {author} {\bibfnamefont {J.}~\bibnamefont
  {Garc\'\i{}a-Bellido}},\ }\bibfield  {title} {\bibinfo {title} {{Massive
  Primordial Black Holes from Hybrid Inflation as Dark Matter and the seeds of
  Galaxies}},\ }\href {https://doi.org/10.1103/PhysRevD.92.023524} {\bibfield
  {journal} {\bibinfo  {journal} {Phys. Rev. D}\ }\textbf {\bibinfo {volume}
  {92}},\ \bibinfo {pages} {023524} (\bibinfo {year} {2015})},\ \Eprint
  {https://arxiv.org/abs/1501.07565} {arXiv:1501.07565 [astro-ph.CO]}
  \BibitemShut {NoStop}%
\bibitem [{\citenamefont {Cheng}\ \emph {et~al.}(2017)\citenamefont {Cheng},
  \citenamefont {Lee},\ and\ \citenamefont {Ng}}]{Cheng:2016qzb}%
  \BibitemOpen
  \bibfield  {author} {\bibinfo {author} {\bibfnamefont {S.-L.}\ \bibnamefont
  {Cheng}}, \bibinfo {author} {\bibfnamefont {W.}~\bibnamefont {Lee}},\ and\
  \bibinfo {author} {\bibfnamefont {K.-W.}\ \bibnamefont {Ng}},\ }\bibfield
  {title} {\bibinfo {title} {{Production of high stellar-mass primordial black
  holes in trapped inflation}},\ }\href
  {https://doi.org/10.1007/JHEP02(2017)008} {\bibfield  {journal} {\bibinfo
  {journal} {JHEP}\ }\textbf {\bibinfo {volume} {02}},\ \bibinfo {pages}
  {008}},\ \Eprint {https://arxiv.org/abs/1606.00206} {arXiv:1606.00206
  [astro-ph.CO]} \BibitemShut {NoStop}%
\bibitem [{\citenamefont {Espinosa}\ \emph
  {et~al.}(2018{\natexlab{a}})\citenamefont {Espinosa}, \citenamefont {Racco},\
  and\ \citenamefont {Riotto}}]{Espinosa:2017sgp}%
  \BibitemOpen
  \bibfield  {author} {\bibinfo {author} {\bibfnamefont {J.~R.}\ \bibnamefont
  {Espinosa}}, \bibinfo {author} {\bibfnamefont {D.}~\bibnamefont {Racco}},\
  and\ \bibinfo {author} {\bibfnamefont {A.}~\bibnamefont {Riotto}},\
  }\bibfield  {title} {\bibinfo {title} {{Cosmological Signature of the
  Standard Model Higgs Vacuum Instability: Primordial Black Holes as Dark
  Matter}},\ }\href {https://doi.org/10.1103/PhysRevLett.120.121301} {\bibfield
   {journal} {\bibinfo  {journal} {Phys. Rev. Lett.}\ }\textbf {\bibinfo
  {volume} {120}},\ \bibinfo {pages} {121301} (\bibinfo {year}
  {2018}{\natexlab{a}})},\ \Eprint {https://arxiv.org/abs/1710.11196}
  {arXiv:1710.11196 [hep-ph]} \BibitemShut {NoStop}%
\bibitem [{\citenamefont {Kannike}\ \emph {et~al.}(2017)\citenamefont
  {Kannike}, \citenamefont {Marzola}, \citenamefont {Raidal},\ and\
  \citenamefont {Veerm\"ae}}]{Kannike:2017bxn}%
  \BibitemOpen
  \bibfield  {author} {\bibinfo {author} {\bibfnamefont {K.}~\bibnamefont
  {Kannike}}, \bibinfo {author} {\bibfnamefont {L.}~\bibnamefont {Marzola}},
  \bibinfo {author} {\bibfnamefont {M.}~\bibnamefont {Raidal}},\ and\ \bibinfo
  {author} {\bibfnamefont {H.}~\bibnamefont {Veerm\"ae}},\ }\bibfield  {title}
  {\bibinfo {title} {{Single Field Double Inflation and Primordial Black
  Holes}},\ }\href {https://doi.org/10.1088/1475-7516/2017/09/020} {\bibfield
  {journal} {\bibinfo  {journal} {JCAP}\ }\textbf {\bibinfo {volume} {09}},\
  \bibinfo {pages} {020}},\ \Eprint {https://arxiv.org/abs/1705.06225}
  {arXiv:1705.06225 [astro-ph.CO]} \BibitemShut {NoStop}%
\bibitem [{\citenamefont {Garcia-Bellido}\ and\ \citenamefont
  {Ruiz~Morales}(2017)}]{Garcia-Bellido:2017mdw}%
  \BibitemOpen
  \bibfield  {author} {\bibinfo {author} {\bibfnamefont {J.}~\bibnamefont
  {Garcia-Bellido}}\ and\ \bibinfo {author} {\bibfnamefont {E.}~\bibnamefont
  {Ruiz~Morales}},\ }\bibfield  {title} {\bibinfo {title} {{Primordial black
  holes from single field models of inflation}},\ }\href
  {https://doi.org/10.1016/j.dark.2017.09.007} {\bibfield  {journal} {\bibinfo
  {journal} {Phys. Dark Univ.}\ }\textbf {\bibinfo {volume} {18}},\ \bibinfo
  {pages} {47} (\bibinfo {year} {2017})},\ \Eprint
  {https://arxiv.org/abs/1702.03901} {arXiv:1702.03901 [astro-ph.CO]}
  \BibitemShut {NoStop}%
\bibitem [{\citenamefont {Cheng}\ \emph {et~al.}(2018)\citenamefont {Cheng},
  \citenamefont {Lee},\ and\ \citenamefont {Ng}}]{Cheng:2018yyr}%
  \BibitemOpen
  \bibfield  {author} {\bibinfo {author} {\bibfnamefont {S.-L.}\ \bibnamefont
  {Cheng}}, \bibinfo {author} {\bibfnamefont {W.}~\bibnamefont {Lee}},\ and\
  \bibinfo {author} {\bibfnamefont {K.-W.}\ \bibnamefont {Ng}},\ }\bibfield
  {title} {\bibinfo {title} {{Primordial black holes and associated
  gravitational waves in axion monodromy inflation}},\ }\href
  {https://doi.org/10.1088/1475-7516/2018/07/001} {\bibfield  {journal}
  {\bibinfo  {journal} {JCAP}\ }\textbf {\bibinfo {volume} {07}},\ \bibinfo
  {pages} {001}},\ \Eprint {https://arxiv.org/abs/1801.09050} {arXiv:1801.09050
  [astro-ph.CO]} \BibitemShut {NoStop}%
\bibitem [{\citenamefont {Passaglia}\ \emph {et~al.}(2019)\citenamefont
  {Passaglia}, \citenamefont {Hu},\ and\ \citenamefont
  {Motohashi}}]{Passaglia:2018ixg}%
  \BibitemOpen
  \bibfield  {author} {\bibinfo {author} {\bibfnamefont {S.}~\bibnamefont
  {Passaglia}}, \bibinfo {author} {\bibfnamefont {W.}~\bibnamefont {Hu}},\ and\
  \bibinfo {author} {\bibfnamefont {H.}~\bibnamefont {Motohashi}},\ }\bibfield
  {title} {\bibinfo {title} {{Primordial black holes and local non-Gaussianity
  in canonical inflation}},\ }\href
  {https://doi.org/10.1103/PhysRevD.99.043536} {\bibfield  {journal} {\bibinfo
  {journal} {Phys. Rev. D}\ }\textbf {\bibinfo {volume} {99}},\ \bibinfo
  {pages} {043536} (\bibinfo {year} {2019})},\ \Eprint
  {https://arxiv.org/abs/1812.08243} {arXiv:1812.08243 [astro-ph.CO]}
  \BibitemShut {NoStop}%
\bibitem [{\citenamefont {Ando}\ \emph {et~al.}(2018)\citenamefont {Ando},
  \citenamefont {Kawasaki},\ and\ \citenamefont {Nakatsuka}}]{Ando:2018nge}%
  \BibitemOpen
  \bibfield  {author} {\bibinfo {author} {\bibfnamefont {K.}~\bibnamefont
  {Ando}}, \bibinfo {author} {\bibfnamefont {M.}~\bibnamefont {Kawasaki}},\
  and\ \bibinfo {author} {\bibfnamefont {H.}~\bibnamefont {Nakatsuka}},\
  }\bibfield  {title} {\bibinfo {title} {{Formation of primordial black holes
  in an axionlike curvaton model}},\ }\href
  {https://doi.org/10.1103/PhysRevD.98.083508} {\bibfield  {journal} {\bibinfo
  {journal} {Phys. Rev. D}\ }\textbf {\bibinfo {volume} {98}},\ \bibinfo
  {pages} {083508} (\bibinfo {year} {2018})},\ \Eprint
  {https://arxiv.org/abs/1805.07757} {arXiv:1805.07757 [astro-ph.CO]}
  \BibitemShut {NoStop}%
\bibitem [{\citenamefont {Espinosa}\ \emph
  {et~al.}(2018{\natexlab{b}})\citenamefont {Espinosa}, \citenamefont {Racco},\
  and\ \citenamefont {Riotto}}]{Espinosa:2018eve}%
  \BibitemOpen
  \bibfield  {author} {\bibinfo {author} {\bibfnamefont {J.~R.}\ \bibnamefont
  {Espinosa}}, \bibinfo {author} {\bibfnamefont {D.}~\bibnamefont {Racco}},\
  and\ \bibinfo {author} {\bibfnamefont {A.}~\bibnamefont {Riotto}},\
  }\bibfield  {title} {\bibinfo {title} {{A Cosmological Signature of the SM
  Higgs Instability: Gravitational Waves}},\ }\href
  {https://doi.org/10.1088/1475-7516/2018/09/012} {\bibfield  {journal}
  {\bibinfo  {journal} {JCAP}\ }\textbf {\bibinfo {volume} {09}},\ \bibinfo
  {pages} {012}},\ \Eprint {https://arxiv.org/abs/1804.07732} {arXiv:1804.07732
  [hep-ph]} \BibitemShut {NoStop}%
\bibitem [{\citenamefont {Inomata}\ \emph {et~al.}(2018)\citenamefont
  {Inomata}, \citenamefont {Kawasaki}, \citenamefont {Mukaida},\ and\
  \citenamefont {Yanagida}}]{Inomata:2018cht}%
  \BibitemOpen
  \bibfield  {author} {\bibinfo {author} {\bibfnamefont {K.}~\bibnamefont
  {Inomata}}, \bibinfo {author} {\bibfnamefont {M.}~\bibnamefont {Kawasaki}},
  \bibinfo {author} {\bibfnamefont {K.}~\bibnamefont {Mukaida}},\ and\ \bibinfo
  {author} {\bibfnamefont {T.~T.}\ \bibnamefont {Yanagida}},\ }\bibfield
  {title} {\bibinfo {title} {{Double inflation as a single origin of primordial
  black holes for all dark matter and LIGO observations}},\ }\href
  {https://doi.org/10.1103/PhysRevD.97.043514} {\bibfield  {journal} {\bibinfo
  {journal} {Phys. Rev. D}\ }\textbf {\bibinfo {volume} {97}},\ \bibinfo
  {pages} {043514} (\bibinfo {year} {2018})},\ \Eprint
  {https://arxiv.org/abs/1711.06129} {arXiv:1711.06129 [astro-ph.CO]}
  \BibitemShut {NoStop}%
\bibitem [{\citenamefont {Braglia}\ \emph
  {et~al.}(2020{\natexlab{a}})\citenamefont {Braglia}, \citenamefont {Hazra},
  \citenamefont {Finelli}, \citenamefont {Smoot}, \citenamefont {Sriramkumar},\
  and\ \citenamefont {Starobinsky}}]{Braglia:2020eai}%
  \BibitemOpen
  \bibfield  {author} {\bibinfo {author} {\bibfnamefont {M.}~\bibnamefont
  {Braglia}}, \bibinfo {author} {\bibfnamefont {D.~K.}\ \bibnamefont {Hazra}},
  \bibinfo {author} {\bibfnamefont {F.}~\bibnamefont {Finelli}}, \bibinfo
  {author} {\bibfnamefont {G.~F.}\ \bibnamefont {Smoot}}, \bibinfo {author}
  {\bibfnamefont {L.}~\bibnamefont {Sriramkumar}},\ and\ \bibinfo {author}
  {\bibfnamefont {A.~A.}\ \bibnamefont {Starobinsky}},\ }\bibfield  {title}
  {\bibinfo {title} {{Generating PBHs and small-scale GWs in two-field models
  of inflation}},\ }\href {https://doi.org/10.1088/1475-7516/2020/08/001}
  {\bibfield  {journal} {\bibinfo  {journal} {JCAP}\ }\textbf {\bibinfo
  {volume} {08}},\ \bibinfo {pages} {001}},\ \Eprint
  {https://arxiv.org/abs/2005.02895} {arXiv:2005.02895 [astro-ph.CO]}
  \BibitemShut {NoStop}%
\bibitem [{\citenamefont {Atal}\ and\ \citenamefont
  {Germani}(2019)}]{Atal:2018neu}%
  \BibitemOpen
  \bibfield  {author} {\bibinfo {author} {\bibfnamefont {V.}~\bibnamefont
  {Atal}}\ and\ \bibinfo {author} {\bibfnamefont {C.}~\bibnamefont {Germani}},\
  }\bibfield  {title} {\bibinfo {title} {{The role of non-gaussianities in
  Primordial Black Hole formation}},\ }\href
  {https://doi.org/10.1016/j.dark.2019.100275} {\bibfield  {journal} {\bibinfo
  {journal} {Phys. Dark Univ.}\ }\textbf {\bibinfo {volume} {24}},\ \bibinfo
  {pages} {100275} (\bibinfo {year} {2019})},\ \Eprint
  {https://arxiv.org/abs/1811.07857} {arXiv:1811.07857 [astro-ph.CO]}
  \BibitemShut {NoStop}%
\bibitem [{\citenamefont {Liu}\ \emph {et~al.}(2020)\citenamefont {Liu},
  \citenamefont {Guo},\ and\ \citenamefont {Cai}}]{Liu:2020oqe}%
  \BibitemOpen
  \bibfield  {author} {\bibinfo {author} {\bibfnamefont {J.}~\bibnamefont
  {Liu}}, \bibinfo {author} {\bibfnamefont {Z.-K.}\ \bibnamefont {Guo}},\ and\
  \bibinfo {author} {\bibfnamefont {R.-G.}\ \bibnamefont {Cai}},\ }\bibfield
  {title} {\bibinfo {title} {{Analytical approximation of the scalar spectrum
  in the ultraslow-roll inflationary models}},\ }\href
  {https://doi.org/10.1103/PhysRevD.101.083535} {\bibfield  {journal} {\bibinfo
   {journal} {Phys. Rev. D}\ }\textbf {\bibinfo {volume} {101}},\ \bibinfo
  {pages} {083535} (\bibinfo {year} {2020})},\ \Eprint
  {https://arxiv.org/abs/2003.02075} {arXiv:2003.02075 [astro-ph.CO]}
  \BibitemShut {NoStop}%
\bibitem [{\citenamefont {Ng}\ and\ \citenamefont {Wu}(2021)}]{Ng:2021hll}%
  \BibitemOpen
  \bibfield  {author} {\bibinfo {author} {\bibfnamefont {K.-W.}\ \bibnamefont
  {Ng}}\ and\ \bibinfo {author} {\bibfnamefont {Y.-P.}\ \bibnamefont {Wu}},\
  }\bibfield  {title} {\bibinfo {title} {{Constant-rate inflation: primordial
  black holes from conformal weight transitions}},\ }\href
  {https://doi.org/10.1007/JHEP11(2021)076} {\bibfield  {journal} {\bibinfo
  {journal} {JHEP}\ }\textbf {\bibinfo {volume} {11}},\ \bibinfo {pages}
  {076}},\ \Eprint {https://arxiv.org/abs/2102.05620} {arXiv:2102.05620
  [astro-ph.CO]} \BibitemShut {NoStop}%
\bibitem [{\citenamefont {Palma}\ \emph {et~al.}(2020)\citenamefont {Palma},
  \citenamefont {Sypsas},\ and\ \citenamefont {Zenteno}}]{Palma:2020ejf}%
  \BibitemOpen
  \bibfield  {author} {\bibinfo {author} {\bibfnamefont {G.~A.}\ \bibnamefont
  {Palma}}, \bibinfo {author} {\bibfnamefont {S.}~\bibnamefont {Sypsas}},\ and\
  \bibinfo {author} {\bibfnamefont {C.}~\bibnamefont {Zenteno}},\ }\bibfield
  {title} {\bibinfo {title} {{Seeding primordial black holes in multifield
  inflation}},\ }\href {https://doi.org/10.1103/PhysRevLett.125.121301}
  {\bibfield  {journal} {\bibinfo  {journal} {Phys. Rev. Lett.}\ }\textbf
  {\bibinfo {volume} {125}},\ \bibinfo {pages} {121301} (\bibinfo {year}
  {2020})},\ \Eprint {https://arxiv.org/abs/2004.06106} {arXiv:2004.06106
  [astro-ph.CO]} \BibitemShut {NoStop}%
\bibitem [{\citenamefont {Fumagalli}\ \emph
  {et~al.}(2020{\natexlab{a}})\citenamefont {Fumagalli}, \citenamefont
  {Renaux-Petel}, \citenamefont {Ronayne},\ and\ \citenamefont
  {Witkowski}}]{Fumagalli:2020adf}%
  \BibitemOpen
  \bibfield  {author} {\bibinfo {author} {\bibfnamefont {J.}~\bibnamefont
  {Fumagalli}}, \bibinfo {author} {\bibfnamefont {S.}~\bibnamefont
  {Renaux-Petel}}, \bibinfo {author} {\bibfnamefont {J.~W.}\ \bibnamefont
  {Ronayne}},\ and\ \bibinfo {author} {\bibfnamefont {L.~T.}\ \bibnamefont
  {Witkowski}},\ }\bibfield  {title} {\bibinfo {title} {{Turning in the
  landscape: a new mechanism for generating Primordial Black Holes}},\
  }\href@noop {} {\  (\bibinfo {year} {2020}{\natexlab{a}})},\ \Eprint
  {https://arxiv.org/abs/2004.08369} {arXiv:2004.08369 [hep-th]} \BibitemShut
  {NoStop}%
\bibitem [{\citenamefont {Inomata}\ \emph {et~al.}(2022)\citenamefont
  {Inomata}, \citenamefont {McDonough},\ and\ \citenamefont
  {Hu}}]{Inomata:2021tpx}%
  \BibitemOpen
  \bibfield  {author} {\bibinfo {author} {\bibfnamefont {K.}~\bibnamefont
  {Inomata}}, \bibinfo {author} {\bibfnamefont {E.}~\bibnamefont {McDonough}},\
  and\ \bibinfo {author} {\bibfnamefont {W.}~\bibnamefont {Hu}},\ }\bibfield
  {title} {\bibinfo {title} {{Amplification of primordial perturbations from
  the rise or fall of the inflaton}},\ }\href
  {https://doi.org/10.1088/1475-7516/2022/02/031} {\bibfield  {journal}
  {\bibinfo  {journal} {JCAP}\ }\textbf {\bibinfo {volume} {02}}\bibfield
  {number} {\bibinfo  {number} { (02)},\ \bibinfo {pages} {031}},\ }\Eprint
  {https://arxiv.org/abs/2110.14641} {arXiv:2110.14641 [astro-ph.CO]}
  \BibitemShut {NoStop}%
\bibitem [{\citenamefont {Gundhi}\ \emph {et~al.}(2021)\citenamefont {Gundhi},
  \citenamefont {Ketov},\ and\ \citenamefont {Steinwachs}}]{Gundhi:2020kzm}%
  \BibitemOpen
  \bibfield  {author} {\bibinfo {author} {\bibfnamefont {A.}~\bibnamefont
  {Gundhi}}, \bibinfo {author} {\bibfnamefont {S.~V.}\ \bibnamefont {Ketov}},\
  and\ \bibinfo {author} {\bibfnamefont {C.~F.}\ \bibnamefont {Steinwachs}},\
  }\bibfield  {title} {\bibinfo {title} {{Primordial black hole dark matter in
  dilaton-extended two-field Starobinsky inflation}},\ }\href
  {https://doi.org/10.1103/PhysRevD.103.083518} {\bibfield  {journal} {\bibinfo
   {journal} {Phys. Rev. D}\ }\textbf {\bibinfo {volume} {103}},\ \bibinfo
  {pages} {083518} (\bibinfo {year} {2021})},\ \Eprint
  {https://arxiv.org/abs/2011.05999} {arXiv:2011.05999 [hep-th]} \BibitemShut
  {NoStop}%
\bibitem [{\citenamefont {Zhou}\ \emph {et~al.}(2020)\citenamefont {Zhou},
  \citenamefont {Jiang}, \citenamefont {Cai}, \citenamefont {Sasaki},\ and\
  \citenamefont {Pi}}]{Zhou:2020kkf}%
  \BibitemOpen
  \bibfield  {author} {\bibinfo {author} {\bibfnamefont {Z.}~\bibnamefont
  {Zhou}}, \bibinfo {author} {\bibfnamefont {J.}~\bibnamefont {Jiang}},
  \bibinfo {author} {\bibfnamefont {Y.-F.}\ \bibnamefont {Cai}}, \bibinfo
  {author} {\bibfnamefont {M.}~\bibnamefont {Sasaki}},\ and\ \bibinfo {author}
  {\bibfnamefont {S.}~\bibnamefont {Pi}},\ }\bibfield  {title} {\bibinfo
  {title} {{Primordial black holes and gravitational waves from resonant
  amplification during inflation}},\ }\href
  {https://doi.org/10.1103/PhysRevD.102.103527} {\bibfield  {journal} {\bibinfo
   {journal} {Phys. Rev. D}\ }\textbf {\bibinfo {volume} {102}},\ \bibinfo
  {pages} {103527} (\bibinfo {year} {2020})},\ \Eprint
  {https://arxiv.org/abs/2010.03537} {arXiv:2010.03537 [astro-ph.CO]}
  \BibitemShut {NoStop}%
\bibitem [{\citenamefont {Ragavendra}\ \emph {et~al.}(2020)\citenamefont
  {Ragavendra}, \citenamefont {Saha}, \citenamefont {Sriramkumar},\ and\
  \citenamefont {Silk}}]{Ragavendra:2020sop}%
  \BibitemOpen
  \bibfield  {author} {\bibinfo {author} {\bibfnamefont {H.~V.}\ \bibnamefont
  {Ragavendra}}, \bibinfo {author} {\bibfnamefont {P.}~\bibnamefont {Saha}},
  \bibinfo {author} {\bibfnamefont {L.}~\bibnamefont {Sriramkumar}},\ and\
  \bibinfo {author} {\bibfnamefont {J.}~\bibnamefont {Silk}},\ }\bibfield
  {title} {\bibinfo {title} {{PBHs and secondary GWs from ultra slow roll and
  punctuated inflation}},\ }\href@noop {} {\  (\bibinfo {year} {2020})},\
  \Eprint {https://arxiv.org/abs/2008.12202} {arXiv:2008.12202 [astro-ph.CO]}
  \BibitemShut {NoStop}%
\bibitem [{\citenamefont {Cai}\ and\ \citenamefont {Piao}(2021)}]{Cai:2020qpu}%
  \BibitemOpen
  \bibfield  {author} {\bibinfo {author} {\bibfnamefont {Y.}~\bibnamefont
  {Cai}}\ and\ \bibinfo {author} {\bibfnamefont {Y.-S.}\ \bibnamefont {Piao}},\
  }\bibfield  {title} {\bibinfo {title} {{Intermittent null energy condition
  violations during inflation and primordial gravitational waves}},\ }\href
  {https://doi.org/10.1103/PhysRevD.103.083521} {\bibfield  {journal} {\bibinfo
   {journal} {Phys. Rev. D}\ }\textbf {\bibinfo {volume} {103}},\ \bibinfo
  {pages} {083521} (\bibinfo {year} {2021})},\ \Eprint
  {https://arxiv.org/abs/2012.11304} {arXiv:2012.11304 [gr-qc]} \BibitemShut
  {NoStop}%
\bibitem [{\citenamefont {Inomata}\ \emph
  {et~al.}(2023{\natexlab{a}})\citenamefont {Inomata}, \citenamefont {Braglia},
  \citenamefont {Chen},\ and\ \citenamefont {Renaux-Petel}}]{Inomata:2022yte}%
  \BibitemOpen
  \bibfield  {author} {\bibinfo {author} {\bibfnamefont {K.}~\bibnamefont
  {Inomata}}, \bibinfo {author} {\bibfnamefont {M.}~\bibnamefont {Braglia}},
  \bibinfo {author} {\bibfnamefont {X.}~\bibnamefont {Chen}},\ and\ \bibinfo
  {author} {\bibfnamefont {S.}~\bibnamefont {Renaux-Petel}},\ }\bibfield
  {title} {\bibinfo {title} {{Questions on calculation of primordial power
  spectrum with large spikes: the resonance model case}},\ }\href
  {https://doi.org/10.1088/1475-7516/2023/04/011} {\bibfield  {journal}
  {\bibinfo  {journal} {JCAP}\ }\textbf {\bibinfo {volume} {04}},\ \bibinfo
  {pages} {011}},\ \Eprint {https://arxiv.org/abs/2211.02586} {arXiv:2211.02586
  [astro-ph.CO]} \BibitemShut {NoStop}%
\bibitem [{\citenamefont {Cai}\ \emph {et~al.}(2023{\natexlab{a}})\citenamefont
  {Cai}, \citenamefont {Zhu},\ and\ \citenamefont {Piao}}]{Cai:2023uhc}%
  \BibitemOpen
  \bibfield  {author} {\bibinfo {author} {\bibfnamefont {Y.}~\bibnamefont
  {Cai}}, \bibinfo {author} {\bibfnamefont {M.}~\bibnamefont {Zhu}},\ and\
  \bibinfo {author} {\bibfnamefont {Y.-S.}\ \bibnamefont {Piao}},\ }\bibfield
  {title} {\bibinfo {title} {{Primordial black holes from null energy condition
  violation during inflation}},\ }\href@noop {} {\  (\bibinfo {year}
  {2023}{\natexlab{a}})},\ \Eprint {https://arxiv.org/abs/2305.10933}
  {arXiv:2305.10933 [gr-qc]} \BibitemShut {NoStop}%
\bibitem [{\citenamefont {Briaud}\ and\ \citenamefont
  {Vennin}(2023)}]{Briaud:2023eae}%
  \BibitemOpen
  \bibfield  {author} {\bibinfo {author} {\bibfnamefont {V.}~\bibnamefont
  {Briaud}}\ and\ \bibinfo {author} {\bibfnamefont {V.}~\bibnamefont
  {Vennin}},\ }\bibfield  {title} {\bibinfo {title} {{Uphill inflation}},\
  }\href {https://doi.org/10.1088/1475-7516/2023/06/029} {\bibfield  {journal}
  {\bibinfo  {journal} {JCAP}\ }\textbf {\bibinfo {volume} {06}},\ \bibinfo
  {pages} {029}},\ \Eprint {https://arxiv.org/abs/2301.09336} {arXiv:2301.09336
  [astro-ph.CO]} \BibitemShut {NoStop}%
\bibitem [{\citenamefont {Heydari}\ and\ \citenamefont
  {Karami}(2023)}]{Heydari:2023xts}%
  \BibitemOpen
  \bibfield  {author} {\bibinfo {author} {\bibfnamefont {S.}~\bibnamefont
  {Heydari}}\ and\ \bibinfo {author} {\bibfnamefont {K.}~\bibnamefont
  {Karami}},\ }\bibfield  {title} {\bibinfo {title} {{Primordial black holes in
  non-canonical scalar field inflation driven by Higgs potential in the
  presence of bump}},\ }\href@noop {} {\  (\bibinfo {year} {2023})},\ \Eprint
  {https://arxiv.org/abs/2309.01239} {arXiv:2309.01239 [astro-ph.CO]}
  \BibitemShut {NoStop}%
\bibitem [{\citenamefont {Karam}\ \emph {et~al.}(2023)\citenamefont {Karam},
  \citenamefont {Koivunen}, \citenamefont {Tomberg}, \citenamefont {Racioppi},\
  and\ \citenamefont {Veerm\"ae}}]{Karam:2023haj}%
  \BibitemOpen
  \bibfield  {author} {\bibinfo {author} {\bibfnamefont {A.}~\bibnamefont
  {Karam}}, \bibinfo {author} {\bibfnamefont {N.}~\bibnamefont {Koivunen}},
  \bibinfo {author} {\bibfnamefont {E.}~\bibnamefont {Tomberg}}, \bibinfo
  {author} {\bibfnamefont {A.}~\bibnamefont {Racioppi}},\ and\ \bibinfo
  {author} {\bibfnamefont {H.}~\bibnamefont {Veerm\"ae}},\ }\bibfield  {title}
  {\bibinfo {title} {{Primordial black holes and inflation from double-well
  potentials}},\ }\href {https://doi.org/10.1088/1475-7516/2023/09/002}
  {\bibfield  {journal} {\bibinfo  {journal} {JCAP}\ }\textbf {\bibinfo
  {volume} {09}},\ \bibinfo {pages} {002}},\ \Eprint
  {https://arxiv.org/abs/2305.09630} {arXiv:2305.09630 [astro-ph.CO]}
  \BibitemShut {NoStop}%
\bibitem [{\citenamefont {\"Ozsoy}\ and\ \citenamefont
  {Tasinato}(2023)}]{Ozsoy:2023ryl}%
  \BibitemOpen
  \bibfield  {author} {\bibinfo {author} {\bibfnamefont {O.}~\bibnamefont
  {\"Ozsoy}}\ and\ \bibinfo {author} {\bibfnamefont {G.}~\bibnamefont
  {Tasinato}},\ }\bibfield  {title} {\bibinfo {title} {{Inflation and
  Primordial Black Holes}},\ }\href@noop {} {\  (\bibinfo {year} {2023})},\
  \Eprint {https://arxiv.org/abs/2301.03600} {arXiv:2301.03600 [astro-ph.CO]}
  \BibitemShut {NoStop}%
\bibitem [{\citenamefont {Leach}\ \emph {et~al.}(2001)\citenamefont {Leach},
  \citenamefont {Sasaki}, \citenamefont {Wands},\ and\ \citenamefont
  {Liddle}}]{Leach:2001zf}%
  \BibitemOpen
  \bibfield  {author} {\bibinfo {author} {\bibfnamefont {S.~M.}\ \bibnamefont
  {Leach}}, \bibinfo {author} {\bibfnamefont {M.}~\bibnamefont {Sasaki}},
  \bibinfo {author} {\bibfnamefont {D.}~\bibnamefont {Wands}},\ and\ \bibinfo
  {author} {\bibfnamefont {A.~R.}\ \bibnamefont {Liddle}},\ }\bibfield  {title}
  {\bibinfo {title} {{Enhancement of superhorizon scale inflationary curvature
  perturbations}},\ }\href {https://doi.org/10.1103/PhysRevD.64.023512}
  {\bibfield  {journal} {\bibinfo  {journal} {Phys. Rev. D}\ }\textbf {\bibinfo
  {volume} {64}},\ \bibinfo {pages} {023512} (\bibinfo {year} {2001})},\
  \Eprint {https://arxiv.org/abs/astro-ph/0101406} {arXiv:astro-ph/0101406}
  \BibitemShut {NoStop}%
\bibitem [{\citenamefont {Byrnes}\ \emph {et~al.}(2019)\citenamefont {Byrnes},
  \citenamefont {Cole},\ and\ \citenamefont {Patil}}]{Byrnes:2018txb}%
  \BibitemOpen
  \bibfield  {author} {\bibinfo {author} {\bibfnamefont {C.~T.}\ \bibnamefont
  {Byrnes}}, \bibinfo {author} {\bibfnamefont {P.~S.}\ \bibnamefont {Cole}},\
  and\ \bibinfo {author} {\bibfnamefont {S.~P.}\ \bibnamefont {Patil}},\
  }\bibfield  {title} {\bibinfo {title} {{Steepest growth of the power spectrum
  and primordial black holes}},\ }\href
  {https://doi.org/10.1088/1475-7516/2019/06/028} {\bibfield  {journal}
  {\bibinfo  {journal} {JCAP}\ }\textbf {\bibinfo {volume} {06}},\ \bibinfo
  {pages} {028}},\ \Eprint {https://arxiv.org/abs/1811.11158} {arXiv:1811.11158
  [astro-ph.CO]} \BibitemShut {NoStop}%
\bibitem [{\citenamefont {Carrilho}\ \emph {et~al.}(2019)\citenamefont
  {Carrilho}, \citenamefont {Malik},\ and\ \citenamefont
  {Mulryne}}]{Carrilho:2019oqg}%
  \BibitemOpen
  \bibfield  {author} {\bibinfo {author} {\bibfnamefont {P.}~\bibnamefont
  {Carrilho}}, \bibinfo {author} {\bibfnamefont {K.~A.}\ \bibnamefont
  {Malik}},\ and\ \bibinfo {author} {\bibfnamefont {D.~J.}\ \bibnamefont
  {Mulryne}},\ }\bibfield  {title} {\bibinfo {title} {{Dissecting the growth of
  the power spectrum for primordial black holes}},\ }\href
  {https://doi.org/10.1103/PhysRevD.100.103529} {\bibfield  {journal} {\bibinfo
   {journal} {Phys. Rev. D}\ }\textbf {\bibinfo {volume} {100}},\ \bibinfo
  {pages} {103529} (\bibinfo {year} {2019})},\ \Eprint
  {https://arxiv.org/abs/1907.05237} {arXiv:1907.05237 [astro-ph.CO]}
  \BibitemShut {NoStop}%
\bibitem [{\citenamefont {\"Ozsoy}\ and\ \citenamefont
  {Tasinato}(2020)}]{Ozsoy:2019lyy}%
  \BibitemOpen
  \bibfield  {author} {\bibinfo {author} {\bibfnamefont {O.}~\bibnamefont
  {\"Ozsoy}}\ and\ \bibinfo {author} {\bibfnamefont {G.}~\bibnamefont
  {Tasinato}},\ }\bibfield  {title} {\bibinfo {title} {{On the slope of the
  curvature power spectrum in non-attractor inflation}},\ }\href
  {https://doi.org/10.1088/1475-7516/2020/04/048} {\bibfield  {journal}
  {\bibinfo  {journal} {JCAP}\ }\textbf {\bibinfo {volume} {04}},\ \bibinfo
  {pages} {048}},\ \Eprint {https://arxiv.org/abs/1912.01061} {arXiv:1912.01061
  [astro-ph.CO]} \BibitemShut {NoStop}%
\bibitem [{\citenamefont {Cole}\ \emph {et~al.}(2022)\citenamefont {Cole},
  \citenamefont {Gow}, \citenamefont {Byrnes},\ and\ \citenamefont
  {Patil}}]{Cole:2022xqc}%
  \BibitemOpen
  \bibfield  {author} {\bibinfo {author} {\bibfnamefont {P.~S.}\ \bibnamefont
  {Cole}}, \bibinfo {author} {\bibfnamefont {A.~D.}\ \bibnamefont {Gow}},
  \bibinfo {author} {\bibfnamefont {C.~T.}\ \bibnamefont {Byrnes}},\ and\
  \bibinfo {author} {\bibfnamefont {S.~P.}\ \bibnamefont {Patil}},\ }\bibfield
  {title} {\bibinfo {title} {{Steepest growth re-examined: repercussions for
  primordial black hole formation}},\ }\href@noop {} {\  (\bibinfo {year}
  {2022})},\ \Eprint {https://arxiv.org/abs/2204.07573} {arXiv:2204.07573
  [astro-ph.CO]} \BibitemShut {NoStop}%
\bibitem [{\citenamefont {Tasinato}(2023)}]{Tasinato:2023ukp}%
  \BibitemOpen
  \bibfield  {author} {\bibinfo {author} {\bibfnamefont {G.}~\bibnamefont
  {Tasinato}},\ }\bibfield  {title} {\bibinfo {title} {{Large
  |\ensuremath{\eta}| approach to single field inflation}},\ }\href
  {https://doi.org/10.1103/PhysRevD.108.043526} {\bibfield  {journal} {\bibinfo
   {journal} {Phys. Rev. D}\ }\textbf {\bibinfo {volume} {108}},\ \bibinfo
  {pages} {043526} (\bibinfo {year} {2023})},\ \Eprint
  {https://arxiv.org/abs/2305.11568} {arXiv:2305.11568 [hep-th]} \BibitemShut
  {NoStop}%
\bibitem [{\citenamefont {Starobinsky}(1985)}]{Starobinsky:1985ibc}%
  \BibitemOpen
  \bibfield  {author} {\bibinfo {author} {\bibfnamefont {A.~A.}\ \bibnamefont
  {Starobinsky}},\ }\bibfield  {title} {\bibinfo {title} {{Multicomponent de
  Sitter (Inflationary) Stages and the Generation of Perturbations}},\
  }\href@noop {} {\bibfield  {journal} {\bibinfo  {journal} {JETP Lett.}\
  }\textbf {\bibinfo {volume} {42}},\ \bibinfo {pages} {152} (\bibinfo {year}
  {1985})}\BibitemShut {NoStop}%
\bibitem [{\citenamefont {Salopek}\ and\ \citenamefont
  {Bond}(1990)}]{Salopek:1990jq}%
  \BibitemOpen
  \bibfield  {author} {\bibinfo {author} {\bibfnamefont {D.}~\bibnamefont
  {Salopek}}\ and\ \bibinfo {author} {\bibfnamefont {J.}~\bibnamefont {Bond}},\
  }\bibfield  {title} {\bibinfo {title} {{Nonlinear evolution of long
  wavelength metric fluctuations in inflationary models}},\ }\href
  {https://doi.org/10.1103/PhysRevD.42.3936} {\bibfield  {journal} {\bibinfo
  {journal} {Phys. Rev. D}\ }\textbf {\bibinfo {volume} {42}},\ \bibinfo
  {pages} {3936} (\bibinfo {year} {1990})}\BibitemShut {NoStop}%
\bibitem [{\citenamefont {Sasaki}\ and\ \citenamefont
  {Stewart}(1996)}]{Sasaki:1995aw}%
  \BibitemOpen
  \bibfield  {author} {\bibinfo {author} {\bibfnamefont {M.}~\bibnamefont
  {Sasaki}}\ and\ \bibinfo {author} {\bibfnamefont {E.~D.}\ \bibnamefont
  {Stewart}},\ }\bibfield  {title} {\bibinfo {title} {{A General analytic
  formula for the spectral index of the density perturbations produced during
  inflation}},\ }\href {https://doi.org/10.1143/PTP.95.71} {\bibfield
  {journal} {\bibinfo  {journal} {Prog. Theor. Phys.}\ }\textbf {\bibinfo
  {volume} {95}},\ \bibinfo {pages} {71} (\bibinfo {year} {1996})},\ \Eprint
  {https://arxiv.org/abs/astro-ph/9507001} {arXiv:astro-ph/9507001}
  \BibitemShut {NoStop}%
\bibitem [{\citenamefont {Wands}\ \emph {et~al.}(2000)\citenamefont {Wands},
  \citenamefont {Malik}, \citenamefont {Lyth},\ and\ \citenamefont
  {Liddle}}]{Wands:2000dp}%
  \BibitemOpen
  \bibfield  {author} {\bibinfo {author} {\bibfnamefont {D.}~\bibnamefont
  {Wands}}, \bibinfo {author} {\bibfnamefont {K.~A.}\ \bibnamefont {Malik}},
  \bibinfo {author} {\bibfnamefont {D.~H.}\ \bibnamefont {Lyth}},\ and\
  \bibinfo {author} {\bibfnamefont {A.~R.}\ \bibnamefont {Liddle}},\ }\bibfield
   {title} {\bibinfo {title} {{A New approach to the evolution of cosmological
  perturbations on large scales}},\ }\href
  {https://doi.org/10.1103/PhysRevD.62.043527} {\bibfield  {journal} {\bibinfo
  {journal} {Phys. Rev. D}\ }\textbf {\bibinfo {volume} {62}},\ \bibinfo
  {pages} {043527} (\bibinfo {year} {2000})},\ \Eprint
  {https://arxiv.org/abs/astro-ph/0003278} {arXiv:astro-ph/0003278}
  \BibitemShut {NoStop}%
\bibitem [{\citenamefont {Lyth}\ \emph {et~al.}(2005)\citenamefont {Lyth},
  \citenamefont {Malik},\ and\ \citenamefont {Sasaki}}]{Lyth:2004gb}%
  \BibitemOpen
  \bibfield  {author} {\bibinfo {author} {\bibfnamefont {D.~H.}\ \bibnamefont
  {Lyth}}, \bibinfo {author} {\bibfnamefont {K.~A.}\ \bibnamefont {Malik}},\
  and\ \bibinfo {author} {\bibfnamefont {M.}~\bibnamefont {Sasaki}},\
  }\bibfield  {title} {\bibinfo {title} {{A General proof of the conservation
  of the curvature perturbation}},\ }\href
  {https://doi.org/10.1088/1475-7516/2005/05/004} {\bibfield  {journal}
  {\bibinfo  {journal} {JCAP}\ }\textbf {\bibinfo {volume} {05}},\ \bibinfo
  {pages} {004}},\ \Eprint {https://arxiv.org/abs/astro-ph/0411220}
  {arXiv:astro-ph/0411220} \BibitemShut {NoStop}%
\bibitem [{\citenamefont {Abolhasani}\ \emph {et~al.}(2019)\citenamefont
  {Abolhasani}, \citenamefont {Firouzjahi}, \citenamefont {Naruko},\ and\
  \citenamefont {Sasaki}}]{Abolhasani:2019cqw}%
  \BibitemOpen
  \bibfield  {author} {\bibinfo {author} {\bibfnamefont {A.~A.}\ \bibnamefont
  {Abolhasani}}, \bibinfo {author} {\bibfnamefont {H.}~\bibnamefont
  {Firouzjahi}}, \bibinfo {author} {\bibfnamefont {A.}~\bibnamefont {Naruko}},\
  and\ \bibinfo {author} {\bibfnamefont {M.}~\bibnamefont {Sasaki}},\ }\href
  {https://doi.org/10.1142/10953} {\emph {\bibinfo {title} {{Delta N Formalism
  in Cosmological Perturbation Theory}}}}\ (\bibinfo  {publisher} {WSP},\
  \bibinfo {year} {2019})\BibitemShut {NoStop}%
\bibitem [{\citenamefont {Biagetti}\ \emph {et~al.}(2018)\citenamefont
  {Biagetti}, \citenamefont {Franciolini}, \citenamefont {Kehagias},\ and\
  \citenamefont {Riotto}}]{Biagetti:2018pjj}%
  \BibitemOpen
  \bibfield  {author} {\bibinfo {author} {\bibfnamefont {M.}~\bibnamefont
  {Biagetti}}, \bibinfo {author} {\bibfnamefont {G.}~\bibnamefont
  {Franciolini}}, \bibinfo {author} {\bibfnamefont {A.}~\bibnamefont
  {Kehagias}},\ and\ \bibinfo {author} {\bibfnamefont {A.}~\bibnamefont
  {Riotto}},\ }\bibfield  {title} {\bibinfo {title} {{Primordial Black Holes
  from Inflation and Quantum Diffusion}},\ }\href
  {https://doi.org/10.1088/1475-7516/2018/07/032} {\bibfield  {journal}
  {\bibinfo  {journal} {JCAP}\ }\textbf {\bibinfo {volume} {07}},\ \bibinfo
  {pages} {032}},\ \Eprint {https://arxiv.org/abs/1804.07124} {arXiv:1804.07124
  [astro-ph.CO]} \BibitemShut {NoStop}%
\bibitem [{\citenamefont {Atal}\ \emph {et~al.}(2019)\citenamefont {Atal},
  \citenamefont {Garriga},\ and\ \citenamefont
  {Marcos-Caballero}}]{Atal:2019cdz}%
  \BibitemOpen
  \bibfield  {author} {\bibinfo {author} {\bibfnamefont {V.}~\bibnamefont
  {Atal}}, \bibinfo {author} {\bibfnamefont {J.}~\bibnamefont {Garriga}},\ and\
  \bibinfo {author} {\bibfnamefont {A.}~\bibnamefont {Marcos-Caballero}},\
  }\bibfield  {title} {\bibinfo {title} {{Primordial black hole formation with
  non-Gaussian curvature perturbations}},\ }\href
  {https://doi.org/10.1088/1475-7516/2019/09/073} {\bibfield  {journal}
  {\bibinfo  {journal} {JCAP}\ }\textbf {\bibinfo {volume} {09}},\ \bibinfo
  {pages} {073}},\ \Eprint {https://arxiv.org/abs/1905.13202} {arXiv:1905.13202
  [astro-ph.CO]} \BibitemShut {NoStop}%
\bibitem [{\citenamefont {Ezquiaga}\ \emph {et~al.}(2020)\citenamefont
  {Ezquiaga}, \citenamefont {Garc\'\i{}a-Bellido},\ and\ \citenamefont
  {Vennin}}]{Ezquiaga:2019ftu}%
  \BibitemOpen
  \bibfield  {author} {\bibinfo {author} {\bibfnamefont {J.~M.}\ \bibnamefont
  {Ezquiaga}}, \bibinfo {author} {\bibfnamefont {J.}~\bibnamefont
  {Garc\'\i{}a-Bellido}},\ and\ \bibinfo {author} {\bibfnamefont
  {V.}~\bibnamefont {Vennin}},\ }\bibfield  {title} {\bibinfo {title} {{The
  exponential tail of inflationary fluctuations: consequences for primordial
  black holes}},\ }\href {https://doi.org/10.1088/1475-7516/2020/03/029}
  {\bibfield  {journal} {\bibinfo  {journal} {JCAP}\ }\textbf {\bibinfo
  {volume} {03}},\ \bibinfo {pages} {029}},\ \Eprint
  {https://arxiv.org/abs/1912.05399} {arXiv:1912.05399 [astro-ph.CO]}
  \BibitemShut {NoStop}%
\bibitem [{\citenamefont {Atal}\ \emph {et~al.}(2020)\citenamefont {Atal},
  \citenamefont {Cid}, \citenamefont {Escriv\`a},\ and\ \citenamefont
  {Garriga}}]{Atal:2019erb}%
  \BibitemOpen
  \bibfield  {author} {\bibinfo {author} {\bibfnamefont {V.}~\bibnamefont
  {Atal}}, \bibinfo {author} {\bibfnamefont {J.}~\bibnamefont {Cid}}, \bibinfo
  {author} {\bibfnamefont {A.}~\bibnamefont {Escriv\`a}},\ and\ \bibinfo
  {author} {\bibfnamefont {J.}~\bibnamefont {Garriga}},\ }\bibfield  {title}
  {\bibinfo {title} {{PBH in single field inflation: the effect of shape
  dispersion and non-Gaussianities}},\ }\href
  {https://doi.org/10.1088/1475-7516/2020/05/022} {\bibfield  {journal}
  {\bibinfo  {journal} {JCAP}\ }\textbf {\bibinfo {volume} {05}},\ \bibinfo
  {pages} {022}},\ \Eprint {https://arxiv.org/abs/1908.11357} {arXiv:1908.11357
  [astro-ph.CO]} \BibitemShut {NoStop}%
\bibitem [{\citenamefont {Pi}\ and\ \citenamefont {Sasaki}(2021)}]{Pi:2021dft}%
  \BibitemOpen
  \bibfield  {author} {\bibinfo {author} {\bibfnamefont {S.}~\bibnamefont
  {Pi}}\ and\ \bibinfo {author} {\bibfnamefont {M.}~\bibnamefont {Sasaki}},\
  }\bibfield  {title} {\bibinfo {title} {{Primordial Black Hole Formation in
  Non-Minimal Curvaton Scenario}},\ }\href@noop {} {\  (\bibinfo {year}
  {2021})},\ \Eprint {https://arxiv.org/abs/2112.12680} {arXiv:2112.12680
  [astro-ph.CO]} \BibitemShut {NoStop}%
\bibitem [{\citenamefont {Cai}\ \emph {et~al.}(2022)\citenamefont {Cai},
  \citenamefont {Ma}, \citenamefont {Sasaki}, \citenamefont {Wang},\ and\
  \citenamefont {Zhou}}]{Cai:2022erk}%
  \BibitemOpen
  \bibfield  {author} {\bibinfo {author} {\bibfnamefont {Y.-F.}\ \bibnamefont
  {Cai}}, \bibinfo {author} {\bibfnamefont {X.-H.}\ \bibnamefont {Ma}},
  \bibinfo {author} {\bibfnamefont {M.}~\bibnamefont {Sasaki}}, \bibinfo
  {author} {\bibfnamefont {D.-G.}\ \bibnamefont {Wang}},\ and\ \bibinfo
  {author} {\bibfnamefont {Z.}~\bibnamefont {Zhou}},\ }\bibfield  {title}
  {\bibinfo {title} {{Highly non-Gaussian tails and primordial black holes from
  single-field inflation}},\ }\href
  {https://doi.org/10.1088/1475-7516/2022/12/034} {\bibfield  {journal}
  {\bibinfo  {journal} {JCAP}\ }\textbf {\bibinfo {volume} {12}},\ \bibinfo
  {pages} {034}},\ \Eprint {https://arxiv.org/abs/2207.11910} {arXiv:2207.11910
  [astro-ph.CO]} \BibitemShut {NoStop}%
\bibitem [{\citenamefont {Abe}\ \emph {et~al.}(2023)\citenamefont {Abe},
  \citenamefont {Inui}, \citenamefont {Tada},\ and\ \citenamefont
  {Yokoyama}}]{Abe:2022xur}%
  \BibitemOpen
  \bibfield  {author} {\bibinfo {author} {\bibfnamefont {K.~T.}\ \bibnamefont
  {Abe}}, \bibinfo {author} {\bibfnamefont {R.}~\bibnamefont {Inui}}, \bibinfo
  {author} {\bibfnamefont {Y.}~\bibnamefont {Tada}},\ and\ \bibinfo {author}
  {\bibfnamefont {S.}~\bibnamefont {Yokoyama}},\ }\bibfield  {title} {\bibinfo
  {title} {{Primordial black holes and gravitational waves induced by
  exponential-tailed perturbations}},\ }\href
  {https://doi.org/10.1088/1475-7516/2023/05/044} {\bibfield  {journal}
  {\bibinfo  {journal} {JCAP}\ }\textbf {\bibinfo {volume} {05}},\ \bibinfo
  {pages} {044}},\ \Eprint {https://arxiv.org/abs/2209.13891} {arXiv:2209.13891
  [astro-ph.CO]} \BibitemShut {NoStop}%
\bibitem [{\citenamefont {Pi}\ and\ \citenamefont {Sasaki}(2023)}]{Pi:2022ysn}%
  \BibitemOpen
  \bibfield  {author} {\bibinfo {author} {\bibfnamefont {S.}~\bibnamefont
  {Pi}}\ and\ \bibinfo {author} {\bibfnamefont {M.}~\bibnamefont {Sasaki}},\
  }\bibfield  {title} {\bibinfo {title} {{Logarithmic Duality of the Curvature
  Perturbation}},\ }\href {https://doi.org/10.1103/PhysRevLett.131.011002}
  {\bibfield  {journal} {\bibinfo  {journal} {Phys. Rev. Lett.}\ }\textbf
  {\bibinfo {volume} {131}},\ \bibinfo {pages} {011002} (\bibinfo {year}
  {2023})},\ \Eprint {https://arxiv.org/abs/2211.13932} {arXiv:2211.13932
  [astro-ph.CO]} \BibitemShut {NoStop}%
\bibitem [{\citenamefont {Animali}\ and\ \citenamefont
  {Vennin}(2023)}]{Animali:2022otk}%
  \BibitemOpen
  \bibfield  {author} {\bibinfo {author} {\bibfnamefont {C.}~\bibnamefont
  {Animali}}\ and\ \bibinfo {author} {\bibfnamefont {V.}~\bibnamefont
  {Vennin}},\ }\bibfield  {title} {\bibinfo {title} {{Primordial black holes
  from stochastic tunnelling}},\ }\href
  {https://doi.org/10.1088/1475-7516/2023/02/043} {\bibfield  {journal}
  {\bibinfo  {journal} {JCAP}\ }\textbf {\bibinfo {volume} {02}},\ \bibinfo
  {pages} {043}},\ \Eprint {https://arxiv.org/abs/2210.03812} {arXiv:2210.03812
  [astro-ph.CO]} \BibitemShut {NoStop}%
\bibitem [{\citenamefont {Kitajima}\ \emph {et~al.}(2021)\citenamefont
  {Kitajima}, \citenamefont {Tada}, \citenamefont {Yokoyama},\ and\
  \citenamefont {Yoo}}]{Kitajima:2021fpq}%
  \BibitemOpen
  \bibfield  {author} {\bibinfo {author} {\bibfnamefont {N.}~\bibnamefont
  {Kitajima}}, \bibinfo {author} {\bibfnamefont {Y.}~\bibnamefont {Tada}},
  \bibinfo {author} {\bibfnamefont {S.}~\bibnamefont {Yokoyama}},\ and\
  \bibinfo {author} {\bibfnamefont {C.-M.}\ \bibnamefont {Yoo}},\ }\bibfield
  {title} {\bibinfo {title} {{Primordial black holes in peak theory with a
  non-Gaussian tail}},\ }\href {https://doi.org/10.1088/1475-7516/2021/10/053}
  {\bibfield  {journal} {\bibinfo  {journal} {JCAP}\ }\textbf {\bibinfo
  {volume} {10}},\ \bibinfo {pages} {053}},\ \Eprint
  {https://arxiv.org/abs/2109.00791} {arXiv:2109.00791 [astro-ph.CO]}
  \BibitemShut {NoStop}%
\bibitem [{\citenamefont {Starobinsky}(1992)}]{Starobinsky:1992ts}%
  \BibitemOpen
  \bibfield  {author} {\bibinfo {author} {\bibfnamefont {A.~A.}\ \bibnamefont
  {Starobinsky}},\ }\bibfield  {title} {\bibinfo {title} {{Spectrum of
  adiabatic perturbations in the universe when there are singularities in the
  inflation potential}},\ }\href@noop {} {\bibfield  {journal} {\bibinfo
  {journal} {JETP Lett.}\ }\textbf {\bibinfo {volume} {55}},\ \bibinfo {pages}
  {489} (\bibinfo {year} {1992})}\BibitemShut {NoStop}%
\bibitem [{\citenamefont {Ivanov}\ \emph {et~al.}(1994)\citenamefont {Ivanov},
  \citenamefont {Naselsky},\ and\ \citenamefont {Novikov}}]{Ivanov:1994pa}%
  \BibitemOpen
  \bibfield  {author} {\bibinfo {author} {\bibfnamefont {P.}~\bibnamefont
  {Ivanov}}, \bibinfo {author} {\bibfnamefont {P.}~\bibnamefont {Naselsky}},\
  and\ \bibinfo {author} {\bibfnamefont {I.}~\bibnamefont {Novikov}},\
  }\bibfield  {title} {\bibinfo {title} {{Inflation and primordial black holes
  as dark matter}},\ }\href {https://doi.org/10.1103/PhysRevD.50.7173}
  {\bibfield  {journal} {\bibinfo  {journal} {Phys. Rev. D}\ }\textbf {\bibinfo
  {volume} {50}},\ \bibinfo {pages} {7173} (\bibinfo {year}
  {1994})}\BibitemShut {NoStop}%
\bibitem [{\citenamefont {Suyama}\ \emph {et~al.}(2021)\citenamefont {Suyama},
  \citenamefont {Tada},\ and\ \citenamefont {Yamaguchi}}]{Suyama:2021adn}%
  \BibitemOpen
  \bibfield  {author} {\bibinfo {author} {\bibfnamefont {T.}~\bibnamefont
  {Suyama}}, \bibinfo {author} {\bibfnamefont {Y.}~\bibnamefont {Tada}},\ and\
  \bibinfo {author} {\bibfnamefont {M.}~\bibnamefont {Yamaguchi}},\ }\bibfield
  {title} {\bibinfo {title} {{Revisiting non-Gaussianity in non-attractor
  inflation models in the light of the cosmological soft theorem}},\ }\href
  {https://doi.org/10.1093/ptep/ptab063} {\bibfield  {journal} {\bibinfo
  {journal} {PTEP}\ }\textbf {\bibinfo {volume} {2021}},\ \bibinfo {pages}
  {073E02} (\bibinfo {year} {2021})},\ \Eprint
  {https://arxiv.org/abs/2101.10682} {arXiv:2101.10682 [hep-th]} \BibitemShut
  {NoStop}%
\bibitem [{\citenamefont {Pi}\ and\ \citenamefont {Wang}(2023)}]{Pi:2022zxs}%
  \BibitemOpen
  \bibfield  {author} {\bibinfo {author} {\bibfnamefont {S.}~\bibnamefont
  {Pi}}\ and\ \bibinfo {author} {\bibfnamefont {J.}~\bibnamefont {Wang}},\
  }\bibfield  {title} {\bibinfo {title} {{Primordial black hole formation in
  Starobinsky's linear potential model}},\ }\href
  {https://doi.org/10.1088/1475-7516/2023/06/018} {\bibfield  {journal}
  {\bibinfo  {journal} {JCAP}\ }\textbf {\bibinfo {volume} {06}},\ \bibinfo
  {pages} {018}},\ \Eprint {https://arxiv.org/abs/2209.14183} {arXiv:2209.14183
  [astro-ph.CO]} \BibitemShut {NoStop}%
\bibitem [{\citenamefont {Martin}\ \emph {et~al.}(2013)\citenamefont {Martin},
  \citenamefont {Motohashi},\ and\ \citenamefont {Suyama}}]{Martin:2012pe}%
  \BibitemOpen
  \bibfield  {author} {\bibinfo {author} {\bibfnamefont {J.}~\bibnamefont
  {Martin}}, \bibinfo {author} {\bibfnamefont {H.}~\bibnamefont {Motohashi}},\
  and\ \bibinfo {author} {\bibfnamefont {T.}~\bibnamefont {Suyama}},\
  }\bibfield  {title} {\bibinfo {title} {{Ultra Slow-Roll Inflation and the
  non-Gaussianity Consistency Relation}},\ }\href
  {https://doi.org/10.1103/PhysRevD.87.023514} {\bibfield  {journal} {\bibinfo
  {journal} {Phys. Rev. D}\ }\textbf {\bibinfo {volume} {87}},\ \bibinfo
  {pages} {023514} (\bibinfo {year} {2013})},\ \Eprint
  {https://arxiv.org/abs/1211.0083} {arXiv:1211.0083 [astro-ph.CO]}
  \BibitemShut {NoStop}%
\bibitem [{\citenamefont {Motohashi}\ \emph {et~al.}(2015)\citenamefont
  {Motohashi}, \citenamefont {Starobinsky},\ and\ \citenamefont
  {Yokoyama}}]{Motohashi:2014ppa}%
  \BibitemOpen
  \bibfield  {author} {\bibinfo {author} {\bibfnamefont {H.}~\bibnamefont
  {Motohashi}}, \bibinfo {author} {\bibfnamefont {A.~A.}\ \bibnamefont
  {Starobinsky}},\ and\ \bibinfo {author} {\bibfnamefont {J.}~\bibnamefont
  {Yokoyama}},\ }\bibfield  {title} {\bibinfo {title} {{Inflation with a
  constant rate of roll}},\ }\href
  {https://doi.org/10.1088/1475-7516/2015/09/018} {\bibfield  {journal}
  {\bibinfo  {journal} {JCAP}\ }\textbf {\bibinfo {volume} {09}},\ \bibinfo
  {pages} {018}},\ \Eprint {https://arxiv.org/abs/1411.5021} {arXiv:1411.5021
  [astro-ph.CO]} \BibitemShut {NoStop}%
\bibitem [{\citenamefont {Motohashi}\ \emph {et~al.}(2020)\citenamefont
  {Motohashi}, \citenamefont {Mukohyama},\ and\ \citenamefont
  {Oliosi}}]{Motohashi:2019rhu}%
  \BibitemOpen
  \bibfield  {author} {\bibinfo {author} {\bibfnamefont {H.}~\bibnamefont
  {Motohashi}}, \bibinfo {author} {\bibfnamefont {S.}~\bibnamefont
  {Mukohyama}},\ and\ \bibinfo {author} {\bibfnamefont {M.}~\bibnamefont
  {Oliosi}},\ }\bibfield  {title} {\bibinfo {title} {{Constant Roll and
  Primordial Black Holes}},\ }\href
  {https://doi.org/10.1088/1475-7516/2020/03/002} {\bibfield  {journal}
  {\bibinfo  {journal} {JCAP}\ }\textbf {\bibinfo {volume} {03}},\ \bibinfo
  {pages} {002}},\ \Eprint {https://arxiv.org/abs/1910.13235} {arXiv:1910.13235
  [gr-qc]} \BibitemShut {NoStop}%
\bibitem [{\citenamefont {Chataignier}\ \emph {et~al.}(2023)\citenamefont
  {Chataignier}, \citenamefont {Kamenshchik}, \citenamefont {Tronconi},\ and\
  \citenamefont {Venturi}}]{Chataignier:2023ago}%
  \BibitemOpen
  \bibfield  {author} {\bibinfo {author} {\bibfnamefont {L.}~\bibnamefont
  {Chataignier}}, \bibinfo {author} {\bibfnamefont {A.~Y.}\ \bibnamefont
  {Kamenshchik}}, \bibinfo {author} {\bibfnamefont {A.}~\bibnamefont
  {Tronconi}},\ and\ \bibinfo {author} {\bibfnamefont {G.}~\bibnamefont
  {Venturi}},\ }\bibfield  {title} {\bibinfo {title} {{Reconstruction methods
  and the amplification of the inflationary spectrum}},\ }\href
  {https://doi.org/10.1103/PhysRevD.107.083506} {\bibfield  {journal} {\bibinfo
   {journal} {Phys. Rev. D}\ }\textbf {\bibinfo {volume} {107}},\ \bibinfo
  {pages} {083506} (\bibinfo {year} {2023})},\ \Eprint
  {https://arxiv.org/abs/2301.04477} {arXiv:2301.04477 [gr-qc]} \BibitemShut
  {NoStop}%
\bibitem [{\citenamefont {Lucchin}\ and\ \citenamefont
  {Matarrese}(1985)}]{Lucchin:1984yf}%
  \BibitemOpen
  \bibfield  {author} {\bibinfo {author} {\bibfnamefont {F.}~\bibnamefont
  {Lucchin}}\ and\ \bibinfo {author} {\bibfnamefont {S.}~\bibnamefont
  {Matarrese}},\ }\bibfield  {title} {\bibinfo {title} {{Power Law
  Inflation}},\ }\href {https://doi.org/10.1103/PhysRevD.32.1316} {\bibfield
  {journal} {\bibinfo  {journal} {Phys. Rev. D}\ }\textbf {\bibinfo {volume}
  {32}},\ \bibinfo {pages} {1316} (\bibinfo {year} {1985})}\BibitemShut
  {NoStop}%
\bibitem [{\citenamefont {Russo}(2004)}]{Russo:2004ym}%
  \BibitemOpen
  \bibfield  {author} {\bibinfo {author} {\bibfnamefont {J.~G.}\ \bibnamefont
  {Russo}},\ }\bibfield  {title} {\bibinfo {title} {{Exact solution of scalar
  tensor cosmology with exponential potentials and transient acceleration}},\
  }\href {https://doi.org/10.1016/j.physletb.2004.09.007} {\bibfield  {journal}
  {\bibinfo  {journal} {Phys. Lett. B}\ }\textbf {\bibinfo {volume} {600}},\
  \bibinfo {pages} {185} (\bibinfo {year} {2004})},\ \Eprint
  {https://arxiv.org/abs/hep-th/0403010} {arXiv:hep-th/0403010} \BibitemShut
  {NoStop}%
\bibitem [{\citenamefont {Andrianov}\ \emph {et~al.}(2011)\citenamefont
  {Andrianov}, \citenamefont {Cannata},\ and\ \citenamefont
  {Kamenshchik}}]{Andrianov:2011fg}%
  \BibitemOpen
  \bibfield  {author} {\bibinfo {author} {\bibfnamefont {A.~A.}\ \bibnamefont
  {Andrianov}}, \bibinfo {author} {\bibfnamefont {F.}~\bibnamefont {Cannata}},\
  and\ \bibinfo {author} {\bibfnamefont {A.~Y.}\ \bibnamefont {Kamenshchik}},\
  }\bibfield  {title} {\bibinfo {title} {{General solution of scalar field
  cosmology with a (piecewise) exponential potential}},\ }\href
  {https://doi.org/10.1088/1475-7516/2011/10/004} {\bibfield  {journal}
  {\bibinfo  {journal} {JCAP}\ }\textbf {\bibinfo {volume} {10}},\ \bibinfo
  {pages} {004}},\ \Eprint {https://arxiv.org/abs/1105.4515} {arXiv:1105.4515
  [gr-qc]} \BibitemShut {NoStop}%
\bibitem [{\citenamefont {Lyth}\ and\ \citenamefont
  {Stewart}(1992)}]{Lyth:1991bc}%
  \BibitemOpen
  \bibfield  {author} {\bibinfo {author} {\bibfnamefont {D.~H.}\ \bibnamefont
  {Lyth}}\ and\ \bibinfo {author} {\bibfnamefont {E.~D.}\ \bibnamefont
  {Stewart}},\ }\bibfield  {title} {\bibinfo {title} {{The Curvature
  perturbation in power law (e.g. extended) inflation}},\ }\href
  {https://doi.org/10.1016/0370-2693(92)90518-9} {\bibfield  {journal}
  {\bibinfo  {journal} {Phys. Lett. B}\ }\textbf {\bibinfo {volume} {274}},\
  \bibinfo {pages} {168} (\bibinfo {year} {1992})}\BibitemShut {NoStop}%
\bibitem [{\citenamefont {Dudas}\ \emph {et~al.}(2011)\citenamefont {Dudas},
  \citenamefont {Kitazawa},\ and\ \citenamefont {Sagnotti}}]{Dudas:2010gi}%
  \BibitemOpen
  \bibfield  {author} {\bibinfo {author} {\bibfnamefont {E.}~\bibnamefont
  {Dudas}}, \bibinfo {author} {\bibfnamefont {N.}~\bibnamefont {Kitazawa}},\
  and\ \bibinfo {author} {\bibfnamefont {A.}~\bibnamefont {Sagnotti}},\
  }\bibfield  {title} {\bibinfo {title} {{On Climbing Scalars in String
  Theory}},\ }\href {https://doi.org/10.1016/j.physletb.2010.09.040} {\bibfield
   {journal} {\bibinfo  {journal} {Phys. Lett. B}\ }\textbf {\bibinfo {volume}
  {694}},\ \bibinfo {pages} {80} (\bibinfo {year} {2011})},\ \Eprint
  {https://arxiv.org/abs/1009.0874} {arXiv:1009.0874 [hep-th]} \BibitemShut
  {NoStop}%
\bibitem [{\citenamefont {Vanzan}(2021)}]{Vanzan_2023}%
  \BibitemOpen
  \bibfield  {author} {\bibinfo {author} {\bibfnamefont {E.}~\bibnamefont
  {Vanzan}},\ }\emph {\bibinfo {title} {Non-gaussian signatures in a
  string-inspired model for the onset of inflation.}},\ \href
  {https://hdl.handle.net/20.500.12608/21759} {Master's thesis} (\bibinfo
  {year} {2021})\BibitemShut {NoStop}%
\bibitem [{\citenamefont {Bellomo}\ \emph {et~al.}(2018)\citenamefont
  {Bellomo}, \citenamefont {Bernal}, \citenamefont {Raccanelli},\ and\
  \citenamefont {Verde}}]{Bellomo:2017zsr}%
  \BibitemOpen
  \bibfield  {author} {\bibinfo {author} {\bibfnamefont {N.}~\bibnamefont
  {Bellomo}}, \bibinfo {author} {\bibfnamefont {J.~L.}\ \bibnamefont {Bernal}},
  \bibinfo {author} {\bibfnamefont {A.}~\bibnamefont {Raccanelli}},\ and\
  \bibinfo {author} {\bibfnamefont {L.}~\bibnamefont {Verde}},\ }\bibfield
  {title} {\bibinfo {title} {{Primordial Black Holes as Dark Matter: Converting
  Constraints from Monochromatic to Extended Mass Distributions}},\ }\href
  {https://doi.org/10.1088/1475-7516/2018/01/004} {\bibfield  {journal}
  {\bibinfo  {journal} {JCAP}\ }\textbf {\bibinfo {volume} {01}},\ \bibinfo
  {pages} {004}},\ \Eprint {https://arxiv.org/abs/1709.07467} {arXiv:1709.07467
  [astro-ph.CO]} \BibitemShut {NoStop}%
\bibitem [{\citenamefont {Carr}\ \emph {et~al.}(2017)\citenamefont {Carr},
  \citenamefont {Raidal}, \citenamefont {Tenkanen}, \citenamefont {Vaskonen},\
  and\ \citenamefont {Veerm\"ae}}]{Carr:2017jsz}%
  \BibitemOpen
  \bibfield  {author} {\bibinfo {author} {\bibfnamefont {B.}~\bibnamefont
  {Carr}}, \bibinfo {author} {\bibfnamefont {M.}~\bibnamefont {Raidal}},
  \bibinfo {author} {\bibfnamefont {T.}~\bibnamefont {Tenkanen}}, \bibinfo
  {author} {\bibfnamefont {V.}~\bibnamefont {Vaskonen}},\ and\ \bibinfo
  {author} {\bibfnamefont {H.}~\bibnamefont {Veerm\"ae}},\ }\bibfield  {title}
  {\bibinfo {title} {{Primordial black hole constraints for extended mass
  functions}},\ }\href {https://doi.org/10.1103/PhysRevD.96.023514} {\bibfield
  {journal} {\bibinfo  {journal} {Phys. Rev. D}\ }\textbf {\bibinfo {volume}
  {96}},\ \bibinfo {pages} {023514} (\bibinfo {year} {2017})},\ \Eprint
  {https://arxiv.org/abs/1705.05567} {arXiv:1705.05567 [astro-ph.CO]}
  \BibitemShut {NoStop}%
\bibitem [{\citenamefont {Bartolo}\ \emph
  {et~al.}(2019{\natexlab{a}})\citenamefont {Bartolo}, \citenamefont {De~Luca},
  \citenamefont {Franciolini}, \citenamefont {Peloso}, \citenamefont {Racco},\
  and\ \citenamefont {Riotto}}]{Bartolo:2018rku}%
  \BibitemOpen
  \bibfield  {author} {\bibinfo {author} {\bibfnamefont {N.}~\bibnamefont
  {Bartolo}}, \bibinfo {author} {\bibfnamefont {V.}~\bibnamefont {De~Luca}},
  \bibinfo {author} {\bibfnamefont {G.}~\bibnamefont {Franciolini}}, \bibinfo
  {author} {\bibfnamefont {M.}~\bibnamefont {Peloso}}, \bibinfo {author}
  {\bibfnamefont {D.}~\bibnamefont {Racco}},\ and\ \bibinfo {author}
  {\bibfnamefont {A.}~\bibnamefont {Riotto}},\ }\bibfield  {title} {\bibinfo
  {title} {{Testing primordial black holes as dark matter with LISA}},\ }\href
  {https://doi.org/10.1103/PhysRevD.99.103521} {\bibfield  {journal} {\bibinfo
  {journal} {Phys. Rev. D}\ }\textbf {\bibinfo {volume} {99}},\ \bibinfo
  {pages} {103521} (\bibinfo {year} {2019}{\natexlab{a}})},\ \Eprint
  {https://arxiv.org/abs/1810.12224} {arXiv:1810.12224 [astro-ph.CO]}
  \BibitemShut {NoStop}%
\bibitem [{\citenamefont {Bartolo}\ \emph
  {et~al.}(2019{\natexlab{b}})\citenamefont {Bartolo}, \citenamefont {De~Luca},
  \citenamefont {Franciolini}, \citenamefont {Lewis}, \citenamefont {Peloso},\
  and\ \citenamefont {Riotto}}]{Bartolo:2018evs}%
  \BibitemOpen
  \bibfield  {author} {\bibinfo {author} {\bibfnamefont {N.}~\bibnamefont
  {Bartolo}}, \bibinfo {author} {\bibfnamefont {V.}~\bibnamefont {De~Luca}},
  \bibinfo {author} {\bibfnamefont {G.}~\bibnamefont {Franciolini}}, \bibinfo
  {author} {\bibfnamefont {A.}~\bibnamefont {Lewis}}, \bibinfo {author}
  {\bibfnamefont {M.}~\bibnamefont {Peloso}},\ and\ \bibinfo {author}
  {\bibfnamefont {A.}~\bibnamefont {Riotto}},\ }\bibfield  {title} {\bibinfo
  {title} {{Primordial Black Hole Dark Matter: LISA Serendipity}},\ }\href
  {https://doi.org/10.1103/PhysRevLett.122.211301} {\bibfield  {journal}
  {\bibinfo  {journal} {Phys. Rev. Lett.}\ }\textbf {\bibinfo {volume} {122}},\
  \bibinfo {pages} {211301} (\bibinfo {year} {2019}{\natexlab{b}})},\ \Eprint
  {https://arxiv.org/abs/1810.12218} {arXiv:1810.12218 [astro-ph.CO]}
  \BibitemShut {NoStop}%
\bibitem [{\citenamefont {Carr}\ and\ \citenamefont
  {Kuhnel}(2020)}]{Carr:2020xqk}%
  \BibitemOpen
  \bibfield  {author} {\bibinfo {author} {\bibfnamefont {B.}~\bibnamefont
  {Carr}}\ and\ \bibinfo {author} {\bibfnamefont {F.}~\bibnamefont {Kuhnel}},\
  }\bibfield  {title} {\bibinfo {title} {{Primordial Black Holes as Dark
  Matter: Recent Developments}},\ }\href
  {https://doi.org/10.1146/annurev-nucl-050520-125911} {\bibfield  {journal}
  {\bibinfo  {journal} {Ann. Rev. Nucl. Part. Sci.}\ }\textbf {\bibinfo
  {volume} {70}},\ \bibinfo {pages} {355} (\bibinfo {year} {2020})},\ \Eprint
  {https://arxiv.org/abs/2006.02838} {arXiv:2006.02838 [astro-ph.CO]}
  \BibitemShut {NoStop}%
\bibitem [{\citenamefont {Bird}\ \emph {et~al.}(2016)\citenamefont {Bird},
  \citenamefont {Cholis}, \citenamefont {Mu\~noz}, \citenamefont
  {Ali-Ha\"\i{}moud}, \citenamefont {Kamionkowski}, \citenamefont {Kovetz},
  \citenamefont {Raccanelli},\ and\ \citenamefont {Riess}}]{Bird:2016dcv}%
  \BibitemOpen
  \bibfield  {author} {\bibinfo {author} {\bibfnamefont {S.}~\bibnamefont
  {Bird}}, \bibinfo {author} {\bibfnamefont {I.}~\bibnamefont {Cholis}},
  \bibinfo {author} {\bibfnamefont {J.~B.}\ \bibnamefont {Mu\~noz}}, \bibinfo
  {author} {\bibfnamefont {Y.}~\bibnamefont {Ali-Ha\"\i{}moud}}, \bibinfo
  {author} {\bibfnamefont {M.}~\bibnamefont {Kamionkowski}}, \bibinfo {author}
  {\bibfnamefont {E.~D.}\ \bibnamefont {Kovetz}}, \bibinfo {author}
  {\bibfnamefont {A.}~\bibnamefont {Raccanelli}},\ and\ \bibinfo {author}
  {\bibfnamefont {A.~G.}\ \bibnamefont {Riess}},\ }\bibfield  {title} {\bibinfo
  {title} {{Did LIGO detect dark matter?}},\ }\href
  {https://doi.org/10.1103/PhysRevLett.116.201301} {\bibfield  {journal}
  {\bibinfo  {journal} {Phys. Rev. Lett.}\ }\textbf {\bibinfo {volume} {116}},\
  \bibinfo {pages} {201301} (\bibinfo {year} {2016})},\ \Eprint
  {https://arxiv.org/abs/1603.00464} {arXiv:1603.00464 [astro-ph.CO]}
  \BibitemShut {NoStop}%
\bibitem [{\citenamefont {Sasaki}\ \emph {et~al.}(2016)\citenamefont {Sasaki},
  \citenamefont {Suyama}, \citenamefont {Tanaka},\ and\ \citenamefont
  {Yokoyama}}]{Sasaki:2016jop}%
  \BibitemOpen
  \bibfield  {author} {\bibinfo {author} {\bibfnamefont {M.}~\bibnamefont
  {Sasaki}}, \bibinfo {author} {\bibfnamefont {T.}~\bibnamefont {Suyama}},
  \bibinfo {author} {\bibfnamefont {T.}~\bibnamefont {Tanaka}},\ and\ \bibinfo
  {author} {\bibfnamefont {S.}~\bibnamefont {Yokoyama}},\ }\bibfield  {title}
  {\bibinfo {title} {{Primordial Black Hole Scenario for the Gravitational-Wave
  Event GW150914}},\ }\href {https://doi.org/10.1103/PhysRevLett.117.061101}
  {\bibfield  {journal} {\bibinfo  {journal} {Phys. Rev. Lett.}\ }\textbf
  {\bibinfo {volume} {117}},\ \bibinfo {pages} {061101} (\bibinfo {year}
  {2016})},\ \bibinfo {note} {[Erratum: Phys.Rev.Lett. 121, 059901 (2018)]},\
  \Eprint {https://arxiv.org/abs/1603.08338} {arXiv:1603.08338 [astro-ph.CO]}
  \BibitemShut {NoStop}%
\bibitem [{\citenamefont {Wong}\ \emph {et~al.}(2021)\citenamefont {Wong},
  \citenamefont {Franciolini}, \citenamefont {De~Luca}, \citenamefont
  {Baibhav}, \citenamefont {Berti}, \citenamefont {Pani},\ and\ \citenamefont
  {Riotto}}]{Wong:2020yig}%
  \BibitemOpen
  \bibfield  {author} {\bibinfo {author} {\bibfnamefont {K.~W.~K.}\
  \bibnamefont {Wong}}, \bibinfo {author} {\bibfnamefont {G.}~\bibnamefont
  {Franciolini}}, \bibinfo {author} {\bibfnamefont {V.}~\bibnamefont
  {De~Luca}}, \bibinfo {author} {\bibfnamefont {V.}~\bibnamefont {Baibhav}},
  \bibinfo {author} {\bibfnamefont {E.}~\bibnamefont {Berti}}, \bibinfo
  {author} {\bibfnamefont {P.}~\bibnamefont {Pani}},\ and\ \bibinfo {author}
  {\bibfnamefont {A.}~\bibnamefont {Riotto}},\ }\bibfield  {title} {\bibinfo
  {title} {{Constraining the primordial black hole scenario with Bayesian
  inference and machine learning: the GWTC-2 gravitational wave catalog}},\
  }\href {https://doi.org/10.1103/PhysRevD.103.023026} {\bibfield  {journal}
  {\bibinfo  {journal} {Phys. Rev. D}\ }\textbf {\bibinfo {volume} {103}},\
  \bibinfo {pages} {023026} (\bibinfo {year} {2021})},\ \Eprint
  {https://arxiv.org/abs/2011.01865} {arXiv:2011.01865 [gr-qc]} \BibitemShut
  {NoStop}%
\bibitem [{\citenamefont {Franciolini}\ \emph {et~al.}(2022)\citenamefont
  {Franciolini}, \citenamefont {Baibhav}, \citenamefont {De~Luca},
  \citenamefont {Ng}, \citenamefont {Wong}, \citenamefont {Berti},
  \citenamefont {Pani}, \citenamefont {Riotto},\ and\ \citenamefont
  {Vitale}}]{Franciolini:2021tla}%
  \BibitemOpen
  \bibfield  {author} {\bibinfo {author} {\bibfnamefont {G.}~\bibnamefont
  {Franciolini}}, \bibinfo {author} {\bibfnamefont {V.}~\bibnamefont
  {Baibhav}}, \bibinfo {author} {\bibfnamefont {V.}~\bibnamefont {De~Luca}},
  \bibinfo {author} {\bibfnamefont {K.~K.~Y.}\ \bibnamefont {Ng}}, \bibinfo
  {author} {\bibfnamefont {K.~W.~K.}\ \bibnamefont {Wong}}, \bibinfo {author}
  {\bibfnamefont {E.}~\bibnamefont {Berti}}, \bibinfo {author} {\bibfnamefont
  {P.}~\bibnamefont {Pani}}, \bibinfo {author} {\bibfnamefont {A.}~\bibnamefont
  {Riotto}},\ and\ \bibinfo {author} {\bibfnamefont {S.}~\bibnamefont
  {Vitale}},\ }\bibfield  {title} {\bibinfo {title} {{Searching for a
  subpopulation of primordial black holes in LIGO-Virgo gravitational-wave
  data}},\ }\href {https://doi.org/10.1103/PhysRevD.105.083526} {\bibfield
  {journal} {\bibinfo  {journal} {Phys. Rev. D}\ }\textbf {\bibinfo {volume}
  {105}},\ \bibinfo {pages} {083526} (\bibinfo {year} {2022})},\ \Eprint
  {https://arxiv.org/abs/2105.03349} {arXiv:2105.03349 [gr-qc]} \BibitemShut
  {NoStop}%
\bibitem [{\citenamefont {Kawasaki}\ \emph {et~al.}(2012)\citenamefont
  {Kawasaki}, \citenamefont {Kusenko},\ and\ \citenamefont
  {Yanagida}}]{Kawasaki:2012kn}%
  \BibitemOpen
  \bibfield  {author} {\bibinfo {author} {\bibfnamefont {M.}~\bibnamefont
  {Kawasaki}}, \bibinfo {author} {\bibfnamefont {A.}~\bibnamefont {Kusenko}},\
  and\ \bibinfo {author} {\bibfnamefont {T.~T.}\ \bibnamefont {Yanagida}},\
  }\bibfield  {title} {\bibinfo {title} {{Primordial seeds of supermassive
  black holes}},\ }\href {https://doi.org/10.1016/j.physletb.2012.03.056}
  {\bibfield  {journal} {\bibinfo  {journal} {Phys. Lett. B}\ }\textbf
  {\bibinfo {volume} {711}},\ \bibinfo {pages} {1} (\bibinfo {year} {2012})},\
  \Eprint {https://arxiv.org/abs/1202.3848} {arXiv:1202.3848 [astro-ph.CO]}
  \BibitemShut {NoStop}%
\bibitem [{\citenamefont {Carr}\ and\ \citenamefont
  {Silk}(2018)}]{Carr:2018rid}%
  \BibitemOpen
  \bibfield  {author} {\bibinfo {author} {\bibfnamefont {B.}~\bibnamefont
  {Carr}}\ and\ \bibinfo {author} {\bibfnamefont {J.}~\bibnamefont {Silk}},\
  }\bibfield  {title} {\bibinfo {title} {{Primordial Black Holes as Generators
  of Cosmic Structures}},\ }\href {https://doi.org/10.1093/mnras/sty1204}
  {\bibfield  {journal} {\bibinfo  {journal} {Mon. Not. Roy. Astron. Soc.}\
  }\textbf {\bibinfo {volume} {478}},\ \bibinfo {pages} {3756} (\bibinfo {year}
  {2018})},\ \Eprint {https://arxiv.org/abs/1801.00672} {arXiv:1801.00672
  [astro-ph.CO]} \BibitemShut {NoStop}%
\bibitem [{\citenamefont {Crawford}\ and\ \citenamefont
  {Schramm}(1982)}]{Crawford:1982yz}%
  \BibitemOpen
  \bibfield  {author} {\bibinfo {author} {\bibfnamefont {M.}~\bibnamefont
  {Crawford}}\ and\ \bibinfo {author} {\bibfnamefont {D.~N.}\ \bibnamefont
  {Schramm}},\ }\bibfield  {title} {\bibinfo {title} {{Spontaneous Generation
  of Density Perturbations in the Early Universe}},\ }\href
  {https://doi.org/10.1038/298538a0} {\bibfield  {journal} {\bibinfo  {journal}
  {Nature}\ }\textbf {\bibinfo {volume} {298}},\ \bibinfo {pages} {538}
  (\bibinfo {year} {1982})}\BibitemShut {NoStop}%
\bibitem [{\citenamefont {Kodama}\ \emph {et~al.}(1982)\citenamefont {Kodama},
  \citenamefont {Sasaki},\ and\ \citenamefont {Sato}}]{Kodama:1982sf}%
  \BibitemOpen
  \bibfield  {author} {\bibinfo {author} {\bibfnamefont {H.}~\bibnamefont
  {Kodama}}, \bibinfo {author} {\bibfnamefont {M.}~\bibnamefont {Sasaki}},\
  and\ \bibinfo {author} {\bibfnamefont {K.}~\bibnamefont {Sato}},\ }\bibfield
  {title} {\bibinfo {title} {{Abundance of Primordial Holes Produced by
  Cosmological First Order Phase Transition}},\ }\href
  {https://doi.org/10.1143/PTP.68.1979} {\bibfield  {journal} {\bibinfo
  {journal} {Prog. Theor. Phys.}\ }\textbf {\bibinfo {volume} {68}},\ \bibinfo
  {pages} {1979} (\bibinfo {year} {1982})}\BibitemShut {NoStop}%
\bibitem [{\citenamefont {Cotner}\ and\ \citenamefont
  {Kusenko}(2017)}]{Cotner:2016cvr}%
  \BibitemOpen
  \bibfield  {author} {\bibinfo {author} {\bibfnamefont {E.}~\bibnamefont
  {Cotner}}\ and\ \bibinfo {author} {\bibfnamefont {A.}~\bibnamefont
  {Kusenko}},\ }\bibfield  {title} {\bibinfo {title} {{Primordial black holes
  from supersymmetry in the early universe}},\ }\href
  {https://doi.org/10.1103/PhysRevLett.119.031103} {\bibfield  {journal}
  {\bibinfo  {journal} {Phys. Rev. Lett.}\ }\textbf {\bibinfo {volume} {119}},\
  \bibinfo {pages} {031103} (\bibinfo {year} {2017})},\ \Eprint
  {https://arxiv.org/abs/1612.02529} {arXiv:1612.02529 [astro-ph.CO]}
  \BibitemShut {NoStop}%
\bibitem [{\citenamefont {Cotner}\ \emph {et~al.}(2019)\citenamefont {Cotner},
  \citenamefont {Kusenko}, \citenamefont {Sasaki},\ and\ \citenamefont
  {Takhistov}}]{Cotner:2019ykd}%
  \BibitemOpen
  \bibfield  {author} {\bibinfo {author} {\bibfnamefont {E.}~\bibnamefont
  {Cotner}}, \bibinfo {author} {\bibfnamefont {A.}~\bibnamefont {Kusenko}},
  \bibinfo {author} {\bibfnamefont {M.}~\bibnamefont {Sasaki}},\ and\ \bibinfo
  {author} {\bibfnamefont {V.}~\bibnamefont {Takhistov}},\ }\bibfield  {title}
  {\bibinfo {title} {{Analytic Description of Primordial Black Hole Formation
  from Scalar Field Fragmentation}},\ }\href
  {https://doi.org/10.1088/1475-7516/2019/10/077} {\bibfield  {journal}
  {\bibinfo  {journal} {JCAP}\ }\textbf {\bibinfo {volume} {10}},\ \bibinfo
  {pages} {077}},\ \Eprint {https://arxiv.org/abs/1907.10613} {arXiv:1907.10613
  [astro-ph.CO]} \BibitemShut {NoStop}%
\bibitem [{\citenamefont {Flores}\ and\ \citenamefont
  {Kusenko}(2023)}]{Flores:2021jas}%
  \BibitemOpen
  \bibfield  {author} {\bibinfo {author} {\bibfnamefont {M.~M.}\ \bibnamefont
  {Flores}}\ and\ \bibinfo {author} {\bibfnamefont {A.}~\bibnamefont
  {Kusenko}},\ }\bibfield  {title} {\bibinfo {title} {{Primordial black holes
  as a dark matter candidate in theories with supersymmetry and inflation}},\
  }\href {https://doi.org/10.1088/1475-7516/2023/05/013} {\bibfield  {journal}
  {\bibinfo  {journal} {JCAP}\ }\textbf {\bibinfo {volume} {05}},\ \bibinfo
  {pages} {013}},\ \Eprint {https://arxiv.org/abs/2108.08416} {arXiv:2108.08416
  [hep-ph]} \BibitemShut {NoStop}%
\bibitem [{\citenamefont {Amendola}\ \emph {et~al.}(2018)\citenamefont
  {Amendola}, \citenamefont {Rubio},\ and\ \citenamefont
  {Wetterich}}]{Amendola:2017xhl}%
  \BibitemOpen
  \bibfield  {author} {\bibinfo {author} {\bibfnamefont {L.}~\bibnamefont
  {Amendola}}, \bibinfo {author} {\bibfnamefont {J.}~\bibnamefont {Rubio}},\
  and\ \bibinfo {author} {\bibfnamefont {C.}~\bibnamefont {Wetterich}},\
  }\bibfield  {title} {\bibinfo {title} {{Primordial black holes from fifth
  forces}},\ }\href {https://doi.org/10.1103/PhysRevD.97.081302} {\bibfield
  {journal} {\bibinfo  {journal} {Phys. Rev. D}\ }\textbf {\bibinfo {volume}
  {97}},\ \bibinfo {pages} {081302} (\bibinfo {year} {2018})},\ \Eprint
  {https://arxiv.org/abs/1711.09915} {arXiv:1711.09915 [astro-ph.CO]}
  \BibitemShut {NoStop}%
\bibitem [{\citenamefont {Flores}\ and\ \citenamefont
  {Kusenko}(2021)}]{Flores:2020drq}%
  \BibitemOpen
  \bibfield  {author} {\bibinfo {author} {\bibfnamefont {M.~M.}\ \bibnamefont
  {Flores}}\ and\ \bibinfo {author} {\bibfnamefont {A.}~\bibnamefont
  {Kusenko}},\ }\bibfield  {title} {\bibinfo {title} {{Primordial Black Holes
  from Long-Range Scalar Forces and Scalar Radiative Cooling}},\ }\href
  {https://doi.org/10.1103/PhysRevLett.126.041101} {\bibfield  {journal}
  {\bibinfo  {journal} {Phys. Rev. Lett.}\ }\textbf {\bibinfo {volume} {126}},\
  \bibinfo {pages} {041101} (\bibinfo {year} {2021})},\ \Eprint
  {https://arxiv.org/abs/2008.12456} {arXiv:2008.12456 [astro-ph.CO]}
  \BibitemShut {NoStop}%
\bibitem [{\citenamefont {Dom\`enech}\ \emph {et~al.}(2023)\citenamefont
  {Dom\`enech}, \citenamefont {Inman}, \citenamefont {Kusenko},\ and\
  \citenamefont {Sasaki}}]{Domenech:2023afs}%
  \BibitemOpen
  \bibfield  {author} {\bibinfo {author} {\bibfnamefont {G.}~\bibnamefont
  {Dom\`enech}}, \bibinfo {author} {\bibfnamefont {D.}~\bibnamefont {Inman}},
  \bibinfo {author} {\bibfnamefont {A.}~\bibnamefont {Kusenko}},\ and\ \bibinfo
  {author} {\bibfnamefont {M.}~\bibnamefont {Sasaki}},\ }\bibfield  {title}
  {\bibinfo {title} {{Halo Formation from Yukawa Forces in the Very Early
  Universe}},\ }\href@noop {} {\  (\bibinfo {year} {2023})},\ \Eprint
  {https://arxiv.org/abs/2304.13053} {arXiv:2304.13053 [astro-ph.CO]}
  \BibitemShut {NoStop}%
\bibitem [{\citenamefont {Agazie}\ \emph
  {et~al.}(2023{\natexlab{a}})\citenamefont {Agazie} \emph
  {et~al.}}]{NG15-SGWB}%
  \BibitemOpen
  \bibfield  {author} {\bibinfo {author} {\bibfnamefont {G.}~\bibnamefont
  {Agazie}} \emph {et~al.} (\bibinfo {collaboration} {NANOGrav}),\ }\bibfield
  {title} {\bibinfo {title} {The nanograv 15-year data set: Evidence for a
  gravitational-wave background},\ }\href@noop {} {\  (\bibinfo {year}
  {2023}{\natexlab{a}})},\ \Eprint {https://arxiv.org/abs/2306.16213}
  {arXiv:2306.16213 [astro-ph.HE]} \BibitemShut {NoStop}%
\bibitem [{\citenamefont {Agazie}\ \emph
  {et~al.}(2023{\natexlab{b}})\citenamefont {Agazie} \emph
  {et~al.}}]{NG15-pulsars}%
  \BibitemOpen
  \bibfield  {author} {\bibinfo {author} {\bibfnamefont {G.}~\bibnamefont
  {Agazie}} \emph {et~al.} (\bibinfo {collaboration} {NANOGrav}),\ }\bibfield
  {title} {\bibinfo {title} {The nanograv 15-year data set: Observations and
  timing of 68 millisecond pulsars},\ }\href@noop {} {\  (\bibinfo {year}
  {2023}{\natexlab{b}})},\ \Eprint {https://arxiv.org/abs/2306.16217}
  {arXiv:2306.16217 [astro-ph.HE]} \BibitemShut {NoStop}%
\bibitem [{\citenamefont {Antoniadis}\ \emph
  {et~al.}(2023{\natexlab{a}})\citenamefont {Antoniadis} \emph
  {et~al.}}]{EPTA2-SGWB}%
  \BibitemOpen
  \bibfield  {author} {\bibinfo {author} {\bibfnamefont {J.}~\bibnamefont
  {Antoniadis}} \emph {et~al.} (\bibinfo {collaboration} {EPTA}),\ }\bibfield
  {title} {\bibinfo {title} {The second data release from the european pulsar
  timing array iii. search for gravitational wave signals},\ }\href@noop {} {\
  (\bibinfo {year} {2023}{\natexlab{a}})},\ \Eprint
  {https://arxiv.org/abs/2306.16214} {arXiv:2306.16214 [astro-ph.HE]}
  \BibitemShut {NoStop}%
\bibitem [{\citenamefont {Antoniadis}\ \emph
  {et~al.}(2023{\natexlab{b}})\citenamefont {Antoniadis} \emph
  {et~al.}}]{EPTA2-pulsars}%
  \BibitemOpen
  \bibfield  {author} {\bibinfo {author} {\bibfnamefont {J.}~\bibnamefont
  {Antoniadis}} \emph {et~al.} (\bibinfo {collaboration} {EPTA}),\ }\bibfield
  {title} {\bibinfo {title} {The second data release from the european pulsar
  timing array i. the dataset and timing analysis},\ }\href@noop {} {\
  (\bibinfo {year} {2023}{\natexlab{b}})},\ \Eprint
  {https://arxiv.org/abs/2306.16224} {arXiv:2306.16224 [astro-ph.HE]}
  \BibitemShut {NoStop}%
\bibitem [{\citenamefont {Antoniadis}\ \emph
  {et~al.}(2023{\natexlab{c}})\citenamefont {Antoniadis} \emph
  {et~al.}}]{EPTA2-SMBHB-NP}%
  \BibitemOpen
  \bibfield  {author} {\bibinfo {author} {\bibfnamefont {J.}~\bibnamefont
  {Antoniadis}} \emph {et~al.} (\bibinfo {collaboration} {EPTA}),\ }\bibfield
  {title} {\bibinfo {title} {The second data release from the european pulsar
  timing array: V. implications for massive black holes, dark matter and the
  early universe},\ }\href@noop {} {\  (\bibinfo {year}
  {2023}{\natexlab{c}})},\ \Eprint {https://arxiv.org/abs/2306.16227}
  {arXiv:2306.16227 [astro-ph.HE]} \BibitemShut {NoStop}%
\bibitem [{\citenamefont {Reardon}\ \emph
  {et~al.}(2023{\natexlab{a}})\citenamefont {Reardon} \emph
  {et~al.}}]{PPTA3-SGWB}%
  \BibitemOpen
  \bibfield  {author} {\bibinfo {author} {\bibfnamefont {D.}~\bibnamefont
  {Reardon}} \emph {et~al.} (\bibinfo {collaboration} {PPTA}),\ }\bibfield
  {title} {\bibinfo {title} {Search for an isotropic gravitational-wave
  background with the parkes pulsar timing array},\ }\href@noop {} {\
  (\bibinfo {year} {2023}{\natexlab{a}})},\ \Eprint
  {https://arxiv.org/abs/2306.16215} {arXiv:2306.16215 [astro-ph.HE]}
  \BibitemShut {NoStop}%
\bibitem [{\citenamefont {Zic}\ \emph {et~al.}(2023)\citenamefont {Zic} \emph
  {et~al.}}]{PPTA3-pulsars}%
  \BibitemOpen
  \bibfield  {author} {\bibinfo {author} {\bibfnamefont {A.}~\bibnamefont
  {Zic}} \emph {et~al.} (\bibinfo {collaboration} {PPTA}),\ }\bibfield  {title}
  {\bibinfo {title} {The parkes pulsar timing array third data release},\
  }\href@noop {} {\  (\bibinfo {year} {2023})},\ \Eprint
  {https://arxiv.org/abs/2306.16230} {arXiv:2306.16230 [astro-ph.HE]}
  \BibitemShut {NoStop}%
\bibitem [{\citenamefont {Reardon}\ \emph
  {et~al.}(2023{\natexlab{b}})\citenamefont {Reardon} \emph
  {et~al.}}]{PPTA3-SMBHB}%
  \BibitemOpen
  \bibfield  {author} {\bibinfo {author} {\bibfnamefont {D.}~\bibnamefont
  {Reardon}} \emph {et~al.} (\bibinfo {collaboration} {PPTA}),\ }\bibfield
  {title} {\bibinfo {title} {The gravitational-wave background null hypothesis:
  Characterizing noise in millisecond pulsar arrival times with the parkes
  pulsar timing array},\ }\href@noop {} {\  (\bibinfo {year}
  {2023}{\natexlab{b}})},\ \Eprint {https://arxiv.org/abs/2306.16229}
  {arXiv:2306.16229 [astro-ph.HE]} \BibitemShut {NoStop}%
\bibitem [{\citenamefont {Xu}\ \emph {et~al.}(2023)\citenamefont {Xu} \emph
  {et~al.}}]{CPTA-SGWB}%
  \BibitemOpen
  \bibfield  {author} {\bibinfo {author} {\bibfnamefont {H.}~\bibnamefont {Xu}}
  \emph {et~al.} (\bibinfo {collaboration} {CPTA}),\ }\bibfield  {title}
  {\bibinfo {title} {Searching for the nano-hertz stochastic gravitational wave
  background with the chinese pulsar timing array data release i},\ }\href@noop
  {} {\  (\bibinfo {year} {2023})},\ \Eprint {https://arxiv.org/abs/2306.16216}
  {arXiv:2306.16216 [astro-ph.HE]} \BibitemShut {NoStop}%
\bibitem [{\citenamefont {Agazie}\ \emph
  {et~al.}(2023{\natexlab{c}})\citenamefont {Agazie} \emph
  {et~al.}}]{InternationalPulsarTimingArray:2023mzf}%
  \BibitemOpen
  \bibfield  {author} {\bibinfo {author} {\bibfnamefont {G.}~\bibnamefont
  {Agazie}} \emph {et~al.} (\bibinfo {collaboration} {International Pulsar
  Timing Array}),\ }\bibfield  {title} {\bibinfo {title} {{Comparing recent PTA
  results on the nanohertz stochastic gravitational wave background}},\
  }\href@noop {} {\  (\bibinfo {year} {2023}{\natexlab{c}})},\ \Eprint
  {https://arxiv.org/abs/2309.00693} {arXiv:2309.00693 [astro-ph.HE]}
  \BibitemShut {NoStop}%
\bibitem [{\citenamefont {Franciolini}\ \emph
  {et~al.}(2023{\natexlab{a}})\citenamefont {Franciolini}, \citenamefont
  {Iovino}, \citenamefont {Vaskonen},\ and\ \citenamefont
  {Veermae}}]{Franciolini:2023pbf}%
  \BibitemOpen
  \bibfield  {author} {\bibinfo {author} {\bibfnamefont {G.}~\bibnamefont
  {Franciolini}}, \bibinfo {author} {\bibfnamefont {A.}~\bibnamefont {Iovino},
  \bibfnamefont {Junior.}}, \bibinfo {author} {\bibfnamefont {V.}~\bibnamefont
  {Vaskonen}},\ and\ \bibinfo {author} {\bibfnamefont {H.}~\bibnamefont
  {Veermae}},\ }\bibfield  {title} {\bibinfo {title} {{The recent gravitational
  wave observation by pulsar timing arrays and primordial black holes: the
  importance of non-gaussianities}},\ }\href@noop {} {\  (\bibinfo {year}
  {2023}{\natexlab{a}})},\ \Eprint {https://arxiv.org/abs/2306.17149}
  {arXiv:2306.17149 [astro-ph.CO]} \BibitemShut {NoStop}%
\bibitem [{\citenamefont {Franciolini}\ \emph
  {et~al.}(2023{\natexlab{b}})\citenamefont {Franciolini}, \citenamefont
  {Racco},\ and\ \citenamefont {Rompineve}}]{Franciolini:2023wjm}%
  \BibitemOpen
  \bibfield  {author} {\bibinfo {author} {\bibfnamefont {G.}~\bibnamefont
  {Franciolini}}, \bibinfo {author} {\bibfnamefont {D.}~\bibnamefont {Racco}},\
  and\ \bibinfo {author} {\bibfnamefont {F.}~\bibnamefont {Rompineve}},\
  }\bibfield  {title} {\bibinfo {title} {{Footprints of the QCD Crossover on
  Cosmological Gravitational Waves at Pulsar Timing Arrays}},\ }\href@noop {}
  {\  (\bibinfo {year} {2023}{\natexlab{b}})},\ \Eprint
  {https://arxiv.org/abs/2306.17136} {arXiv:2306.17136 [astro-ph.CO]}
  \BibitemShut {NoStop}%
\bibitem [{\citenamefont {Inomata}\ \emph
  {et~al.}(2023{\natexlab{b}})\citenamefont {Inomata}, \citenamefont {Kohri},\
  and\ \citenamefont {Terada}}]{Inomata:2023zup}%
  \BibitemOpen
  \bibfield  {author} {\bibinfo {author} {\bibfnamefont {K.}~\bibnamefont
  {Inomata}}, \bibinfo {author} {\bibfnamefont {K.}~\bibnamefont {Kohri}},\
  and\ \bibinfo {author} {\bibfnamefont {T.}~\bibnamefont {Terada}},\
  }\bibfield  {title} {\bibinfo {title} {{The Detected Stochastic Gravitational
  Waves and Sub-Solar Primordial Black Holes}},\ }\href@noop {} {\  (\bibinfo
  {year} {2023}{\natexlab{b}})},\ \Eprint {https://arxiv.org/abs/2306.17834}
  {arXiv:2306.17834 [astro-ph.CO]} \BibitemShut {NoStop}%
\bibitem [{\citenamefont {Cai}\ \emph {et~al.}(2023{\natexlab{b}})\citenamefont
  {Cai}, \citenamefont {He}, \citenamefont {Ma}, \citenamefont {Yan},\ and\
  \citenamefont {Yuan}}]{Cai:2023dls}%
  \BibitemOpen
  \bibfield  {author} {\bibinfo {author} {\bibfnamefont {Y.-F.}\ \bibnamefont
  {Cai}}, \bibinfo {author} {\bibfnamefont {X.-C.}\ \bibnamefont {He}},
  \bibinfo {author} {\bibfnamefont {X.}~\bibnamefont {Ma}}, \bibinfo {author}
  {\bibfnamefont {S.-F.}\ \bibnamefont {Yan}},\ and\ \bibinfo {author}
  {\bibfnamefont {G.-W.}\ \bibnamefont {Yuan}},\ }\bibfield  {title} {\bibinfo
  {title} {{Limits on scalar-induced gravitational waves from the stochastic
  background by pulsar timing array observations}},\ }\href@noop {} {\
  (\bibinfo {year} {2023}{\natexlab{b}})},\ \Eprint
  {https://arxiv.org/abs/2306.17822} {arXiv:2306.17822 [gr-qc]} \BibitemShut
  {NoStop}%
\bibitem [{\citenamefont {Wang}\ \emph
  {et~al.}(2023{\natexlab{b}})\citenamefont {Wang}, \citenamefont {Zhao},
  \citenamefont {Li},\ and\ \citenamefont {Zhu}}]{Wang:2023ost}%
  \BibitemOpen
  \bibfield  {author} {\bibinfo {author} {\bibfnamefont {S.}~\bibnamefont
  {Wang}}, \bibinfo {author} {\bibfnamefont {Z.-C.}\ \bibnamefont {Zhao}},
  \bibinfo {author} {\bibfnamefont {J.-P.}\ \bibnamefont {Li}},\ and\ \bibinfo
  {author} {\bibfnamefont {Q.-H.}\ \bibnamefont {Zhu}},\ }\bibfield  {title}
  {\bibinfo {title} {{Exploring the Implications of 2023 Pulsar Timing Array
  Datasets for Scalar-Induced Gravitational Waves and Primordial Black
  Holes}},\ }\href@noop {} {\  (\bibinfo {year} {2023}{\natexlab{b}})},\
  \Eprint {https://arxiv.org/abs/2307.00572} {arXiv:2307.00572 [astro-ph.CO]}
  \BibitemShut {NoStop}%
\bibitem [{\citenamefont {Liu}\ \emph {et~al.}(2023{\natexlab{a}})\citenamefont
  {Liu}, \citenamefont {Chen},\ and\ \citenamefont {Huang}}]{Liu:2023ymk}%
  \BibitemOpen
  \bibfield  {author} {\bibinfo {author} {\bibfnamefont {L.}~\bibnamefont
  {Liu}}, \bibinfo {author} {\bibfnamefont {Z.-C.}\ \bibnamefont {Chen}},\ and\
  \bibinfo {author} {\bibfnamefont {Q.-G.}\ \bibnamefont {Huang}},\ }\bibfield
  {title} {\bibinfo {title} {{Implications for the non-Gaussianity of curvature
  perturbation from pulsar timing arrays}},\ }\href@noop {} {\  (\bibinfo
  {year} {2023}{\natexlab{a}})},\ \Eprint {https://arxiv.org/abs/2307.01102}
  {arXiv:2307.01102 [astro-ph.CO]} \BibitemShut {NoStop}%
\bibitem [{\citenamefont {Unal}\ \emph {et~al.}(2023)\citenamefont {Unal},
  \citenamefont {Papageorgiou},\ and\ \citenamefont {Obata}}]{Unal:2023srk}%
  \BibitemOpen
  \bibfield  {author} {\bibinfo {author} {\bibfnamefont {C.}~\bibnamefont
  {Unal}}, \bibinfo {author} {\bibfnamefont {A.}~\bibnamefont {Papageorgiou}},\
  and\ \bibinfo {author} {\bibfnamefont {I.}~\bibnamefont {Obata}},\ }\bibfield
   {title} {\bibinfo {title} {{Axion-Gauge Dynamics During Inflation as the
  Origin of Pulsar Timing Array Signals and Primordial Black Holes}},\
  }\href@noop {} {\  (\bibinfo {year} {2023})},\ \Eprint
  {https://arxiv.org/abs/2307.02322} {arXiv:2307.02322 [astro-ph.CO]}
  \BibitemShut {NoStop}%
\bibitem [{\citenamefont {Figueroa}\ \emph {et~al.}(2023)\citenamefont
  {Figueroa}, \citenamefont {Pieroni}, \citenamefont {Ricciardone},\ and\
  \citenamefont {Simakachorn}}]{Figueroa:2023zhu}%
  \BibitemOpen
  \bibfield  {author} {\bibinfo {author} {\bibfnamefont {D.~G.}\ \bibnamefont
  {Figueroa}}, \bibinfo {author} {\bibfnamefont {M.}~\bibnamefont {Pieroni}},
  \bibinfo {author} {\bibfnamefont {A.}~\bibnamefont {Ricciardone}},\ and\
  \bibinfo {author} {\bibfnamefont {P.}~\bibnamefont {Simakachorn}},\
  }\bibfield  {title} {\bibinfo {title} {{Cosmological Background
  Interpretation of Pulsar Timing Array Data}},\ }\href@noop {} {\  (\bibinfo
  {year} {2023})},\ \Eprint {https://arxiv.org/abs/2307.02399}
  {arXiv:2307.02399 [astro-ph.CO]} \BibitemShut {NoStop}%
\bibitem [{\citenamefont {Yi}\ \emph {et~al.}(2023{\natexlab{a}})\citenamefont
  {Yi}, \citenamefont {Gao}, \citenamefont {Gong}, \citenamefont {Wang},\ and\
  \citenamefont {Zhang}}]{Yi:2023mbm}%
  \BibitemOpen
  \bibfield  {author} {\bibinfo {author} {\bibfnamefont {Z.}~\bibnamefont
  {Yi}}, \bibinfo {author} {\bibfnamefont {Q.}~\bibnamefont {Gao}}, \bibinfo
  {author} {\bibfnamefont {Y.}~\bibnamefont {Gong}}, \bibinfo {author}
  {\bibfnamefont {Y.}~\bibnamefont {Wang}},\ and\ \bibinfo {author}
  {\bibfnamefont {F.}~\bibnamefont {Zhang}},\ }\bibfield  {title} {\bibinfo
  {title} {{The waveform of the scalar induced gravitational waves in light of
  Pulsar Timing Array data}},\ }\href@noop {} {\  (\bibinfo {year}
  {2023}{\natexlab{a}})},\ \Eprint {https://arxiv.org/abs/2307.02467}
  {arXiv:2307.02467 [gr-qc]} \BibitemShut {NoStop}%
\bibitem [{\citenamefont {Zhu}\ \emph {et~al.}(2023)\citenamefont {Zhu},
  \citenamefont {Zhao},\ and\ \citenamefont {Wang}}]{Zhu:2023faa}%
  \BibitemOpen
  \bibfield  {author} {\bibinfo {author} {\bibfnamefont {Q.-H.}\ \bibnamefont
  {Zhu}}, \bibinfo {author} {\bibfnamefont {Z.-C.}\ \bibnamefont {Zhao}},\ and\
  \bibinfo {author} {\bibfnamefont {S.}~\bibnamefont {Wang}},\ }\bibfield
  {title} {\bibinfo {title} {{Joint implications of BBN, CMB, and PTA Datasets
  for Scalar-Induced Gravitational Waves of Second and Third orders}},\
  }\href@noop {} {\  (\bibinfo {year} {2023})},\ \Eprint
  {https://arxiv.org/abs/2307.03095} {arXiv:2307.03095 [astro-ph.CO]}
  \BibitemShut {NoStop}%
\bibitem [{\citenamefont {Firouzjahi}\ and\ \citenamefont
  {Talebian}(2023)}]{Firouzjahi:2023lzg}%
  \BibitemOpen
  \bibfield  {author} {\bibinfo {author} {\bibfnamefont {H.}~\bibnamefont
  {Firouzjahi}}\ and\ \bibinfo {author} {\bibfnamefont {A.}~\bibnamefont
  {Talebian}},\ }\bibfield  {title} {\bibinfo {title} {{Induced Gravitational
  Waves from Ultra Slow-Roll Inflation and Pulsar Timing Arrays
  Observations}},\ }\href@noop {} {\  (\bibinfo {year} {2023})},\ \Eprint
  {https://arxiv.org/abs/2307.03164} {arXiv:2307.03164 [gr-qc]} \BibitemShut
  {NoStop}%
\bibitem [{\citenamefont {Li}\ \emph {et~al.}(2023)\citenamefont {Li},
  \citenamefont {Wang}, \citenamefont {Zhao},\ and\ \citenamefont
  {Kohri}}]{Li:2023qua}%
  \BibitemOpen
  \bibfield  {author} {\bibinfo {author} {\bibfnamefont {J.-P.}\ \bibnamefont
  {Li}}, \bibinfo {author} {\bibfnamefont {S.}~\bibnamefont {Wang}}, \bibinfo
  {author} {\bibfnamefont {Z.-C.}\ \bibnamefont {Zhao}},\ and\ \bibinfo
  {author} {\bibfnamefont {K.}~\bibnamefont {Kohri}},\ }\bibfield  {title}
  {\bibinfo {title} {{Primordial Non-Gaussianity and Anisotropies in
  Gravitational Waves induced by Scalar Perturbations}},\ }\href@noop {} {\
  (\bibinfo {year} {2023})},\ \Eprint {https://arxiv.org/abs/2305.19950}
  {arXiv:2305.19950 [astro-ph.CO]} \BibitemShut {NoStop}%
\bibitem [{\citenamefont {You}\ \emph {et~al.}(2023)\citenamefont {You},
  \citenamefont {Yi},\ and\ \citenamefont {Wu}}]{You:2023rmn}%
  \BibitemOpen
  \bibfield  {author} {\bibinfo {author} {\bibfnamefont {Z.-Q.}\ \bibnamefont
  {You}}, \bibinfo {author} {\bibfnamefont {Z.}~\bibnamefont {Yi}},\ and\
  \bibinfo {author} {\bibfnamefont {Y.}~\bibnamefont {Wu}},\ }\bibfield
  {title} {\bibinfo {title} {{Constraints on primordial curvature power
  spectrum with pulsar timing arrays}},\ }\href@noop {} {\  (\bibinfo {year}
  {2023})},\ \Eprint {https://arxiv.org/abs/2307.04419} {arXiv:2307.04419
  [gr-qc]} \BibitemShut {NoStop}%
\bibitem [{\citenamefont {Balaji}\ \emph {et~al.}(2023)\citenamefont {Balaji},
  \citenamefont {Dom\`enech},\ and\ \citenamefont
  {Franciolini}}]{Balaji:2023ehk}%
  \BibitemOpen
  \bibfield  {author} {\bibinfo {author} {\bibfnamefont {S.}~\bibnamefont
  {Balaji}}, \bibinfo {author} {\bibfnamefont {G.}~\bibnamefont {Dom\`enech}},\
  and\ \bibinfo {author} {\bibfnamefont {G.}~\bibnamefont {Franciolini}},\
  }\bibfield  {title} {\bibinfo {title} {{Scalar-induced gravitational wave
  interpretation of PTA data: the role of scalar fluctuation propagation
  speed}},\ }\href@noop {} {\  (\bibinfo {year} {2023})},\ \Eprint
  {https://arxiv.org/abs/2307.08552} {arXiv:2307.08552 [gr-qc]} \BibitemShut
  {NoStop}%
\bibitem [{\citenamefont {Hosseini~Mansoori}\ \emph {et~al.}(2023)\citenamefont
  {Hosseini~Mansoori}, \citenamefont {Felegray}, \citenamefont {Talebian},\
  and\ \citenamefont {Sami}}]{HosseiniMansoori:2023mqh}%
  \BibitemOpen
  \bibfield  {author} {\bibinfo {author} {\bibfnamefont {S.~A.}\ \bibnamefont
  {Hosseini~Mansoori}}, \bibinfo {author} {\bibfnamefont {F.}~\bibnamefont
  {Felegray}}, \bibinfo {author} {\bibfnamefont {A.}~\bibnamefont {Talebian}},\
  and\ \bibinfo {author} {\bibfnamefont {M.}~\bibnamefont {Sami}},\ }\bibfield
  {title} {\bibinfo {title} {{PBHs and GWs from
  \ensuremath{\mathbb{T}}$^{2}$-inflation and NANOGrav 15-year data}},\ }\href
  {https://doi.org/10.1088/1475-7516/2023/08/067} {\bibfield  {journal}
  {\bibinfo  {journal} {JCAP}\ }\textbf {\bibinfo {volume} {08}},\ \bibinfo
  {pages} {067}},\ \Eprint {https://arxiv.org/abs/2307.06757} {arXiv:2307.06757
  [astro-ph.CO]} \BibitemShut {NoStop}%
\bibitem [{\citenamefont {Zhao}\ \emph {et~al.}(2023)\citenamefont {Zhao},
  \citenamefont {Zhu}, \citenamefont {Wang},\ and\ \citenamefont
  {Zhang}}]{Zhao:2023joc}%
  \BibitemOpen
  \bibfield  {author} {\bibinfo {author} {\bibfnamefont {Z.-C.}\ \bibnamefont
  {Zhao}}, \bibinfo {author} {\bibfnamefont {Q.-H.}\ \bibnamefont {Zhu}},
  \bibinfo {author} {\bibfnamefont {S.}~\bibnamefont {Wang}},\ and\ \bibinfo
  {author} {\bibfnamefont {X.}~\bibnamefont {Zhang}},\ }\bibfield  {title}
  {\bibinfo {title} {{Exploring the Equation of State of the Early Universe:
  Insights from BBN, CMB, and PTA Observations}},\ }\href@noop {} {\  (\bibinfo
  {year} {2023})},\ \Eprint {https://arxiv.org/abs/2307.13574}
  {arXiv:2307.13574 [astro-ph.CO]} \BibitemShut {NoStop}%
\bibitem [{\citenamefont {Liu}\ \emph {et~al.}(2023{\natexlab{b}})\citenamefont
  {Liu}, \citenamefont {Chen},\ and\ \citenamefont {Huang}}]{Liu:2023pau}%
  \BibitemOpen
  \bibfield  {author} {\bibinfo {author} {\bibfnamefont {L.}~\bibnamefont
  {Liu}}, \bibinfo {author} {\bibfnamefont {Z.-C.}\ \bibnamefont {Chen}},\ and\
  \bibinfo {author} {\bibfnamefont {Q.-G.}\ \bibnamefont {Huang}},\ }\bibfield
  {title} {\bibinfo {title} {{Probing the equation of state of the early
  Universe with pulsar timing arrays}},\ }\href@noop {} {\  (\bibinfo {year}
  {2023}{\natexlab{b}})},\ \Eprint {https://arxiv.org/abs/2307.14911}
  {arXiv:2307.14911 [astro-ph.CO]} \BibitemShut {NoStop}%
\bibitem [{\citenamefont {Yi}\ \emph {et~al.}(2023{\natexlab{b}})\citenamefont
  {Yi}, \citenamefont {You},\ and\ \citenamefont {Wu}}]{Yi:2023tdk}%
  \BibitemOpen
  \bibfield  {author} {\bibinfo {author} {\bibfnamefont {Z.}~\bibnamefont
  {Yi}}, \bibinfo {author} {\bibfnamefont {Z.-Q.}\ \bibnamefont {You}},\ and\
  \bibinfo {author} {\bibfnamefont {Y.}~\bibnamefont {Wu}},\ }\bibfield
  {title} {\bibinfo {title} {{Model-independent reconstruction of the
  primordial curvature power spectrum from PTA data}},\ }\href@noop {} {\
  (\bibinfo {year} {2023}{\natexlab{b}})},\ \Eprint
  {https://arxiv.org/abs/2308.05632} {arXiv:2308.05632 [astro-ph.CO]}
  \BibitemShut {NoStop}%
\bibitem [{\citenamefont {Bhaumik}\ \emph {et~al.}(2023)\citenamefont
  {Bhaumik}, \citenamefont {Jain},\ and\ \citenamefont
  {Lewicki}}]{Bhaumik:2023wmw}%
  \BibitemOpen
  \bibfield  {author} {\bibinfo {author} {\bibfnamefont {N.}~\bibnamefont
  {Bhaumik}}, \bibinfo {author} {\bibfnamefont {R.~K.}\ \bibnamefont {Jain}},\
  and\ \bibinfo {author} {\bibfnamefont {M.}~\bibnamefont {Lewicki}},\
  }\bibfield  {title} {\bibinfo {title} {{Ultra-low mass PBHs in the early
  universe can explain the PTA signal}},\ }\href@noop {} {\  (\bibinfo {year}
  {2023})},\ \Eprint {https://arxiv.org/abs/2308.07912} {arXiv:2308.07912
  [astro-ph.CO]} \BibitemShut {NoStop}%
\bibitem [{\citenamefont {Choudhury}\ \emph {et~al.}(2023)\citenamefont
  {Choudhury}, \citenamefont {Karde}, \citenamefont {Panda},\ and\
  \citenamefont {Sami}}]{Choudhury:2023hfm}%
  \BibitemOpen
  \bibfield  {author} {\bibinfo {author} {\bibfnamefont {S.}~\bibnamefont
  {Choudhury}}, \bibinfo {author} {\bibfnamefont {A.}~\bibnamefont {Karde}},
  \bibinfo {author} {\bibfnamefont {S.}~\bibnamefont {Panda}},\ and\ \bibinfo
  {author} {\bibfnamefont {M.}~\bibnamefont {Sami}},\ }\bibfield  {title}
  {\bibinfo {title} {{Scalar induced gravity waves from ultra slow-roll
  Galileon inflation}},\ }\href@noop {} {\  (\bibinfo {year} {2023})},\ \Eprint
  {https://arxiv.org/abs/2308.09273} {arXiv:2308.09273 [astro-ph.CO]}
  \BibitemShut {NoStop}%
\bibitem [{\citenamefont {Yi}\ \emph {et~al.}(2023{\natexlab{c}})\citenamefont
  {Yi}, \citenamefont {You}, \citenamefont {Wu}, \citenamefont {Chen},\ and\
  \citenamefont {Liu}}]{Yi:2023npi}%
  \BibitemOpen
  \bibfield  {author} {\bibinfo {author} {\bibfnamefont {Z.}~\bibnamefont
  {Yi}}, \bibinfo {author} {\bibfnamefont {Z.-Q.}\ \bibnamefont {You}},
  \bibinfo {author} {\bibfnamefont {Y.}~\bibnamefont {Wu}}, \bibinfo {author}
  {\bibfnamefont {Z.-C.}\ \bibnamefont {Chen}},\ and\ \bibinfo {author}
  {\bibfnamefont {L.}~\bibnamefont {Liu}},\ }\bibfield  {title} {\bibinfo
  {title} {{Exploring the NANOGrav Signal and Planet-mass Primordial Black
  Holes through Higgs Inflation}},\ }\href@noop {} {\  (\bibinfo {year}
  {2023}{\natexlab{c}})},\ \Eprint {https://arxiv.org/abs/2308.14688}
  {arXiv:2308.14688 [astro-ph.CO]} \BibitemShut {NoStop}%
\bibitem [{\citenamefont {Harigaya}\ \emph {et~al.}(2023)\citenamefont
  {Harigaya}, \citenamefont {Inomata},\ and\ \citenamefont
  {Terada}}]{Harigaya:2023pmw}%
  \BibitemOpen
  \bibfield  {author} {\bibinfo {author} {\bibfnamefont {K.}~\bibnamefont
  {Harigaya}}, \bibinfo {author} {\bibfnamefont {K.}~\bibnamefont {Inomata}},\
  and\ \bibinfo {author} {\bibfnamefont {T.}~\bibnamefont {Terada}},\
  }\bibfield  {title} {\bibinfo {title} {{Induced Gravitational Waves with
  Kination Era for Recent Pulsar Timing Array Signals}},\ }\href@noop {} {\
  (\bibinfo {year} {2023})},\ \Eprint {https://arxiv.org/abs/2309.00228}
  {arXiv:2309.00228 [astro-ph.CO]} \BibitemShut {NoStop}%
\bibitem [{\citenamefont {Basilakos}\ \emph {et~al.}(2023)\citenamefont
  {Basilakos}, \citenamefont {Nanopoulos}, \citenamefont {Papanikolaou},
  \citenamefont {Saridakis},\ and\ \citenamefont
  {Tzerefos}}]{Basilakos:2023xof}%
  \BibitemOpen
  \bibfield  {author} {\bibinfo {author} {\bibfnamefont {S.}~\bibnamefont
  {Basilakos}}, \bibinfo {author} {\bibfnamefont {D.~V.}\ \bibnamefont
  {Nanopoulos}}, \bibinfo {author} {\bibfnamefont {T.}~\bibnamefont
  {Papanikolaou}}, \bibinfo {author} {\bibfnamefont {E.~N.}\ \bibnamefont
  {Saridakis}},\ and\ \bibinfo {author} {\bibfnamefont {C.}~\bibnamefont
  {Tzerefos}},\ }\bibfield  {title} {\bibinfo {title} {{Signatures of
  Superstring theory in NANOGrav}},\ }\href@noop {} {\  (\bibinfo {year}
  {2023})},\ \Eprint {https://arxiv.org/abs/2307.08601} {arXiv:2307.08601
  [hep-th]} \BibitemShut {NoStop}%
\bibitem [{\citenamefont {Huang}\ \emph {et~al.}(2023)\citenamefont {Huang},
  \citenamefont {Cai}, \citenamefont {Jiang}, \citenamefont {Zhang},\ and\
  \citenamefont {Piao}}]{Huang:2023chx}%
  \BibitemOpen
  \bibfield  {author} {\bibinfo {author} {\bibfnamefont {H.-L.}\ \bibnamefont
  {Huang}}, \bibinfo {author} {\bibfnamefont {Y.}~\bibnamefont {Cai}}, \bibinfo
  {author} {\bibfnamefont {J.-Q.}\ \bibnamefont {Jiang}}, \bibinfo {author}
  {\bibfnamefont {J.}~\bibnamefont {Zhang}},\ and\ \bibinfo {author}
  {\bibfnamefont {Y.-S.}\ \bibnamefont {Piao}},\ }\bibfield  {title} {\bibinfo
  {title} {{Supermassive primordial black holes in multiverse: for nano-Hertz
  gravitational wave and high-redshift JWST galaxies}},\ }\href@noop {} {\
  (\bibinfo {year} {2023})},\ \Eprint {https://arxiv.org/abs/2306.17577}
  {arXiv:2306.17577 [gr-qc]} \BibitemShut {NoStop}%
\bibitem [{\citenamefont {Gouttenoire}\ \emph {et~al.}(2023)\citenamefont
  {Gouttenoire}, \citenamefont {Trifinopoulos}, \citenamefont {Valogiannis},\
  and\ \citenamefont {Vanvlasselaer}}]{Gouttenoire:2023nzr}%
  \BibitemOpen
  \bibfield  {author} {\bibinfo {author} {\bibfnamefont {Y.}~\bibnamefont
  {Gouttenoire}}, \bibinfo {author} {\bibfnamefont {S.}~\bibnamefont
  {Trifinopoulos}}, \bibinfo {author} {\bibfnamefont {G.}~\bibnamefont
  {Valogiannis}},\ and\ \bibinfo {author} {\bibfnamefont {M.}~\bibnamefont
  {Vanvlasselaer}},\ }\bibfield  {title} {\bibinfo {title} {{Scrutinizing the
  Primordial Black Holes Interpretation of PTA Gravitational Waves and JWST
  Early Galaxies}},\ }\href@noop {} {\  (\bibinfo {year} {2023})},\ \Eprint
  {https://arxiv.org/abs/2307.01457} {arXiv:2307.01457 [astro-ph.CO]}
  \BibitemShut {NoStop}%
\bibitem [{\citenamefont {Depta}\ \emph {et~al.}(2023)\citenamefont {Depta},
  \citenamefont {Schmidt-Hoberg},\ and\ \citenamefont
  {Tasillo}}]{Depta:2023qst}%
  \BibitemOpen
  \bibfield  {author} {\bibinfo {author} {\bibfnamefont {P.~F.}\ \bibnamefont
  {Depta}}, \bibinfo {author} {\bibfnamefont {K.}~\bibnamefont
  {Schmidt-Hoberg}},\ and\ \bibinfo {author} {\bibfnamefont {C.}~\bibnamefont
  {Tasillo}},\ }\bibfield  {title} {\bibinfo {title} {{Do pulsar timing arrays
  observe merging primordial black holes?}},\ }\href@noop {} {\  (\bibinfo
  {year} {2023})},\ \Eprint {https://arxiv.org/abs/2306.17836}
  {arXiv:2306.17836 [astro-ph.CO]} \BibitemShut {NoStop}%
\bibitem [{\citenamefont {Kodama}\ and\ \citenamefont
  {Sasaki}(1984)}]{Kodama:1985bj}%
  \BibitemOpen
  \bibfield  {author} {\bibinfo {author} {\bibfnamefont {H.}~\bibnamefont
  {Kodama}}\ and\ \bibinfo {author} {\bibfnamefont {M.}~\bibnamefont
  {Sasaki}},\ }\bibfield  {title} {\bibinfo {title} {{Cosmological Perturbation
  Theory}},\ }\href {https://doi.org/10.1143/PTPS.78.1} {\bibfield  {journal}
  {\bibinfo  {journal} {Prog. Theor. Phys. Suppl.}\ }\textbf {\bibinfo {volume}
  {78}},\ \bibinfo {pages} {1} (\bibinfo {year} {1984})}\BibitemShut {NoStop}%
\bibitem [{\citenamefont {Mukhanov}\ \emph {et~al.}(1992)\citenamefont
  {Mukhanov}, \citenamefont {Feldman},\ and\ \citenamefont
  {Brandenberger}}]{Mukhanov:1990me}%
  \BibitemOpen
  \bibfield  {author} {\bibinfo {author} {\bibfnamefont {V.~F.}\ \bibnamefont
  {Mukhanov}}, \bibinfo {author} {\bibfnamefont {H.~A.}\ \bibnamefont
  {Feldman}},\ and\ \bibinfo {author} {\bibfnamefont {R.~H.}\ \bibnamefont
  {Brandenberger}},\ }\bibfield  {title} {\bibinfo {title} {{Theory of
  cosmological perturbations. Part 1. Classical perturbations. Part 2. Quantum
  theory of perturbations. Part 3. Extensions}},\ }\href
  {https://doi.org/10.1016/0370-1573(92)90044-Z} {\bibfield  {journal}
  {\bibinfo  {journal} {Phys. Rept.}\ }\textbf {\bibinfo {volume} {215}},\
  \bibinfo {pages} {203} (\bibinfo {year} {1992})}\BibitemShut {NoStop}%
\bibitem [{\citenamefont {Brandenberger}(2004)}]{Brandenberger:2003vk}%
  \BibitemOpen
  \bibfield  {author} {\bibinfo {author} {\bibfnamefont {R.~H.}\ \bibnamefont
  {Brandenberger}},\ }\bibfield  {title} {\bibinfo {title} {{Lectures on the
  theory of cosmological perturbations}},\ }\href
  {https://doi.org/10.1007/978-3-540-40918-2_5} {\bibfield  {journal} {\bibinfo
   {journal} {Lect. Notes Phys.}\ }\textbf {\bibinfo {volume} {646}},\ \bibinfo
  {pages} {127} (\bibinfo {year} {2004})},\ \Eprint
  {https://arxiv.org/abs/hep-th/0306071} {arXiv:hep-th/0306071} \BibitemShut
  {NoStop}%
\bibitem [{\citenamefont {Durrer}(2004)}]{Durrer:2004fx}%
  \BibitemOpen
  \bibfield  {author} {\bibinfo {author} {\bibfnamefont {R.}~\bibnamefont
  {Durrer}},\ }\bibfield  {title} {\bibinfo {title} {{Cosmological perturbation
  theory}},\ }\href {https://doi.org/10.1007/978-3-540-31535-3_2} {\bibfield
  {journal} {\bibinfo  {journal} {Lect. Notes Phys.}\ }\textbf {\bibinfo
  {volume} {653}},\ \bibinfo {pages} {31} (\bibinfo {year} {2004})},\ \Eprint
  {https://arxiv.org/abs/astro-ph/0402129} {arXiv:astro-ph/0402129}
  \BibitemShut {NoStop}%
\bibitem [{\citenamefont {Baumann}(2009)}]{Baumann:2009ds}%
  \BibitemOpen
  \bibfield  {author} {\bibinfo {author} {\bibfnamefont {D.}~\bibnamefont
  {Baumann}},\ }\bibfield  {title} {\bibinfo {title} {{Inflation}},\ }in\ \href
  {https://doi.org/10.1142/9789814327183_0010} {\emph {\bibinfo {booktitle}
  {{Theoretical Advanced Study Institute in Elementary Particle Physics}}}}\
  (\bibinfo {year} {2009})\ \Eprint {https://arxiv.org/abs/0907.5424}
  {arXiv:0907.5424 [hep-th]} \BibitemShut {NoStop}%
\bibitem [{\citenamefont {Koyama}(2010)}]{Koyama:2010xj}%
  \BibitemOpen
  \bibfield  {author} {\bibinfo {author} {\bibfnamefont {K.}~\bibnamefont
  {Koyama}},\ }\bibfield  {title} {\bibinfo {title} {{Non-Gaussianity of
  quantum fields during inflation}},\ }\href
  {https://doi.org/10.1088/0264-9381/27/12/124001} {\bibfield  {journal}
  {\bibinfo  {journal} {Class. Quant. Grav.}\ }\textbf {\bibinfo {volume}
  {27}},\ \bibinfo {pages} {124001} (\bibinfo {year} {2010})},\ \Eprint
  {https://arxiv.org/abs/1002.0600} {arXiv:1002.0600 [hep-th]} \BibitemShut
  {NoStop}%
\bibitem [{\citenamefont {{Witkowski}}(2022)}]{2022arXiv220905296W}%
  \BibitemOpen
  \bibfield  {author} {\bibinfo {author} {\bibfnamefont {L.~T.}\ \bibnamefont
  {{Witkowski}}},\ }\bibfield  {title} {\bibinfo {title} {{SIGWfast: a python
  package for the computation of scalar-induced gravitational wave spectra}},\
  }\bibfield  {journal} {\bibinfo  {journal} {arXiv e-prints}\ }\href
  {https://doi.org/10.48550/arXiv.2209.05296} {10.48550/arXiv.2209.05296}
  (\bibinfo {year} {2022}),\ \Eprint {https://arxiv.org/abs/2209.05296}
  {arXiv:2209.05296 [astro-ph.CO]} \BibitemShut {NoStop}%
\bibitem [{\citenamefont {Cai}\ \emph {et~al.}(2019{\natexlab{b}})\citenamefont
  {Cai}, \citenamefont {Pi},\ and\ \citenamefont {Sasaki}}]{Cai:2018dig}%
  \BibitemOpen
  \bibfield  {author} {\bibinfo {author} {\bibfnamefont {R.-g.}\ \bibnamefont
  {Cai}}, \bibinfo {author} {\bibfnamefont {S.}~\bibnamefont {Pi}},\ and\
  \bibinfo {author} {\bibfnamefont {M.}~\bibnamefont {Sasaki}},\ }\bibfield
  {title} {\bibinfo {title} {{Gravitational Waves Induced by non-Gaussian
  Scalar Perturbations}},\ }\href
  {https://doi.org/10.1103/PhysRevLett.122.201101} {\bibfield  {journal}
  {\bibinfo  {journal} {Phys. Rev. Lett.}\ }\textbf {\bibinfo {volume} {122}},\
  \bibinfo {pages} {201101} (\bibinfo {year} {2019}{\natexlab{b}})},\ \Eprint
  {https://arxiv.org/abs/1810.11000} {arXiv:1810.11000 [astro-ph.CO]}
  \BibitemShut {NoStop}%
\bibitem [{\citenamefont {Yuan}\ \emph {et~al.}(2020)\citenamefont {Yuan},
  \citenamefont {Chen},\ and\ \citenamefont {Huang}}]{Yuan:2019wwo}%
  \BibitemOpen
  \bibfield  {author} {\bibinfo {author} {\bibfnamefont {C.}~\bibnamefont
  {Yuan}}, \bibinfo {author} {\bibfnamefont {Z.-C.}\ \bibnamefont {Chen}},\
  and\ \bibinfo {author} {\bibfnamefont {Q.-G.}\ \bibnamefont {Huang}},\
  }\bibfield  {title} {\bibinfo {title} {{Log-dependent slope of scalar induced
  gravitational waves in the infrared regions}},\ }\href
  {https://doi.org/10.1103/PhysRevD.101.043019} {\bibfield  {journal} {\bibinfo
   {journal} {Phys. Rev. D}\ }\textbf {\bibinfo {volume} {101}},\ \bibinfo
  {pages} {043019} (\bibinfo {year} {2020})},\ \Eprint
  {https://arxiv.org/abs/1910.09099} {arXiv:1910.09099 [astro-ph.CO]}
  \BibitemShut {NoStop}%
\bibitem [{\citenamefont {Espinosa}\ \emph
  {et~al.}(2018{\natexlab{c}})\citenamefont {Espinosa}, \citenamefont {Racco},\
  and\ \citenamefont {Riotto}}]{Espinosa_2018}%
  \BibitemOpen
  \bibfield  {author} {\bibinfo {author} {\bibfnamefont {J.}~\bibnamefont
  {Espinosa}}, \bibinfo {author} {\bibfnamefont {D.}~\bibnamefont {Racco}},\
  and\ \bibinfo {author} {\bibfnamefont {A.}~\bibnamefont {Riotto}},\
  }\bibfield  {title} {\bibinfo {title} {A cosmological signature of the {SM}
  higgs instability: gravitational waves},\ }\href
  {https://doi.org/10.1088/1475-7516/2018/09/012} {\bibfield  {journal}
  {\bibinfo  {journal} {Journal of Cosmology and Astroparticle Physics}\
  }\textbf {\bibinfo {volume} {2018}}\bibinfo  {number} { (09)},\ \bibinfo
  {pages} {012}}\BibitemShut {NoStop}%
\bibitem [{\citenamefont {Kohri}\ and\ \citenamefont
  {Terada}(2018)}]{Kohri:2018awv}%
  \BibitemOpen
\bibfield  {number} {  }\bibfield  {author} {\bibinfo {author} {\bibfnamefont
  {K.}~\bibnamefont {Kohri}}\ and\ \bibinfo {author} {\bibfnamefont
  {T.}~\bibnamefont {Terada}},\ }\bibfield  {title} {\bibinfo {title}
  {{Semianalytic calculation of gravitational wave spectrum nonlinearly induced
  from primordial curvature perturbations}},\ }\href
  {https://doi.org/10.1103/PhysRevD.97.123532} {\bibfield  {journal} {\bibinfo
  {journal} {Phys. Rev. D}\ }\textbf {\bibinfo {volume} {97}},\ \bibinfo
  {pages} {123532} (\bibinfo {year} {2018})},\ \Eprint
  {https://arxiv.org/abs/1804.08577} {arXiv:1804.08577 [gr-qc]} \BibitemShut
  {NoStop}%
\bibitem [{\citenamefont {Dom\`enech}(2020)}]{Domenech:2019quo}%
  \BibitemOpen
  \bibfield  {author} {\bibinfo {author} {\bibfnamefont {G.}~\bibnamefont
  {Dom\`enech}},\ }\bibfield  {title} {\bibinfo {title} {{Induced gravitational
  waves in a general cosmological background}},\ }\href
  {https://doi.org/10.1142/S0218271820500285} {\bibfield  {journal} {\bibinfo
  {journal} {Int. J. Mod. Phys. D}\ }\textbf {\bibinfo {volume} {29}},\
  \bibinfo {pages} {2050028} (\bibinfo {year} {2020})},\ \Eprint
  {https://arxiv.org/abs/1912.05583} {arXiv:1912.05583 [gr-qc]} \BibitemShut
  {NoStop}%
\bibitem [{\citenamefont {Dom\`enech}\ \emph {et~al.}(2020)\citenamefont
  {Dom\`enech}, \citenamefont {Pi},\ and\ \citenamefont
  {Sasaki}}]{Domenech:2020kqm}%
  \BibitemOpen
  \bibfield  {author} {\bibinfo {author} {\bibfnamefont {G.}~\bibnamefont
  {Dom\`enech}}, \bibinfo {author} {\bibfnamefont {S.}~\bibnamefont {Pi}},\
  and\ \bibinfo {author} {\bibfnamefont {M.}~\bibnamefont {Sasaki}},\
  }\bibfield  {title} {\bibinfo {title} {{Induced gravitational waves as a
  probe of thermal history of the universe}},\ }\href
  {https://doi.org/10.1088/1475-7516/2020/08/017} {\bibfield  {journal}
  {\bibinfo  {journal} {JCAP}\ }\textbf {\bibinfo {volume} {08}},\ \bibinfo
  {pages} {017}},\ \Eprint {https://arxiv.org/abs/2005.12314} {arXiv:2005.12314
  [gr-qc]} \BibitemShut {NoStop}%
\bibitem [{\citenamefont {Inomata}(2021)}]{Inomata:2021zel}%
  \BibitemOpen
  \bibfield  {author} {\bibinfo {author} {\bibfnamefont {K.}~\bibnamefont
  {Inomata}},\ }\bibfield  {title} {\bibinfo {title} {{Bound on induced
  gravitational waves during inflation era}},\ }\href
  {https://doi.org/10.1103/PhysRevD.104.123525} {\bibfield  {journal} {\bibinfo
   {journal} {Phys. Rev. D}\ }\textbf {\bibinfo {volume} {104}},\ \bibinfo
  {pages} {123525} (\bibinfo {year} {2021})},\ \Eprint
  {https://arxiv.org/abs/2109.06192} {arXiv:2109.06192 [astro-ph.CO]}
  \BibitemShut {NoStop}%
\bibitem [{\citenamefont {Fumagalli}\ \emph {et~al.}(2022)\citenamefont
  {Fumagalli}, \citenamefont {Palma}, \citenamefont {Renaux-Petel},
  \citenamefont {Sypsas}, \citenamefont {Witkowski},\ and\ \citenamefont
  {Zenteno}}]{Fumagalli:2021mpc}%
  \BibitemOpen
  \bibfield  {author} {\bibinfo {author} {\bibfnamefont {J.}~\bibnamefont
  {Fumagalli}}, \bibinfo {author} {\bibfnamefont {G.~A.}\ \bibnamefont
  {Palma}}, \bibinfo {author} {\bibfnamefont {S.}~\bibnamefont {Renaux-Petel}},
  \bibinfo {author} {\bibfnamefont {S.}~\bibnamefont {Sypsas}}, \bibinfo
  {author} {\bibfnamefont {L.~T.}\ \bibnamefont {Witkowski}},\ and\ \bibinfo
  {author} {\bibfnamefont {C.}~\bibnamefont {Zenteno}},\ }\bibfield  {title}
  {\bibinfo {title} {{Primordial gravitational waves from excited states}},\
  }\href {https://doi.org/10.1007/JHEP03(2022)196} {\bibfield  {journal}
  {\bibinfo  {journal} {JHEP}\ }\textbf {\bibinfo {volume} {03}},\ \bibinfo
  {pages} {196}},\ \Eprint {https://arxiv.org/abs/2111.14664} {arXiv:2111.14664
  [astro-ph.CO]} \BibitemShut {NoStop}%
\bibitem [{\citenamefont {Cai}\ \emph {et~al.}(2020)\citenamefont {Cai},
  \citenamefont {Pi},\ and\ \citenamefont {Sasaki}}]{Cai:2019cdl}%
  \BibitemOpen
  \bibfield  {author} {\bibinfo {author} {\bibfnamefont {R.-G.}\ \bibnamefont
  {Cai}}, \bibinfo {author} {\bibfnamefont {S.}~\bibnamefont {Pi}},\ and\
  \bibinfo {author} {\bibfnamefont {M.}~\bibnamefont {Sasaki}},\ }\bibfield
  {title} {\bibinfo {title} {{Universal infrared scaling of gravitational wave
  background spectra}},\ }\href {https://doi.org/10.1103/PhysRevD.102.083528}
  {\bibfield  {journal} {\bibinfo  {journal} {Phys. Rev. D}\ }\textbf {\bibinfo
  {volume} {102}},\ \bibinfo {pages} {083528} (\bibinfo {year} {2020})},\
  \Eprint {https://arxiv.org/abs/1909.13728} {arXiv:1909.13728 [astro-ph.CO]}
  \BibitemShut {NoStop}%
\bibitem [{\citenamefont {Atal}\ and\ \citenamefont
  {Dom\`enech}(2021)}]{Atal:2021jyo}%
  \BibitemOpen
  \bibfield  {author} {\bibinfo {author} {\bibfnamefont {V.}~\bibnamefont
  {Atal}}\ and\ \bibinfo {author} {\bibfnamefont {G.}~\bibnamefont
  {Dom\`enech}},\ }\bibfield  {title} {\bibinfo {title} {{Probing
  non-Gaussianities with the high frequency tail of induced gravitational
  waves}},\ }\href {https://doi.org/10.1088/1475-7516/2021/06/001} {\bibfield
  {journal} {\bibinfo  {journal} {JCAP}\ }\textbf {\bibinfo {volume} {06}},\
  \bibinfo {pages} {001}},\ \Eprint {https://arxiv.org/abs/2103.01056}
  {arXiv:2103.01056 [astro-ph.CO]} \BibitemShut {NoStop}%
\bibitem [{\citenamefont {Balaji}\ \emph {et~al.}(2022)\citenamefont {Balaji},
  \citenamefont {Domenech},\ and\ \citenamefont {Silk}}]{Balaji:2022dbi}%
  \BibitemOpen
  \bibfield  {author} {\bibinfo {author} {\bibfnamefont {S.}~\bibnamefont
  {Balaji}}, \bibinfo {author} {\bibfnamefont {G.}~\bibnamefont {Domenech}},\
  and\ \bibinfo {author} {\bibfnamefont {J.}~\bibnamefont {Silk}},\ }\bibfield
  {title} {\bibinfo {title} {{Induced gravitational waves from slow-roll
  inflation after an enhancing phase}},\ }\href
  {https://doi.org/10.1088/1475-7516/2022/09/016} {\bibfield  {journal}
  {\bibinfo  {journal} {JCAP}\ }\textbf {\bibinfo {volume} {09}},\ \bibinfo
  {pages} {016}},\ \Eprint {https://arxiv.org/abs/2205.01696} {arXiv:2205.01696
  [astro-ph.CO]} \BibitemShut {NoStop}%
\bibitem [{\citenamefont {Fumagalli}\ \emph
  {et~al.}(2020{\natexlab{b}})\citenamefont {Fumagalli}, \citenamefont
  {Renaux-Petel},\ and\ \citenamefont {Witkowski}}]{Fumagalli:2020nvq}%
  \BibitemOpen
  \bibfield  {author} {\bibinfo {author} {\bibfnamefont {J.}~\bibnamefont
  {Fumagalli}}, \bibinfo {author} {\bibfnamefont {S.}~\bibnamefont
  {Renaux-Petel}},\ and\ \bibinfo {author} {\bibfnamefont {L.~T.}\ \bibnamefont
  {Witkowski}},\ }\bibfield  {title} {\bibinfo {title} {{Oscillations in the
  stochastic gravitational wave background from sharp features and particle
  production during inflation}},\ }\href@noop {} {\  (\bibinfo {year}
  {2020}{\natexlab{b}})},\ \Eprint {https://arxiv.org/abs/2012.02761}
  {arXiv:2012.02761 [astro-ph.CO]} \BibitemShut {NoStop}%
\bibitem [{\citenamefont {Fumagalli}\ \emph {et~al.}(2021)\citenamefont
  {Fumagalli}, \citenamefont {Renaux-Petel},\ and\ \citenamefont
  {Witkowski}}]{Fumagalli:2021cel}%
  \BibitemOpen
  \bibfield  {author} {\bibinfo {author} {\bibfnamefont {J.}~\bibnamefont
  {Fumagalli}}, \bibinfo {author} {\bibfnamefont {S.}~\bibnamefont
  {Renaux-Petel}},\ and\ \bibinfo {author} {\bibfnamefont {L.~T.}\ \bibnamefont
  {Witkowski}},\ }\bibfield  {title} {\bibinfo {title} {{Resonant features in
  the stochastic gravitational wave background}},\ }\href@noop {} {\  (\bibinfo
  {year} {2021})},\ \Eprint {https://arxiv.org/abs/2105.06481}
  {arXiv:2105.06481 [astro-ph.CO]} \BibitemShut {NoStop}%
\bibitem [{\citenamefont {Witkowski}\ \emph {et~al.}(2022)\citenamefont
  {Witkowski}, \citenamefont {Dom\`enech}, \citenamefont {Fumagalli},\ and\
  \citenamefont {Renaux-Petel}}]{Witkowski:2021raz}%
  \BibitemOpen
  \bibfield  {author} {\bibinfo {author} {\bibfnamefont {L.~T.}\ \bibnamefont
  {Witkowski}}, \bibinfo {author} {\bibfnamefont {G.}~\bibnamefont
  {Dom\`enech}}, \bibinfo {author} {\bibfnamefont {J.}~\bibnamefont
  {Fumagalli}},\ and\ \bibinfo {author} {\bibfnamefont {S.}~\bibnamefont
  {Renaux-Petel}},\ }\bibfield  {title} {\bibinfo {title} {{Expansion
  history-dependent oscillations in the scalar-induced gravitational wave
  background}},\ }\href {https://doi.org/10.1088/1475-7516/2022/05/028}
  {\bibfield  {journal} {\bibinfo  {journal} {JCAP}\ }\textbf {\bibinfo
  {volume} {05}}\bibfield  {number} {\bibinfo  {number} { (05)},\ \bibinfo
  {pages} {028}},\ }\Eprint {https://arxiv.org/abs/2110.09480}
  {arXiv:2110.09480 [astro-ph.CO]} \BibitemShut {NoStop}%
\bibitem [{\citenamefont {Braglia}\ \emph
  {et~al.}(2020{\natexlab{b}})\citenamefont {Braglia}, \citenamefont {Chen},\
  and\ \citenamefont {Hazra}}]{Braglia:2020taf}%
  \BibitemOpen
  \bibfield  {author} {\bibinfo {author} {\bibfnamefont {M.}~\bibnamefont
  {Braglia}}, \bibinfo {author} {\bibfnamefont {X.}~\bibnamefont {Chen}},\ and\
  \bibinfo {author} {\bibfnamefont {D.~K.}\ \bibnamefont {Hazra}},\ }\bibfield
  {title} {\bibinfo {title} {{Probing Primordial Features with the Stochastic
  Gravitational Wave Background}},\ }\href@noop {} {\  (\bibinfo {year}
  {2020}{\natexlab{b}})},\ \Eprint {https://arxiv.org/abs/2012.05821}
  {arXiv:2012.05821 [astro-ph.CO]} \BibitemShut {NoStop}%
\bibitem [{\citenamefont {Franciolini}\ \emph
  {et~al.}(2023{\natexlab{c}})\citenamefont {Franciolini}, \citenamefont
  {Iovino}, \citenamefont {Taoso},\ and\ \citenamefont
  {Urbano}}]{Franciolini:2023lgy}%
  \BibitemOpen
  \bibfield  {author} {\bibinfo {author} {\bibfnamefont {G.}~\bibnamefont
  {Franciolini}}, \bibinfo {author} {\bibfnamefont {A.}~\bibnamefont {Iovino},
  \bibfnamefont {Junior.}}, \bibinfo {author} {\bibfnamefont {M.}~\bibnamefont
  {Taoso}},\ and\ \bibinfo {author} {\bibfnamefont {A.}~\bibnamefont
  {Urbano}},\ }\bibfield  {title} {\bibinfo {title} {{One loop to rule them
  all: Perturbativity in the presence of ultra slow-roll dynamics}},\
  }\href@noop {} {\  (\bibinfo {year} {2023}{\natexlab{c}})},\ \Eprint
  {https://arxiv.org/abs/2305.03491} {arXiv:2305.03491 [astro-ph.CO]}
  \BibitemShut {NoStop}%
\bibitem [{\citenamefont {Fumagalli}\ \emph {et~al.}(2023)\citenamefont
  {Fumagalli}, \citenamefont {Bhattacharya}, \citenamefont {Peloso},
  \citenamefont {Renaux-Petel},\ and\ \citenamefont
  {Witkowski}}]{Fumagalli:2023loc}%
  \BibitemOpen
  \bibfield  {author} {\bibinfo {author} {\bibfnamefont {J.}~\bibnamefont
  {Fumagalli}}, \bibinfo {author} {\bibfnamefont {S.}~\bibnamefont
  {Bhattacharya}}, \bibinfo {author} {\bibfnamefont {M.}~\bibnamefont
  {Peloso}}, \bibinfo {author} {\bibfnamefont {S.}~\bibnamefont
  {Renaux-Petel}},\ and\ \bibinfo {author} {\bibfnamefont {L.~T.}\ \bibnamefont
  {Witkowski}},\ }\bibfield  {title} {\bibinfo {title} {{One-loop infrared
  rescattering by enhanced scalar fluctuations during inflation}},\ }\href@noop
  {} {\  (\bibinfo {year} {2023})},\ \Eprint {https://arxiv.org/abs/2307.08358}
  {arXiv:2307.08358 [astro-ph.CO]} \BibitemShut {NoStop}%
\bibitem [{\citenamefont {Kristiano}\ and\ \citenamefont
  {Yokoyama}(2022)}]{Kristiano:2022maq}%
  \BibitemOpen
  \bibfield  {author} {\bibinfo {author} {\bibfnamefont {J.}~\bibnamefont
  {Kristiano}}\ and\ \bibinfo {author} {\bibfnamefont {J.}~\bibnamefont
  {Yokoyama}},\ }\bibfield  {title} {\bibinfo {title} {{Ruling Out Primordial
  Black Hole Formation From Single-Field Inflation}},\ }\href@noop {} {\
  (\bibinfo {year} {2022})},\ \Eprint {https://arxiv.org/abs/2211.03395}
  {arXiv:2211.03395 [hep-th]} \BibitemShut {NoStop}%
\bibitem [{\citenamefont {Riotto}(2023{\natexlab{a}})}]{Riotto:2023hoz}%
  \BibitemOpen
  \bibfield  {author} {\bibinfo {author} {\bibfnamefont {A.}~\bibnamefont
  {Riotto}},\ }\bibfield  {title} {\bibinfo {title} {{The Primordial Black Hole
  Formation from Single-Field Inflation is Not Ruled Out}},\ }\href@noop {} {\
  (\bibinfo {year} {2023}{\natexlab{a}})},\ \Eprint
  {https://arxiv.org/abs/2301.00599} {arXiv:2301.00599 [astro-ph.CO]}
  \BibitemShut {NoStop}%
\bibitem [{\citenamefont {Kristiano}\ and\ \citenamefont
  {Yokoyama}(2023)}]{Kristiano:2023scm}%
  \BibitemOpen
  \bibfield  {author} {\bibinfo {author} {\bibfnamefont {J.}~\bibnamefont
  {Kristiano}}\ and\ \bibinfo {author} {\bibfnamefont {J.}~\bibnamefont
  {Yokoyama}},\ }\bibfield  {title} {\bibinfo {title} {{Response to criticism
  on ''Ruling Out Primordial Black Hole Formation From Single-Field
  Inflation'': A note on bispectrum and one-loop correction in single-field
  inflation with primordial black hole formation}},\ }\href@noop {} {\
  (\bibinfo {year} {2023})},\ \Eprint {https://arxiv.org/abs/2303.00341}
  {arXiv:2303.00341 [hep-th]} \BibitemShut {NoStop}%
\bibitem [{\citenamefont {Riotto}(2023{\natexlab{b}})}]{Riotto:2023gpm}%
  \BibitemOpen
  \bibfield  {author} {\bibinfo {author} {\bibfnamefont {A.}~\bibnamefont
  {Riotto}},\ }\bibfield  {title} {\bibinfo {title} {{The Primordial Black Hole
  Formation from Single-Field Inflation is Still Not Ruled Out}},\ }\href@noop
  {} {\  (\bibinfo {year} {2023}{\natexlab{b}})},\ \Eprint
  {https://arxiv.org/abs/2303.01727} {arXiv:2303.01727 [astro-ph.CO]}
  \BibitemShut {NoStop}%
\bibitem [{\citenamefont {Firouzjahi}(2023)}]{Firouzjahi:2023aum}%
  \BibitemOpen
  \bibfield  {author} {\bibinfo {author} {\bibfnamefont {H.}~\bibnamefont
  {Firouzjahi}},\ }\bibfield  {title} {\bibinfo {title} {{One-loop Corrections
  in Power Spectrum in Single Field Inflation}},\ }\href@noop {} {\  (\bibinfo
  {year} {2023})},\ \Eprint {https://arxiv.org/abs/2303.12025}
  {arXiv:2303.12025 [astro-ph.CO]} \BibitemShut {NoStop}%
\bibitem [{\citenamefont {Firouzjahi}\ and\ \citenamefont
  {Riotto}(2023)}]{Firouzjahi:2023ahg}%
  \BibitemOpen
  \bibfield  {author} {\bibinfo {author} {\bibfnamefont {H.}~\bibnamefont
  {Firouzjahi}}\ and\ \bibinfo {author} {\bibfnamefont {A.}~\bibnamefont
  {Riotto}},\ }\bibfield  {title} {\bibinfo {title} {{Primordial Black Holes
  and Loops in Single-Field Inflation}},\ }\href@noop {} {\  (\bibinfo {year}
  {2023})},\ \Eprint {https://arxiv.org/abs/2304.07801} {arXiv:2304.07801
  [astro-ph.CO]} \BibitemShut {NoStop}%
\bibitem [{\citenamefont {Cheng}\ \emph {et~al.}(2023)\citenamefont {Cheng},
  \citenamefont {Lee},\ and\ \citenamefont {Ng}}]{Cheng:2023ikq}%
  \BibitemOpen
  \bibfield  {author} {\bibinfo {author} {\bibfnamefont {S.-L.}\ \bibnamefont
  {Cheng}}, \bibinfo {author} {\bibfnamefont {D.-S.}\ \bibnamefont {Lee}},\
  and\ \bibinfo {author} {\bibfnamefont {K.-W.}\ \bibnamefont {Ng}},\
  }\bibfield  {title} {\bibinfo {title} {{Primordial perturbations from
  ultra-slow-roll single-field inflation with quantum loop effects}},\
  }\href@noop {} {\  (\bibinfo {year} {2023})},\ \Eprint
  {https://arxiv.org/abs/2305.16810} {arXiv:2305.16810 [astro-ph.CO]}
  \BibitemShut {NoStop}%
\bibitem [{\citenamefont {Fumagalli}(2023)}]{Fumagalli:2023hpa}%
  \BibitemOpen
  \bibfield  {author} {\bibinfo {author} {\bibfnamefont {J.}~\bibnamefont
  {Fumagalli}},\ }\bibfield  {title} {\bibinfo {title} {{Absence of one-loop
  effects on large scales from small scales in non-slow-roll dynamics}},\
  }\href@noop {} {\  (\bibinfo {year} {2023})},\ \Eprint
  {https://arxiv.org/abs/2305.19263} {arXiv:2305.19263 [astro-ph.CO]}
  \BibitemShut {NoStop}%
\bibitem [{\citenamefont {Domenech}\ \emph {et~al.}(2017)\citenamefont
  {Domenech}, \citenamefont {Gong},\ and\ \citenamefont
  {Sasaki}}]{Domenech:2016zxn}%
  \BibitemOpen
  \bibfield  {author} {\bibinfo {author} {\bibfnamefont {G.}~\bibnamefont
  {Domenech}}, \bibinfo {author} {\bibfnamefont {J.-O.}\ \bibnamefont {Gong}},\
  and\ \bibinfo {author} {\bibfnamefont {M.}~\bibnamefont {Sasaki}},\
  }\bibfield  {title} {\bibinfo {title} {{Consistency relation and inflaton
  field redefinition in the \ensuremath{\delta}N formalism}},\ }\href
  {https://doi.org/10.1016/j.physletb.2017.04.014} {\bibfield  {journal}
  {\bibinfo  {journal} {Phys. Lett. B}\ }\textbf {\bibinfo {volume} {769}},\
  \bibinfo {pages} {413} (\bibinfo {year} {2017})},\ \Eprint
  {https://arxiv.org/abs/1606.03343} {arXiv:1606.03343 [astro-ph.CO]}
  \BibitemShut {NoStop}%
\bibitem [{\citenamefont {Maldacena}(2003)}]{Maldacena:2002vr}%
  \BibitemOpen
  \bibfield  {author} {\bibinfo {author} {\bibfnamefont {J.~M.}\ \bibnamefont
  {Maldacena}},\ }\bibfield  {title} {\bibinfo {title} {{Non-Gaussian features
  of primordial fluctuations in single field inflationary models}},\ }\href
  {https://doi.org/10.1088/1126-6708/2003/05/013} {\bibfield  {journal}
  {\bibinfo  {journal} {JHEP}\ }\textbf {\bibinfo {volume} {05}},\ \bibinfo
  {pages} {013}},\ \Eprint {https://arxiv.org/abs/astro-ph/0210603}
  {arXiv:astro-ph/0210603} \BibitemShut {NoStop}%
\bibitem [{\citenamefont {Adshead}\ and\ \citenamefont
  {Hu}(2014)}]{Adshead:2014sga}%
  \BibitemOpen
  \bibfield  {author} {\bibinfo {author} {\bibfnamefont {P.}~\bibnamefont
  {Adshead}}\ and\ \bibinfo {author} {\bibfnamefont {W.}~\bibnamefont {Hu}},\
  }\bibfield  {title} {\bibinfo {title} {{Bounds on nonadiabatic evolution in
  single-field inflation}},\ }\href
  {https://doi.org/10.1103/PhysRevD.89.083531} {\bibfield  {journal} {\bibinfo
  {journal} {Phys. Rev. D}\ }\textbf {\bibinfo {volume} {89}},\ \bibinfo
  {pages} {083531} (\bibinfo {year} {2014})},\ \Eprint
  {https://arxiv.org/abs/1402.1677} {arXiv:1402.1677 [astro-ph.CO]}
  \BibitemShut {NoStop}%
\bibitem [{\citenamefont {Cannone}\ \emph {et~al.}(2014)\citenamefont
  {Cannone}, \citenamefont {Bartolo},\ and\ \citenamefont
  {Matarrese}}]{Cannone:2014qna}%
  \BibitemOpen
  \bibfield  {author} {\bibinfo {author} {\bibfnamefont {D.}~\bibnamefont
  {Cannone}}, \bibinfo {author} {\bibfnamefont {N.}~\bibnamefont {Bartolo}},\
  and\ \bibinfo {author} {\bibfnamefont {S.}~\bibnamefont {Matarrese}},\
  }\bibfield  {title} {\bibinfo {title} {{Perturbative Unitarity of
  Inflationary Models with Features}},\ }\href
  {https://doi.org/10.1103/PhysRevD.89.127301} {\bibfield  {journal} {\bibinfo
  {journal} {Phys. Rev. D}\ }\textbf {\bibinfo {volume} {89}},\ \bibinfo
  {pages} {127301} (\bibinfo {year} {2014})},\ \Eprint
  {https://arxiv.org/abs/1402.2258} {arXiv:1402.2258 [astro-ph.CO]}
  \BibitemShut {NoStop}%
\bibitem [{\citenamefont {Achucarro}\ \emph {et~al.}(2014)\citenamefont
  {Achucarro}, \citenamefont {Atal}, \citenamefont {Hu}, \citenamefont
  {Ortiz},\ and\ \citenamefont {Torrado}}]{Achucarro:2014msa}%
  \BibitemOpen
  \bibfield  {author} {\bibinfo {author} {\bibfnamefont {A.}~\bibnamefont
  {Achucarro}}, \bibinfo {author} {\bibfnamefont {V.}~\bibnamefont {Atal}},
  \bibinfo {author} {\bibfnamefont {B.}~\bibnamefont {Hu}}, \bibinfo {author}
  {\bibfnamefont {P.}~\bibnamefont {Ortiz}},\ and\ \bibinfo {author}
  {\bibfnamefont {J.}~\bibnamefont {Torrado}},\ }\bibfield  {title} {\bibinfo
  {title} {{Inflation with moderately sharp features in the speed of sound:
  Generalized slow roll and in-in formalism for power spectrum and
  bispectrum}},\ }\href {https://doi.org/10.1103/PhysRevD.90.023511} {\bibfield
   {journal} {\bibinfo  {journal} {Phys. Rev. D}\ }\textbf {\bibinfo {volume}
  {90}},\ \bibinfo {pages} {023511} (\bibinfo {year} {2014})},\ \Eprint
  {https://arxiv.org/abs/1404.7522} {arXiv:1404.7522 [astro-ph.CO]}
  \BibitemShut {NoStop}%
\bibitem [{\citenamefont {Pattison}\ \emph {et~al.}(2021)\citenamefont
  {Pattison}, \citenamefont {Vennin}, \citenamefont {Wands},\ and\
  \citenamefont {Assadullahi}}]{Pattison:2021oen}%
  \BibitemOpen
  \bibfield  {author} {\bibinfo {author} {\bibfnamefont {C.}~\bibnamefont
  {Pattison}}, \bibinfo {author} {\bibfnamefont {V.}~\bibnamefont {Vennin}},
  \bibinfo {author} {\bibfnamefont {D.}~\bibnamefont {Wands}},\ and\ \bibinfo
  {author} {\bibfnamefont {H.}~\bibnamefont {Assadullahi}},\ }\bibfield
  {title} {\bibinfo {title} {{Ultra-slow-roll inflation with quantum
  diffusion}},\ }\href {https://doi.org/10.1088/1475-7516/2021/04/080}
  {\bibfield  {journal} {\bibinfo  {journal} {JCAP}\ }\textbf {\bibinfo
  {volume} {04}},\ \bibinfo {pages} {080}},\ \Eprint
  {https://arxiv.org/abs/2101.05741} {arXiv:2101.05741 [astro-ph.CO]}
  \BibitemShut {NoStop}%
\bibitem [{\citenamefont {Lyth}\ and\ \citenamefont
  {Wands}(2002)}]{Lyth:2001nq}%
  \BibitemOpen
  \bibfield  {author} {\bibinfo {author} {\bibfnamefont {D.~H.}\ \bibnamefont
  {Lyth}}\ and\ \bibinfo {author} {\bibfnamefont {D.}~\bibnamefont {Wands}},\
  }\bibfield  {title} {\bibinfo {title} {{Generating the curvature perturbation
  without an inflaton}},\ }\href
  {https://doi.org/10.1016/S0370-2693(01)01366-1} {\bibfield  {journal}
  {\bibinfo  {journal} {Phys. Lett. B}\ }\textbf {\bibinfo {volume} {524}},\
  \bibinfo {pages} {5} (\bibinfo {year} {2002})},\ \Eprint
  {https://arxiv.org/abs/hep-ph/0110002} {arXiv:hep-ph/0110002} \BibitemShut
  {NoStop}%
\bibitem [{\citenamefont {Enqvist}\ and\ \citenamefont
  {Sloth}(2002)}]{Enqvist:2001zp}%
  \BibitemOpen
  \bibfield  {author} {\bibinfo {author} {\bibfnamefont {K.}~\bibnamefont
  {Enqvist}}\ and\ \bibinfo {author} {\bibfnamefont {M.~S.}\ \bibnamefont
  {Sloth}},\ }\bibfield  {title} {\bibinfo {title} {{Adiabatic CMB
  perturbations in pre - big bang string cosmology}},\ }\href
  {https://doi.org/10.1016/S0550-3213(02)00043-3} {\bibfield  {journal}
  {\bibinfo  {journal} {Nucl. Phys. B}\ }\textbf {\bibinfo {volume} {626}},\
  \bibinfo {pages} {395} (\bibinfo {year} {2002})},\ \Eprint
  {https://arxiv.org/abs/hep-ph/0109214} {arXiv:hep-ph/0109214} \BibitemShut
  {NoStop}%
\bibitem [{\citenamefont {Moroi}\ and\ \citenamefont
  {Takahashi}(2002)}]{Moroi:2002rd}%
  \BibitemOpen
  \bibfield  {author} {\bibinfo {author} {\bibfnamefont {T.}~\bibnamefont
  {Moroi}}\ and\ \bibinfo {author} {\bibfnamefont {T.}~\bibnamefont
  {Takahashi}},\ }\bibfield  {title} {\bibinfo {title} {{Cosmic density
  perturbations from late decaying scalar condensations}},\ }\href
  {https://doi.org/10.1103/PhysRevD.66.063501} {\bibfield  {journal} {\bibinfo
  {journal} {Phys. Rev. D}\ }\textbf {\bibinfo {volume} {66}},\ \bibinfo
  {pages} {063501} (\bibinfo {year} {2002})},\ \Eprint
  {https://arxiv.org/abs/hep-ph/0206026} {arXiv:hep-ph/0206026} \BibitemShut
  {NoStop}%
\end{thebibliography}%

\end{document}